\newenvironment{resumo} {
  \cleardoublepage
  \setsinglecolumn
  \chapter*{\centering \Large Resumo}
  \thispagestyle{empty}
}

\newcommand\blankpage{%
    \null
    \thispagestyle{empty}%
    \addtocounter{page}{-1}%
    \newpage}

\documentclass[a4paper,12pt,numbered,twoside,print,index]{Classes/PhDThesisPSnPDF}

\usepackage[stable]{footmisc}
\usepackage{algorithm}
\usepackage[noend]{algpseudocode}

\input{Preamble/preamble}

\title{Entity Retrieval \\and\\ Text Mining\\for\\ \hspace{ 1cm}Online Reputation Monitoring}

\author{Pedro dos Santos Saleiro da Cruz}

\dept{Departamento de Engenharia Inform\'atica}

\university{Faculdade de Engenharia da Universidade do Porto}
\crest{\includegraphics[width=0.6\textwidth]{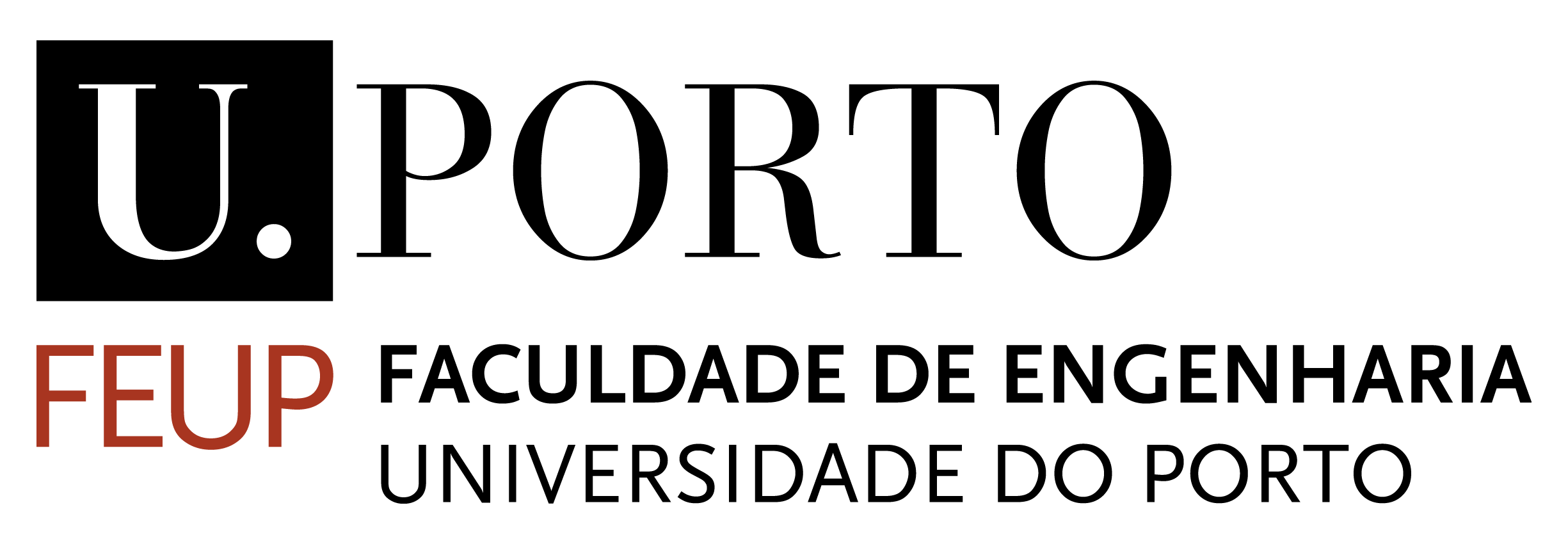}}

\renewcommand{\submissiontext}{In partial fulfillment of requirements for the degree of}

\degreetitle{Doctor in Informatics Engineering}

\college{FEUP}

\degreedate{2017} 

\ifdefineAbstract
\fi

\ifdefineChapter
\fi

\begin{document}

\frontmatter

\maketitle

\blankpage
\blankpage
\begin{flushleft}
\ \\
Supervisor: Dr. Carlos Soares \\
Co-Supervisor: Dr. Eduarda Mendes Rodrigues \\
\hfill \break
Faculdade de Engenharia \\
Universidade do Porto \\
Rua Dr. Roberto Frias, s/n \\
4200-465 Porto, Portugal
\end{flushleft}
\ 
\\[20\baselineskip]
\begin{center}
Copyright \copyright \ 2017 by Pedro Saleiro
\end{center}
\blankpage
\blankpage
\begin{flushleft}
\ \\
\textbf{Doctoral Committee}:\\
\ \\
\ \\
Dr. Eugénio Oliveira, Full Professor at FEUP, University of Porto\\
\ \\
Dr. Mark Carman, Senior Lecturer at Monash University\\
\ \\
Dr. Bruno Martins, Assistant Professor at IST, University of Lisbon\\
\ \\
Dr. Luís Torgo, Associate Professor at FCUP, University of Porto\\
\ \\
Dr. Carlos Soares, Associate Professor at FEUP, University of Porto\\
\hfill \break
\end{flushleft}
\blankpage
\blankpage


\begin{dedication} 

\textit{Esta tese é dedicada à minha Mãe,\\ Maria de Lurdes,\\
pelo seu amor e dedicação constantes.\\ }
\end{dedication}

\begin{acknowledgements}

First, I would like to thank everybody that contributed somehow to this work, from co-authors to reviewers, colleagues and faculty staff. I will probably forget to mention someone in particular and I sincerely apologize for that. This work was funded by SAPO Labs, FCT and Microsoft Research. Without their financial support, I would not be able to conclude this thesis. 

I am deeply grateful to my supervisor, Carlos Soares. Although I was not working in meta-learning :) Carlos always showed a genuine  enthusiasm about this work. Thank you for giving me the freedom to grow independently as a researcher and to pursuit my own ideas, even in the moments I was not delivering at the rhythm you expected. I believe you made me more pragmatic after all. We had really interesting and thorough discussions during the last 5 years.  We have not tried all the cool ideas but I hope we will do it someday.  Last, I would like to thank you for all the support and protection, as well as, for believing that I should expand my horizons to the US! I will send you a postcard from Chicago!

I am also thankful to Eduarda Mendes Rodrigues, co-supervisor of this work. It all started with you. Thank you for receiving me at FEUP back in October 2012. I still remember the day we drew the first draft of our framework for ORM.  You always have encouraged me and your positive feedback was a source of inspiration and motivation. Even at distance you have always been available when I needed. I must also mention your decisive role in helping me pursuing a Summer internship in a top notch place such as Microsoft Research.

Being a graduate student is an opportunity to collaborate with new and inspirational people and that was what happened when I had the chance to start working with Natasa Milic-Frayling at Microsoft Research. When we are around Natasa we believe we can make things happen. Thank you for your patience, support and motivation. I hope we can keep our collaboration for many years.

I show my gratitude to Prof. Eugénio Oliveira who helped me with a smooth transition to LIACC and always had the door open to discuss my issues. Furthermore, Prof. Eugénio was very enthusiastic about my work even not being my supervisor and I sincerely appreciate that. Thank you for your advices and I will miss our conversations about the past and future of AI. 

I wish to thank Luís Sarmento for introducing me to the world of Data Science, and Text Mining in particular. I just regret we did not had the chance to collaborate more often.   

I must thank two special friends that are also graduate students, Jorge Teixeira and Damião Rodrigues. Jorge, you were my ``brother in arms'' throughout this journey and I will always be thankful. Damião, my friend and colleague for more than 10 years, is the friend I searched for sharing the ups and downs of being graduate student. Thank you for your support and motivation.

I would like to thank Cristina Ribeiro for the administrative support regarding my funding throughout these years. I must also address Rosaldo Rossetti for believing in my abilities and for starting our productive collaboration, and of course, for being a really enjoyable and funny colleague. Arian Pasquali also deserves a personal mention for all the support, as well as, Luís Gomes, Luís Rei, Sílvio Amir, Tiago Cunha and Gustavo Laboreiro. I would like also to thank all the members of the POPSTAR project, specially Pedro Magalhães.

Nesta hora não poderia deixar de estender os agradecimentos aos amigos e à família. Em especial queria referir o grupo do Rumo ao Penta onde todos os \textit{passarões} me proporcionam grandes momentos de boa disposição sem os quais seria impossível abstrair-me dos problemas do dia a dia. Um grande abraço para o André, João Miguel, Jorge e Márcio. Queria também deixar aqui uma palavra para a minha prima Xana pela amizade desde sempre.

Como é óbvio tenho imenso a agradecer à minha Mãe, a quem dedico esta tese. Obrigado pelo amor, dedicação e pela liberdade que sempre me deste em todas as minhas escolhas. Como não deixaria de ser, sempre foste uma entusiasta deste meu desafio que agora chega ao fim. Parabéns a ti! Deixo também um beijinho à minha irmã Guida, ao querido Tomás e ao bebé Diogo que ainda não conheci mas a vida de emigrante tem destas coisas.

E por fim, deixo o meu sentido agradecimento à minha namorada, Maria. Foste crucial nesta caminhada. Ter-te ao meu lado deu-me força todos os dias para avançar mais um pouco. Sei que este trabalho comprometeu muito o nosso tempo a dois mas agradeço-te a compreensão e os incentivos constantes para levar isto até ao fim. Prometo compensar no futuro :) Ah e claro tenho que agradecer ao Bobby pela companhia que me fez durante a escrita e por me conseguir fazer sorrir mesmo quando a vida é madrasta.

\end{acknowledgements}

\begin{abstract}
Online Reputation Monitoring (ORM) is concerned with the use of computational tools to measure the reputation of entities online, such as politicians or companies. In practice, current ORM methods are constrained to the generation of data analytics reports, which aggregate statistics of popularity and sentiment on social media.
We argue that this format is too restrictive as end users often like to have the flexibility to search for entity-centric information that is not available in predefined charts. 

As such, we propose the inclusion of entity retrieval capabilities as a first step towards the extension of current ORM capabilities. However, an entity's reputation is also influenced by the entity's relationships with other entities. Therefore, we address the problem of Entity-Relationship (E-R) retrieval in which the goal is to search for multiple connected entities. This is a challenging problem which traditional entity search systems cannot cope with.

Besides E-R retrieval we also believe ORM would benefit of text-based entity-centric prediction capabilities, such as predicting entity popularity on social media based on news events or the outcome of political surveys. However, none of these tasks can provide useful results if there is no effective entity disambiguation and sentiment analysis tailored to the context of ORM. 

Consequently, this thesis address two computational problems in Online Reputation Monitoring: Entity Retrieval and Text Mining. We researched and developed methods to extract, retrieve and predict entity-centric information spread across the Web. 

We proposed a new probabilistic modeling of the problem of E-R retrieval together with two fusion-based design patterns for creating representations of both entities and relationships. Furthermore, we propose the Entity-Relationship Dependence Model, a novel early-fusion supervised model based on the Markov Random Field framework for Retrieval. Together with a new semi-automatic method to create test collections for E-R retrieval, we released a new test collection for that purpose that will foster research in this area. We performed experiments at scale with results showing that it is possible to perform E-R retrieval without using fix and pre-defined entity and relationship types, enabling a wide range of queries to be addressed.

We tackled Entity Filtering and Financial Sentiment Analysis using a supervised learning approach and studied several possible features for that purpose. We participated in two well known external competitions on both tasks, obtaining state-of-the-art performance. Moreover, we performed analysis of the predictive power of a wide set of signals extracted from online news to predict the popularity of entities on Twitter. We also studied several sentiment aggregate functions on Twitter to study the feasibility of using entity-centric sentiment on social media to predict political opinion polls. 

Finally, we created and released an adaptable Entity Retrieval and Text Mining framework that puts together all the building blocks necessary to perform ORM and can be reused in multiple application scenarios, from computational journalism to politics and finance. This framework is able to collect texts from online media, identify entities of interest, perform entity and E-R retrieval as well as classify sentiment polarity and intensity. It supports multiple data aggregation methods together with visualization and modeling techniques that can be used for both descriptive and predictive analytics.

\end{abstract}

\begin{resumo}
A Monitorização da Reputação Online (MRO) consiste na utilização de ferramentas computacionais para medir a reputação de entidades online, como por exemplo, políticos ou empresas. Na prática, os métodos actuais de MRO estão restringidos à produção de relatórios constituídos por análises de dados, tais como estatísticas agregadas da popularidade e do sentimento nos media sociais. Consideramos que esta prática é demasiado restritiva uma vez que os utilizadores finais das plataformas MRO desejam frequentemente ter a flexibilidade que lhes permita pesquisar por informação centrada nas entidades que vai além da disponibilizada nos gráficos pré-definidos.

Por conseguinte, propomos a inclusão da capacidade de recuperação de entidades como um primeiro passo no sentido de estender as o estado atual das ferramentas de MRO. No entanto, a reputação de uma dada entidade também é influenciada pelas relações desta com outras entidades. Neste sentido, propomo-nos a tratar do problema de recuperação de entidade-relações (E-R) onde o objectivo consiste na pesquisa por múltiplas entidades relacionadas entre si. Trata-se de um desafio que os sitemas tradicionais de recuperação de entidades ainda não são capazes de lidar. 

Para além da recuperação E-R, também acreditamos que a MRO iria beneficiar da capacidade de efectuar previsões baseadas em texto e centradas nas entidades, como por exemplo a previsão da popularidade de entidades nos media sociais utilizando eventos retratados nas notícias ou o resultado de sondagens. No entanto, nenhuma destas tarefas terá sucesso e utilidade se não houver a capacidade efetiva de desambiguar entidades mencionadas nos textos, assim como uma análise de sentimento específica para o contexto da MRO.

Consequentemente, esta tese trata dois problemas computacionais da Monitorização da Reputação Online: Recuperação de Entidades e Prospeção de Texto. Investigámos e desenvolvemos métodos para extrair, recuperar e prever informação centrada em entidades e espalhada pela Internet.

Propomos um novo modelo probabilístico do problema de recuperação E-R conjuntamente com dois padrões de desenho baseados em fusão de texto para criar representações de entidades e relações. Propomos também o Modelo de Dependência Entitdade-Relação (MDER), um novo modelo supervisionado de fusão antecipada baseado no Campo Aleatório de Markov para a Recuperação de Informação. Conjutamente com um novo método semi-automático de geração de coleções de teste para recuperação E-R, lançamos uma nova coleção de teste com esse propósito que irá fomentar a investigação nesta área. Efetuamos experiências de grande escala e os resultados mostram que é possível realizar recuperação E-R sem utilizar tipos fixos e pré-definidos de entidades e relações, o que permite atuar sobre o conjunto alargado de pesquisas.

Tratamos também das tarefas de Filtragem de Entidades e Análise de Sentimento Financeiro utilizando uma abordagem de aprendizagem supervisionada em que estudamos várias características para esse fim. Participámos em duas competições exterrnas em ambas as tarefas, atingindo resultados ao nível do estado da arte. Além disso, realizámos uma análise do poder preditivo de um grande conjunto de sinais extraídos das notícias online para parever a popularidade de entidades no Twitter. Assim como, um estudo de várias funções de agregação de sentimento do Twitter para estudar a praticabilidade de utilizar informação de sentimento nos media sociais para prever sondagens eleitorais.

Finalmente, criámos e disponibilizámos uma plataforma de recuperação de entidades e prospeção de texto que conjuga todos os blocos necessários para a realização de MRO. Pode ser reutilizada em diversos cenários de aplicação, desde o jornalismo computacional à política e finança. Esta plataforma é capaz de recolher textos dos media online, identificar entidades alvo, efectuar recuperação de entidades e relaçãos, assim como classificar sentimento e intensidade associada. Suporta vários métodos de agregação de dados e juntamente com métodos de visualização e previsão pode ser utilizada tanto para análises descritivas como preditivas.

\end{resumo}


\tableofcontents

\listoffigures

\listoftables



\mainmatter
\chapter{Introduction}
\label{ch:intro}

Nowadays, people have pervasive access to connected devices, applications and services that enable them to obtain and share information almost instantly, on a 24/7 basis. With Social Media growing at an astonishing speed, user opinions about people, companies and products quickly spread over large communities. Consequently, companies and personalities are under thorough scrutiny, with every event and every statement potentially observed and evaluated by a global audience, which reflects one's perceived reputation. 

Van Riel and Fombrun~\cite{van2007essentials} define reputation as the ``overall assessment of organizations by their stakeholders.'' The authors use the term \emph{organization} in the definition, but it may as well apply to individuals (e.g. politicians) or products (e.g. mobile phone brands). A stakeholder is someone who has some relationship with the organization, such as employees, customers or shareholders. This definition and other similar ones \cite{atvesson1990organization}, focus on the perspective that reputation represents perceptions that others have on the target entity. 

However, the rise of Social Media and online news publishing has brought about wider public awareness about the entities' activities, influencing people's perceptions about their reputation.
While traditional reputation analysis is mostly manual and focused on particular entities, with online media it is possible to automate much of the process of collecting, preparing and understanding large streams of content, to identify facts and opinions about a much wider set of entities. Online Reputation Monitoring (ORM) addresses this challenge: the use of computational tools to measure the reputation of entities from online media content. Early ORM started with counting occurrences of a brand name in Social Media as a channel to estimate the knowledge/reach of a brand.

There are several challenges to collect, process and mine online media data for these purposes~\cite{chall1}. Social Media texts are short, informal, with many abbreviations, slang, jargon and idioms. Often, the users do not care about the correct use of grammar and therefore the text tends to have misspellings, incomplete and unstructured sentences. Furthermore, the lack of context poses a very difficult problem for tasks relevant in the context of Text Mining, such as Named Entity Disambiguation or Sentiment Analysis. Once we classify the sentiment polarity of a given document (e.g. tweet or news title), it is necessary to aggregate several document scores to create meaningful hourly/daily indicators. These tasks are technically complex for most of the people interested in tracking entities on the web.  For this reason, most research has focused on investigating parts of this problem leading to the development of tools that only address sub-tasks of this endeavor. 

Text data usually includes a large number of entities and relationships between them. We broadly define an entity to be a thing or concept that exists in the world, such as a person, a company, organization, an event or a film. Entities exist as mentions across documents and in external knowledge resources. In recent years, entities have gained increased importance as the basic unit of information to answer particular information needs, instead of entire documents or text snippets \citep{demartini2010finding, pound2010ad}. The volume of entity-centric data is rapidly increasing on the Web, including RDF and Linked Data, Schema.org, Facebook's Open Graph, and Google's Knowledge Graph, describing entities (e.g., footballers and coaches) and relationships between them (e.g., ``manages''). 

These developments have a great impact in Online Reputation Monitoring as it is mainly focused on entities. More specifically, the ORM process consists in searching and tracking an entity of interest: the personality, the company, organization or brand/product under analysis. On the other hand, news stories, topics and events discussed in the news or Social Media usually contain mentions of entities or concepts represented in a Knowledge Base. Thus, we can say that entities are the gravitational force that drives the Online Reputation Monitoring process.

\section{Thesis Statement}

The ultimate goal of ORM is to track everything that is said on the Web about a given target entity and consequently, to assess/predict the impact on its reputation. From our perspective, this goal is very hard to achieve for two reasons. The first reason has to do with the difficulty of computationally processing, interpreting and accessing the huge amount of information published online everyday. The second reason is inherent to the definition of reputation as being intangible but having tangible outcomes. More specifically, Fombrun and Van Riel \citep{fombrun2004fame} and later Stacks \citep{stacks2010practioner} found a correlation between several indicators, such as reputation or trust, and financial indicators, such as sales or profits. However, this finding does not imply causality, as financial indicators can be influenced by many factors, besides stakeholders' perceived reputation. In conclusion, there is no consensus on how to measure reputation, neither intrinsically nor extrinsically.  

\begin{figure}[H]
    \centering
    \includegraphics[scale=0.7]{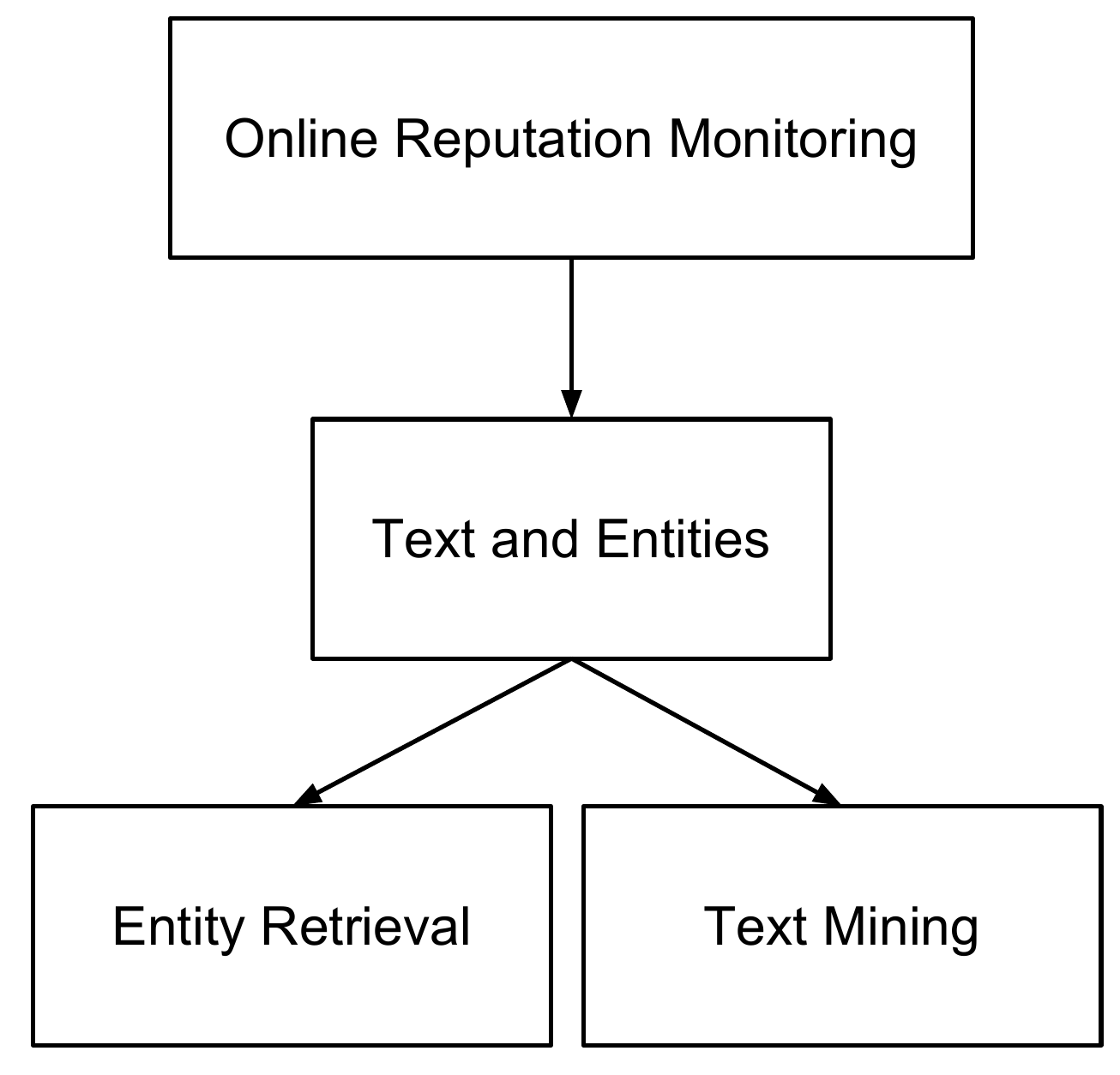}
    \caption{Entity Retrieval and Text Mining as computational problems of ORM.}
    \label{fig:abstraction}
\end{figure}

To the best of our knowledge, current ORM is still very limited and naive. The most standard approach consists in counting mentions of entity names and applying sentiment analysis to produce descriptive reports of aggregated entity popularity and overall sentiment. We propose to make progress in ORM by tackling two computational problems: Entity Retrieval and Text Mining (Figure \ref{fig:abstraction}).  

We believe that a ORM platform, besides providing aggregated statistics and trends about entity popularity and sentiment on the news and social media, would benefit from providing entity retrieval capabilities. End users often like to have the flexibility to search for specific information that is not available in predefined charts. However, ORM has some specificities that traditional entity search systems cannot cope with. More specifically, an entity's reputation is also influenced by the entity's relationships with other entities. 

For instance, the reputation of Apple Inc. was severely damaged with the so called ``Apple Foxconn scandal''. Foxconn was one of the several contractor companies in Apple's supply chain that was accused of exploiting Chinese workers. Although the facts were not directly concerned with Apple itself, its relationship with Foxconn triggered bad public opinion about Apple. The same happened recently with the ``Weinstein sex scandal'', as accusations of sexual harassment aimed at Harvey Weinstein created a wave of damage to companies and personalities associated with the disgraced Hollywood producer. 

Therefore, a ORM platform should provide entity-relationship search capabilities. Entity-Relationship (E-R) Retrieval is a complex case of entity retrieval where the goal is to search for multiple unknown entities and relationships connecting them. Contrary to traditional entity queries, E-R queries expect tuples of connected entities as answers. For instance, ``US technology companies contracts Chinese electronics manufacturers" can be answered by tuples $<$\textit{Apple}, \textit{Foxconn}$>$, while ``Companies founded by disgraced Hollywood producer" is expecting tuples $<$\textit{Miramax}, \textit{Harvey Weinstein}$>$. In essence, an E-R query can be decomposed into a set of sub-queries that specify types of entities and types of relationships between entities.

On the other hand, ORM requires accurate and robust text processing and data analysis methods. Text Mining plays an essential enabling role in developing better ORM. There are several challenges with collecting and extracting relevant entity-centric information from raw text data. It is necessary to filter noisy data otherwise downstream processing tasks, such as sentiment analysis, will be compromised. More specifically, it is essential to develop named entity disambiguation approaches that can distinguish relevant text passages from non-relevant. Named entities are often ambiguous, for example, the word ``bush'' is a surface form for two former U.S. presidents, a music band and a shrub. The ambiguity of named entities is particularly problematic in social media texts, where users often mention entities using a single term. 

ORM platforms would be even more useful if they would be able to predict if social media users will talk a lot about the target entities or not.  For instance, on April 4th 2016, the UK Prime-minister, David Cameron, was mentioned on the news regarding the Panama Papers story. He did not acknowledge the story in detail on that day. However, the news cycle kept mentioning him about this topic in the following days and his mentions on social media kept very high. He had to publicly address the issue on April 9th, when his reputation had  already been severely damaged, blaming himself for not providing further details earlier. Thus we also want to study the feasibility of using entity-centric knowledge extracted from Social Media and online news to predict real world surveys results, such as political polls.

\section{Objectives}

The work reported on this dissertation aimed to understand, formalize and explore the scientific challenges inherent to the problem of using unstructured text data from different Web sources for Online Reputation Monitoring.
We now describe the specific research challenges we proposed to overcome.

\ \\
\textbf{Entity-Relationship Retrieval:} Existing strategies for entity search can be divided in IR-centric and Semantic-Web-based approaches. The former usually rely on statistical language models to match and rank co-occurring terms in the proximity of the target entity \cite{balog2012expertise}. The latter consists in creating a SPARQL query and using it over a structured knowledge base to retrieve relevant RDF triples \cite{heath2011linked}. Neither of these paradigms provide good support for entity-relationship (E-R) retrieval.

Recent work in Semantic-Web search tackled E-R retrieval by extending SPARQL to support joins of multiple query results and creating an extended knowledge graph \cite{yahya2016relationship}. Extracted entities and relationships are typically stored in a knowledge graph. However, it is not always convenient to rely on a structured knowledge graph with predefined and constraining entity types. 

In particular, ORM is interested in transient information sources, such as online news or social media. General purpose knowledge graphs are usually fed with more stable and reliable data sources (e.g. Wikipedia). Furthermore, predefining and constraining entity and relationship types, such as in Semantic Web-based approaches, reduces the range of queries that can be answered and therefore limits the usefulness of entity search, particularly when one wants to leverage free-text.

To the best of our knowledge, E-R retrieval using IR-centric approaches is a new and unexplored research problem within the Information Retrieval research community. One of the objectives of our research is to explore to what degree we can leverage the textual context of entities and relationships, i.e., co-occurring terminology, to relax the notion of an entity or relationship type. 

Instead of being characterized by a fixed type, e.g., \textit{person}, \textit{country}, \textit{place}, the entity would be characterized by any contextual term. The same applies to the relationships. Traditional knowledge graphs have fixed schema of relationships, e.g. \textit{child of}, \textit{created by}, \textit{works for} while our approach relies on contextual terms in the text proximity of every two co-occurring entities in a raw document. Relationships descriptions such as ``criticizes'', ``hits back'', ``meets'' or ``interested in'' would be possible to search for. This is expected to significantly reduce the limitations which structured approaches suffer from, enabling a wider range of queries to be addressed.

\ \\
\textbf{Entity Filtering and Sentiment Analysis:} Entity Filtering is a sub-problem of Named Entity Disambiguation (NED) in which we have a named entity mention and we want to classify it as related or not related with the given target entity. This is a relatively easy problem in well formed texts such as news articles. However, social media texts pose several problems to this task. We are particularly interested in Entity Filtering of tweets and we aim to study a large set of features that can be generated to describe the relationship between a given target entity and a tweet, as well as exploring different learning algorithms to create supervised models for this task.

Sentiment Analysis has been thoroughly studied in the last decade \citep{Giachanou2016}. There have been several PhD thesis entirely dedicated to this subject. It is a broad problem with several ramifications depending on the text source and specific application. Within the context of ORM, we will focus in a particular domain: finance. Sentiment Analysis on financial texts has received increased attention in recent years~\citep{nardo2016}. Neverthless, there are some challenges yet to overcome~\citep{smailovic2014stream}. Financial texts, such as microblogs or newswire, usually contain highly technical and specific vocabulary or jargon, making the development of specific lexical and machine learning approaches necessary.

\ \\
\textbf{Text-based Entity-centric Prediction:} We hypothesize that for entities that are frequently mentioned on the news (e.g. politicians) it is possible to establish a predictive link between online news and popularity on social media. We cast the problem as a supervised learning classification approach: to decide whether popularity will be high or low based on features extracted from the news cycle. We aim to assess if online news are valuable as source of information to effectively predict entity popularity on Twitter. More specifically, we want to find if online news carry different predictive power based on the nature of the entity under study and how predictive performance varies with different times of prediction. We propose to explore different text-based features and how particular ones affect the overall predictive power and specific entities in particular.

On the other hand, we will study if it is possible to use knowledge extracted from social media texts to predict the outcome of public opinion surveys. The automatic content analysis of mass media in the social sciences has become necessary and possible with the rise of social media and computational power. One particularly promising avenue of research concerns the use of sentiment analysis in microblog streams. However, one of the main challenges consists in aggregating sentiment polarity in a timely fashion that can be fed to the prediction method.

\ \\
\textbf{A Framework for ORM:} The majority of the work in ORM consists in ad-hoc studies where researchers collect data from a given social network and produce their specific analysis or predictions, often unreproducible. The availability of open source platforms in this area is scarse. Researchers typically use specific APIs and software modules to produce their studies. However, there has been some effort among the research community to address these issues through open source research platforms. We therefore aim to create an adaptable text mining framework specifically tailored for ORM that can be reused in multiple application scenarios, from politics to finance. This framework is able to collect texts from online media, such as Twitter, and identify entities of interest and classify sentiment polarity and intensity. The framework supports multiple data aggregation methods, as well as visualization and modeling techniques that can be used for both descriptive analytics, such as analyze how political polls evolve over time, and predictive analytics, such as predict elections.

\section{Research Methodology}

We adopted distinct research methodologies in the process of developing the research work described in this thesis. The origin of this work was the POPSTAR project. POPSTAR (Public Opinion and Sentiment Tracking, Analysis, and Research) was a project that developed methods for the collection, measurement and aggregation of political opinions voiced in microblogs (Twitter), in blogs and online news. A first prototype of the framework for ORM was implemented and served as the backend of the POPSTAR website (\url{http://www.popstar.pt/}). The ground work concerned with the development of a framework for ORM was carried in the scope of the project. Therefore, the POPSTAR website served as use case for validating the effectiveness and adaptability of the framework.  

The Entity Filtering and Sentiment Analysis modules of the framework were evaluated using well known external benchmarks resulting in state-of-the-art performance. We participated in RepLab 2013 Filtering Task and evaluated our Entity Filtering method using the dataset created for the competition. One of our submissions obtained the first place at the competition. We also participated in SemEval 2017 Task 5: Fine-grained Sentiment Analysis on Financial Microblogs and News. We were ranked 4th using one of the metrics at the sub-task 5.1 Microblogs.

We performed two experiments regarding the text-based entity centric predictions. For predicting entity popularity on Twitter based on the news cycle we collected tweets and news articles from Portugal using the SocialBus twitter collector and online news from 51 different news outlets collected by SAPO. We used the number of entity mentions on Twitter as target variable and we extracted text-based features from the news datasets. Both datasets were aligned in time. We used the same Twitter dataset for studying different sentiment aggregate functions to serve as features for predicting political polls of a private opinion studies company, Eurosondagem. 

Improvements of Entity-Relationship (E-R) retrieval techniques have been hampered by a lack of test collections, particularly for complex queries involving multiple entities and relationships. We created a method for generating E-R test queries to support comprehensive E-R search experiments. Queries and relevance judgments were created from content that exists in a tabular form where columns represent entity types and the table structure implies one or more relationships among the entities. Editorial work involved creating natural language queries based on relationships represented by the entries in the table. We have publicly released the RELink test collection comprising 600 queries and relevance judgments obtained from a sample of Wikipedia List-of-lists-of-lists tables. 

We evaluated the new methods proposed for E-R retrieval using the RELink query collection together with two other smaller query collections created by research work in Semantic Web-based E-R retrieval. We used a large web corpus, the ClueWeb-09B containing 50 million web pages for creating E-R retrieval tailored indexes for running our experiments. Moreover, we implemented a demo using a large news collection of 12 million Portuguese news articles, resulting in the best demo award at ECIR 2016.

\section{Contributions and Applications}

This work resulted in the following contributions:

\begin{enumerate}

\item  A Text Mining framework that puts together all the building blocks required to perform ORM. The framework is adaptable and can be reused in different application scenarios, such as finance and politics. The framework provides entity-specific Text Mining functionalities that enable the collection, disambiguation, sentiment analysis, aggregation, prediction and visualization of entity-centric information from heterogeneous Web data sources. Furthermore, given that it is built using a modular architecture providing abstraction layers and well defined interfaces, new functionalities can easily be integrated.

\item Generalization of the problem of entity-relationship search to cover entity types and relationships represented by any attribute and predicate, respectively, rather than a pre-defined set.

\item A general probabilistic model for E-R retrieval using Bayesian Networks.

\item Proposal of two design patterns that support retrieval approaches using the E-R model.

\item Proposal of a Entity-Relationship Dependence model that builds on the basic Sequential Dependence Model (SDM) to provide extensible entity-relationship representations and dependencies, suitable for complex, multi-relations queries. 

\item An Entity-relationship indexing and retrieval approach including learning to rank/data fusion methods that can handle entity and relationships ranking and merging of results.

\item The proposal of a method and strategy for automatically obtaining relevance judgments for entity-relationship queries. 

\item We make publicly available queries and relevance judgments for the previous task.

\item  Entity Filtering and Financial Sentiment Analysis methods tailored for Twitter that is able to cope with short informal texts constraints.

\item Analysis of the predictive power of online news regarding entity-centric metrics on Twitter, such as popularity or sentiment. 

\item Analysis of how to combine entity-centric knowledge obtained from heterogeneous sources for survey-like prediction tasks.

\end{enumerate}

We believe this work can be useful in a wide range of applications from which we highlight six:

\begin{itemize}

\item[] \textbf{Reputation Management} is concerned with influencing and controlling company or individual reputation and consequently tracking what is said about entities online is one of the main concerns of this area. For instance, knowing if a given news article will have a negative impact on entity's reputation would be crucial for damage control.

\item[] \textbf{Digital Libraries} are special libraries comprising a collection of digital objects (e.g. text or images) stored in a electronic media format. They are ubiquitous nowadays, from academic repositories, to biomedical databases, law enforcement repositories, etc. We believe the contributions we make to the Entity-Relationship Retrieval research problem can be applied to any digital library enabling a new wide range of search capabilities. 

\item[] \textbf{Fraud Detection} and inside trading detection is an area where information about entities (individuals and companies) and relationships between entities is very useful to discover hidden relationships and contexts of entities that might represent conflicts of interests or even fraud.   

\item[] \textbf{Journalism}, or more specifically, computational journalism would benefit of a powerful entity-relationship search tool in which journalists could investigate how entities were previously mentioned on the Web, including online news through time, as well as relationships among entities and their semantics. 

\item[] \textbf{Political Science} has given a lot of attention to Social Media in recent years due to the sheer amount of people reactions and opinions regarding politically relevant events. Being able to analyze the interplay between online news and Social Media from a political entity perspective can be very interesting for political scientists. On the other hand, it is becoming increasingly difficult to obtain pollsresponses via telephone and it is necessary to start testing alternative approaches.

\item[] \textbf{Social Media Marketing} focuses on communicating through social networks with company potential and effective customers. Evaluating the success of a given campaign is a key aspect of this area. Therefore assessing the volume and polarity of mentions of a given company before and after a campaign would be very useful.

\end{itemize}

\section{Foundations}

Most of the material of this thesis was previously published in journal, conference and workshop publications: 

\begin{itemize}

\item P.Saleiro,  E. M. Rodrigues, C. Soares, E. Oliveira, ``TexRep: A Text Mining Framework for Online Reputation Monitoring'', New Generation Computing, Volume 35, Number 4 2017  \citep{saleiro-ngc}

\item P. Saleiro, N. Milic-Frayling, E. M. Rodrigues, C. Soares, ``RELink: A Research Framework and Test Collection for Entity-Relationship Retrieval'', 40th International ACM SIGIR Conference on Research and Development in Information Retrieval  (SIGIR 2017) \citep{saleiro-sigir}

\item P. Saleiro, N. Milic-Frayling, E. M. Rodrigues, C. Soares, ``Early Fusion Strategy for Entity-Relationship Retrieval'', The First Workshop on Knowledge Graphs and Semantics for Text Retrieval and Analysis (KG4IR@SIGIR 2017) \citep{saleiro-kg4ir}

\item P. Saleiro, L. Sarmento, E. M. Rodrigues, C. Soares, E. Oliveira, ``Learning Word Embeddings from the Portuguese Twitter Stream: A Study of some Practical Aspects'', Progress in Artificial Intelligence (EPIA 2017) \citep{saleiro-epia} 

\item P. Saleiro, E. M. Rodrigues, C. Soares, E. Oliveira, ``FEUP at SemEval-2017 Task 5: Predicting Sentiment Polarity and Intensity with Financial Word Embeddings'', International Workshop on Semantic Evaluation (SemEval@ACL 2017) \citep{saleiro-semeval}

\item P. Saleiro and C. Soares, ``Learning from the News: Predicting Entity Popularity on Twitter'' in Advances in Intelligent Data Analysis XV (IDA 2016) \citep{saleiro-ida}

\item P. Saleiro, J. Teixeira, C. Soares, E. Oliveira, ``TimeMachine: Entity-centric Search and Visualization of News Archives'' in Advances in Information Retrieval: 38th European Conference on IR Research (ECIR 2016) \citep{saleiro-ecir}

\item P. Saleiro, L. Gomes, C. Soares, ``Sentiment Aggregate Functions for Political Opinion Polling using Microblog Streams'' in International C* Conference on Computer Science and Software Engineering (C3S2E 2016) \citep{saleiro-c3s2e}

\item P. Saleiro, S. Amir, M. J. Silva, C. Soares , ``POPmine: Tracking Political Opinion on the Web'' in IEEE International Conference on Computer and Information Technology; Ubiquitous Computing and Communications; Dependable, Autonomic and Secure Computing; Pervasive Intelligence and Computing (IUCC 2015) \citep{saleiro-iucc}

\item P. Saleiro, L. Rei, A. Pasquali, C. Soares, et al., ``POPSTAR at RepLab 2013: Name ambiguity resolution on Twitter'' in Fourth International Conference of the CLEF
initiative (CLEF 2013) \citep{saleiro-clef}

\end{itemize}

\section{Thesis Outline}

In Chapter \ref{ch:sota} we discuss related work to this thesis. In Chapter \ref{ch:er1} we present a formalization of the problem of E-R retrieval using a IR-centric approach. We provide two design patterns for fusion-based E-R retrieval: Early Fusion and Late Fusion. We end the chapter by introducing a new supervised early fusion-based Entity Relationship Dependence Model (ERDM) that can be seen as an extension of the MRF framework for retrieval adapted to E-R retrieval. In Chapter \ref{ch:er2} we describe a set of experiments on E-R retrieval over a Web corpus. First we introduce a new query collection, RELink QC, specifically tailored to this problem. We developed a semi-automatic approach to collect relevance judgments from tabular data and the editorial work consisted in creating E-R queries answered by those relevance judgments. We run experiments using the ClueWeb09-B as dataset and provide evaluation results for the new proposed methods for E-R retrieval.

Chapter \ref{ch:text} is dedicated to Entity Filtering and Financial Sentiment Analysis. We evaluate our approaches using well known external benchmarks, namely, RepLab 2013 and SemEval 2017. In Chapter \ref{ch:predict}, we present two experiments of text-based entity-centric predictions. In the first experiment, we try to predict the popularity of entities on social media using solely features extracted from the news cycle. On the second experiment, we try to assess which sentiment aggregate functions are useful in predicting political polls results. 

In Chapter \ref{ch:framework}, we present an unified framework of ORM. The framework is divided in two major containers: RELink (Entity Retrieval) and TexRep (Text Mining). We present the data flow within the framework and how it can be used as a reference open source framework for researching in ORM. We also present some case studies of using this framework. We end this thesis with Chapter \ref{ch:conclusions} which is dedicated to the conclusions.

\chapter{Background and Related Work} \label{ch:sota}

This chapter introduces an overview of the background concepts and previous research work on the tasks addressed in this dissertation. We start by presenting a brief description of the task of Online Reputation Monitoring (ORM), including related frameworks for ORM. We then survey previous research work in Entity Retrieval and Semantic Search, including a detailed explanation of the Markov Random Field model for retrieval and its variations. We describe the tasks of Named Entity Disambiguation, Sentiment Analysis and previous work on training word embeddings. We end this chapter by providing an overview of related work on text-based predictions, including predicting social media attention or the outcome of political elections.

\section{Online Reputation Monitoring}

The reputation of a company is important for the company itself but as well for the stakeholders. More specifically, stakeholders make decisions about the company and its products faster if they are aware of the image of the company \citep{poiesz1989image}. From the company perspective, reputation is an asset as it attracts stakeholders and it can represent economic profit at the end \citep{jones2000reputation,fombrun2004fame}

In 2001, Newell and Goldsmith used questionnaire and survey methodologies to introduce the first standardized and reliable measure of credibility of companies from a consumer perspective \citep{newell2001development}. There have been also studies that find a correlation between company indicators such as reputation, trust and credibility, and financial indicators, such as sales and profits \citep{fombrun2004fame,stacks2010practioner}. These studies found that although reputations are intangible they influence tangible assets. Following this reasoning, Fombrum created a very successful measurement framework, named RepTrak \citep{fombrun2006reptrak}.

A different methodology compared to questionnaires is media analysis (news, TV and radio broadcasts). Typically, the analysis involves consuming and categorizing media according to stakeholder and polarity (positive, negative) towards the company. Recently, Social Media analysis is becoming an important proxy of people opinion, originating the field of Online Reputation Monitoring \citep{kurniawati2013business}. While traditional reputation monitoring is mostly manual, online media pose the opportunity to process, understand and aggregate large streams of facts about about a company or individual.

ORM requires some level of continuous monitoring~\cite{kaufmann2013concept}. It is crucial to detect early the changes in the perception of a company or personality conveyed in Social Media. Online buzz may be good or bad and consequently, companies must react and address negative trends~\cite{portmann2012fora,gonzalo2016monitoring}. It also creates an opportunity to monitor the reputation of competitors. In this context, Text Mining plays a key, enabling role as it offers methods for deriving high-quality information from textual content~\cite{amigo2013overview}. For instance, Gonzalo \cite{gonzalo2016monitoring} identifies 5 different Text Mining research areas relevant to ORM: entity filtering, topic tracking, reputation priority detection, user profiling and automatic reporting/summarization.

Social Media as a new way of communication and collaboration is an influence for every stakeholder of society, such as personalities, companies or individuals \citep{matevsic2010should}. Social Media users share every aspect of their lives and that includes information about events, news stories, politicians, brands or organizations. Companies have access to all this sharing which opens new horizons for obtaining insights that can be valuable to them and their online reputation. Companies also invest a big share of their public relations on Social Media. Building a strong reputation can take long time and effort but destroying it can take place overnight. Therefore, as the importance of Social Media increased, so did the importance of having powerful tools that deal with this enormous amount of data.

\subsection{Related Frameworks}

The great majority of work in ORM consists in ad-hoc studies and platforms for ORM are usually developed by private companies that do not share internal information. However, there are some open source research projects that can be considered as related frameworks to this work.

Trendminer~\cite{samangooei2012trendminer} is one of such platforms that enables real time analysis of Twitter data, but has a very simple sentiment analysis using word counts and lacks flexibility in order to support entity-centric data processing. A framework for ORM should be entity-centric, i.e., collect, process and aggregate texts and information extracted from those texts in relation to the entities being monitored. 

conTEXT \cite{khalili2014context} addresses adaptability and reusability by allowing a modular interface and allowing plugin components to extend their framework, specially from the perspective of the data sources and text analysis modules. For instance, it does not support Sentiment Analysis module by default but it could be plugged in.  Neverthless, conTEXT does not support the plugin of aggregation and prediction modules which makes it not suitable for ORM. The FORA framework \cite{portmann2012fora} is specifically tailored for ORM. It creates an ontology based on fuzzy clustering of texts but it is only concerned with extracting relevant linguistic units regarding the target entities and does not include automatic sentiment analysis and it does not allow the plugin of new modules.

POPmine~\cite{saleiro2015popmine} was the first version of our Text Mining framework for ORM and it was developed specifically in the context of a project in political data science. It comprises a richer set of modules, including cross media data collection (Twitter, blog posts and online news) and real-time trend analysis based on entity filtering and sentiment analysis modules. In fact, our current version of TexRep, our Text Mining framework for ORM, can be seen as an extension of the POPmine architecture by creating a more general purpose framework for ORM which is not restricted to political analysis. While it would be possible to adapt POPmine's entity disambiguation and sentiment analysis modules, its aggregations are specific to the political scenarios. On the other hand, TexRep supports users to define and plug custom-specific aggregate functions. Moreover, POPmine has limited user configurations (e.g. lacks support for pre-trained word embeddings) and does not include predictive capabilities.

\section{Entity Retrieval and Semantic Search}
Information Retrieval deals with the ``search for information''. It is defined as the activity of finding relevant information resources (usually documents) that meet an information need (usually a query), from within a large collection of resources of an unstructured nature (usually text)  \citep{manning2008introduction}.

In early boolean retrieval systems, documents were retrieved if the exact query term was present and they were represented as a list of terms \citep{manning2008introduction}. With the introduction of the Vector Space Model, each term represents a dimension in a multi-dimensional space, and consequently, each document and query are represented as vectors \citep{salton1968automatic}. Values of each dimension of the document vector correspond to the term frequency (TF) of the term in the document. Therefore, the ranking list of documents is produced based on their spatial distance to the query vector. 

The concept of inverse document frequency (IDF) was later introduced to limit the effect of common terms in a collection \citep{sparck1972statistical}. A term that occurs in many documents of the collection has a lower IDF than terms that occur less often. The combination TF-IDF and variants, such as BM25 \citep{robertson1995okapi}, became commonly used weighting statistics for Vector Space Model. 

Recently, it has been observed that when people have focused information needs, entities better satisfy those queries than a list of documents or large text snippets \citep{pound2010ad}. This type of retrieval is called Entity Retrieval or Entity-oriented retrieval and includes extra Information Extraction tasks for processing documents, such as Named Entity Recognition (NER) and Named Entity Disambiguation (NED). Entity Retrieval is closely connected with Question answering (QA) though, QA systems focus on understanding the semantic intent of a natural language query and deciding which sentences represent the answer to the user.

Considering the query ``British politicians in Panama papers'', the expected result would be a list of names rather than documents related to British politics and the ``Panama Papers'' news story. There are two search patterns related to Entity Retrieval \citep{demartini2010finding}. First, the user knows the existence of a certain entity and aims to find related information about it. For example, a user searching for product related information. Second, the user defines a predicate that constrains the search to a certain type of entities, e.g. searching for movies of a certain genre.

Online Reputation Monitoring systems usually focus on reporting statistical insights based on information extracted from Social Media and online news mentioning the target entity. However, this kind of interaction limits the possibility of users to explore all the knowledge extracted about the target entity. We believe Entity Retrieval could enhance Online Reputation Monitoring by allowing free text search over all mentions of the target entity and, consequently, allow users to discover information that descriptive statistical insights might not be able to identify.

Entity Retrieval differs from traditional document retrieval in the retrieval unit. While document retrieval considers a document as the atomic response to a query, in Entity Retrieval document boundaries are not so important and entities need to be identified based on occurrence in documents \citep{adafre2007entity}. The focus level is more granular as the objective is to search and rank entities among documents. However, traditional Entity Retrieval systems does not exploit semantic relationships between terms in the query and in the collection of documents, i.e. if there is no match between query terms and terms describing the entity, relevant entities tend to be missed.
 
Entity Retrieval has been an active research topic in the last decade, including various specialized tracks, such as Expert finding track \citep{chen2006social}, INEX entity ranking track \citep{demartini2009overview}, TREC entity track \citep{balog2010overview} and SIGIR EOS workshop \citep{balog2012first}. Previous research faced two major challenges: entity representation and entity ranking. Entities are complex objects composed by a different number of properties and are mentioned in a variety of contexts through time. Consequently, there is no single definition of the atomic unit (entity) to be retrieved. Additionally, it is a challenge to devise entity rankings that use various entity representations approaches and tackle different information needs. 

There are two main approaches for tackling Entity Retrieval: ``profile based approach'' and ``voting approach'' \citep{balog2006formal}). The ``profile based approach'' starts by applying NER and NED in the collection in order to extract all entity occurrences. Then, for each entity identified, a meta-document is created by concatenating every passage in which the entity occurs. An index of entity meta-documents is created and a standard document ranking method (e.g. BM25) is applied to rank meta-documents with respect to a given query \citep{azzopardi2005language, craswell2005overview}. One of the main challenges of this approach is the transformation of original text documents to an entity-centric meta-document index, including pre-processing the collection in order to extract all entities and their context.

In the ``voting approach'', the query is processed as typical document retrieval to obtain an initial list of documents  \citep{balog2006formal, ru2005trec}. Entities are extracted from these documents using NER and NED techniques. Then, score functions are calculated to estimate the relation of entities captured and the initial query. For instance, counting the frequency of occurrence of the entity in the top documents combined with each document score (relevance to the query)  \citep{balog2006formal}. Another approach consists in taking into account the distance between the entity mention and the query terms in the documents \citep{petkova2007proximity}.

Recently, there is an increasing research interest in Entity Search over Linked Data, also referred as Semantic Search, due to the availability of structured information about entities and relations in the form of Knowledge Bases \citep{bron2013example,zong2015discovering,zhiltsov2015fielded}. Semantic Search exploits rich structured entity related in machine readable RDF format, expressed as a triple (entity, predicate, object). There are two types of search: keyword-based and natural language based search \citep{pound2012interpreting,unger2012template}. Regardless of the search type, the objective is to interpret the semantic structure of queries and translate it to the underlying schema of the target Knowledge Base. Most of the research focus is on interpreting the query intent \citep{pound2012interpreting,unger2012template} while others focus on how to  devise a ranking framework that deals with similarities between different attributes of the entity entry in the KB and the query terms \citep{zhiltsov2015fielded}

\textbf{Relationship Queries:} Li et al. \cite{li2012entity} were the first to study relationship queries for structured querying entities over Wikipedia text with multiple predicates. This work used a query language with typed variables, for both entities and entity pairs, that integrates text conditions. First it computes individual predicates and then aggregates multiple predicate scores into a result score. The proposed method to score predicates relies on redundant co-occurrence contexts.

Yahya et al. \cite{yahya2016relationship} defined relationship queries as SPARQL-like subject-predicate-object (SPO) queries joined by one or more relationships. The authors cast this problem into a structured query language (SPARQL) and extended it to support textual phrases for each of the SPO arguments. Therefore it allows to combine both structured SPARQL-like triples and text simultaneously. It extended the YAGO knowledge base with triples extracted from ClueWeb using an Open Information Extraction approach \cite{schmitz2012open}.

In the scope of relational databases, keyword-based graph search has been widely studied, including ranking \cite{yu2009keyword}. However, these approaches do not consider full documents of graph nodes and are limited to structured data. While searching over structured data is precise it can be limited in various respects. In order to increase the recall when no results are returned and enable prioritization of results when there are too many, Elbassuoni et al. \cite{elbassuoni2009language} propose a language-model for ranking results. Similarly, the models like EntityRank by Cheng et al. \cite{cheng2007entityrank} and Shallow Semantic Queries by Li et al. \cite{li2012entity}, relax the predicate definitions in the structured queries and, instead, implement proximity operators to bind the instances across entity types. Yahya et al. \cite{yahya2016relationship} propose algorithms for application of a set of relaxation rules that yield higher recall.

\textbf{Entity Retrieval and proximity:} Web documents contain term information that can be used to apply pattern heuristics and statistical analysis often used to infer entities as investigated by Conrad and Utt \cite{conrad1994system}, Petkova and Croft \cite{petkova2007proximity}, Rennie and Jaakkola \cite{rennie2005using}. In fact, early work by Conrad and Utt \cite{conrad1994system} demonstrates a method that retrieves entities located in the proximity of a given keyword. They show that using a fixed-size window around proper-names can be effective for supporting search for people and finding relationship among entities. Similar considerations of the co-occurrence statistics have been used to identify salient terminology, i.e. keyword to include in the document index \cite{petkova2007proximity}.

\subsection{Markov Random Field for IR}

In this section we detail the generic Markov Random Field (MRF) model for retrieval and its variation, the Sequential Dependence Model (SDM). As we later show, this model is the basis for our entity-relationship retrieval model.

The Markov Random Field (MRF) model for retrieval was first proposed by Metzler and Croft \cite{metzler2005markov} to model query term and document dependencies. 
In the context of retrieval, the objective is to rank documents by computing the posterior $P(D|Q)$, given a document $D$ and a query $Q$:

\begin{equation}\label{eq:one}
P(D|Q) =  \frac{P(Q,D)}{P(Q)} 
\end{equation}

For that purpose, a MRF is constructed from a graph $G$, which follows the local Markov property: every random variable in $G$ is independent of its non-neighbors given observed values for its neighbors. Therefore, different edge configurations imply different independence assumptions. 
\begin{figure}[h] 
\centering
\includegraphics[width=0.8\linewidth]{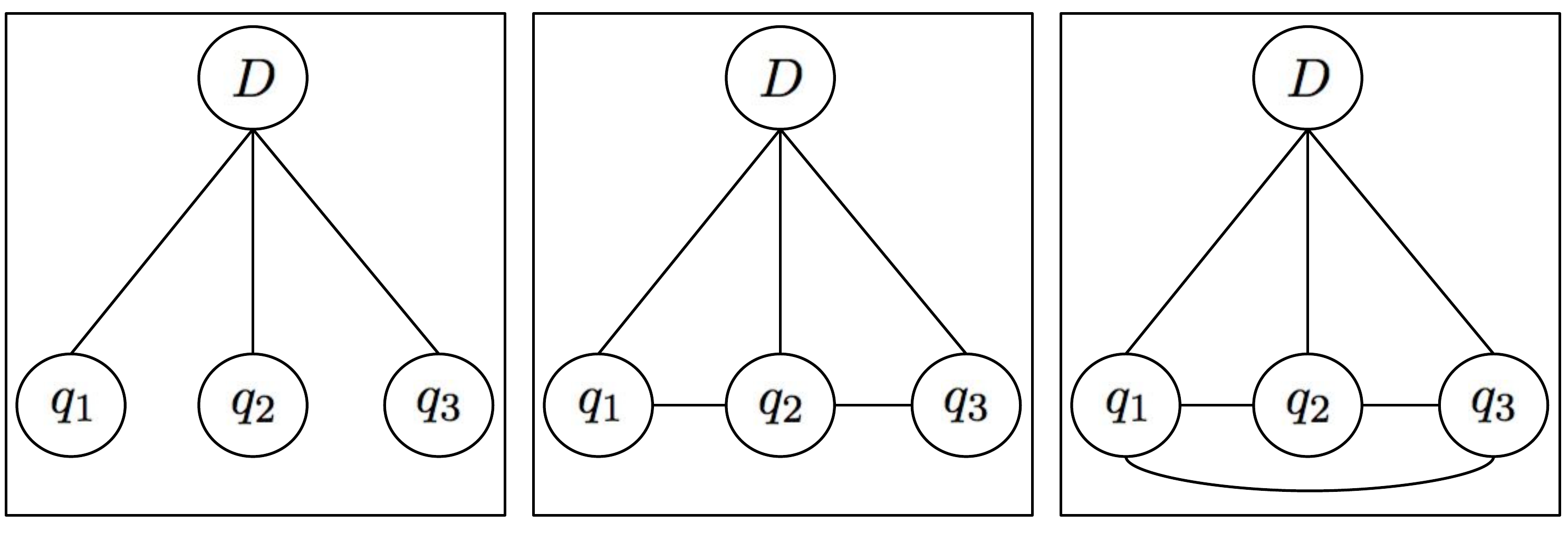}
\caption{Markov Random Field document and term dependencies.}\label{mrf}
\end{figure}

Metzler and Croft \cite{metzler2005markov} defined that $G$ consists of query term nodes $q_i$ and a document node $D$, as depicted in Figure \ref{mrf}. The joint probability mass function over the random variables in $G$ is defined by:

\begin{equation} \label{eq:two}
P_{G, \Lambda} (Q,D) = \frac{1}{Z_ \Lambda } \prod_{c \in C(G)} \psi(c;\Lambda)
\end{equation}

where $Q=q_1,...q_n$ are the query term nodes, $D$ is the document node, $C(G)$ is the set of  maximal cliques in $G$, and $\psi(c;\Lambda)$ is a non-negative potential function over clique configurations. The parameter $Z_ \Lambda = \sum_{Q,D}\prod_{c \in C(G)} \psi(c;\Lambda)$ is the partition function that normalizes the distribution. It is generally unfeasible to compute $Z_ \Lambda$, due to the exponential number of terms in the summation, and it is ignored as it does not influence ranking. 

The potential functions are defined as compatibility functions between nodes in a clique. For instance, a tf-idf score can be measured to reflect the ``aboutness'' between a query term $q_i$ and a document $D$. Metzler and Croft \cite{metzler2005markov} propose to associate one or more real valued feature function with each clique in the graph. The non-negative potential functions are defined using an exponential form $\psi(c;\Lambda) = \text{exp} [\lambda_c f(c)]$, where $\lambda_c$  is a feature weight, which is a free parameter in the model, associated with feature function $f(c)$. The model allows parameter and feature functions sharing across cliques of the same configuration, i.e. same size and type of nodes (e.g. 2-cliques of one query term node and one document node).

For each query $Q$, we construct a graph representing the query term dependencies, define a set of non-negative potential functions over the cliques of this graph and rank documents in descending order of $P_{\Lambda}(D|Q)$:

\begin{align}
\begin{split}\label{eq:3}
   P_{\Lambda}(D|Q)  {} & \stackrel{\text{$rank$}}{=}  \text{log } P_{\Lambda}(D|Q)\\
    & \stackrel{\text{$rank$}}{=} \text{log } P_{\Lambda}(Q,D) - \text{log } P_{\Lambda}(Q)\\
	& \stackrel{\text{$rank$}}{=}  \sum_{c \in C(G)} \text{log } \psi(c;\Lambda)\\
    & \stackrel{\text{$rank$}}{=}  \sum_{c \in C(G)} \text{log } \text{exp} [\lambda_c f(c)]\\
	& \stackrel{\text{$rank$}}{=}  \sum_{c \in C(G)} \lambda_c f(c)\\
\end{split}\\
\end{align}

Metzler and Croft concluded that given its general form, the MRF can emulate most of the retrieval and dependence models, such as language models \cite{song1999general}.

\subsection{Sequential Dependence Model}

The Sequential Dependence Model (SDM) is the most popular variant of the MRF retrieval model  \cite{metzler2005markov}. It defines two clique configurations represented in the following potential functions $ \psi (q_i, D;  \Lambda )$ and $\psi (q_i, q_{i+1}, D;  \Lambda )$. Basically, it considers sequential dependency between adjacent query terms and the document node.

The potential function of the 2-cliques containing a query term node and a document node is represented as $ \psi (q_i, D;  \Lambda ) =  \text{exp} [ \lambda_T f_T(q_i, D)]$. The clique configuration containing contiguous query terms and a document node is represented by two real valued functions. The first considers exact ordered matches of the two query terms in the document, while the second aims to capture unordered matches within $N$ fixed window sizes. Consequently, the second potential function is $\psi (q_i, q_{i+1}, D;  \Lambda ) = \text{exp} [ \lambda_O f_O(q_i,q_{i+1},D) + \lambda_U f_U(q_i,q_{i+1},D)]$.

Replacing $\psi(c;\Lambda)$ by these potential functions in Equation \ref{eq:3} and factoring out the parameters $\lambda$, the SDM can be represented as a mixture model computed over term, phrase and proximity feature classes:

\begin{flalign*}
P(D|Q)   \stackrel{\text{$rank$}}{=} {} &  \lambda_T \text{ } \text{ } \text{ } \text{ } \sum_{q_i \in Q} \text{ } \text{ }  \text{ } f_T(q_i, D) \text{ } + \\
& \lambda_O \sum_{q_i,q_{i+1}  \in Q} f_O(q_i,q_{i+1},D) \text{ } + \\
& \lambda_U  \sum_{q_i,q_{i+1}  \in Q} f_U(q_i,q_{i+1},D)\\
\end{flalign*}

where the free parameters $\lambda$ must follow the constraint $\lambda_T + \lambda_O + \lambda_U = 1$. Coordinate Ascent was chosen to learn the optimal $\lambda$ values that maximize mean average precision using training data \cite{metzler2007linear}.
Considering $tf$ the frequency of the term(s) in the document $D$, $cf$ the frequency of the term(s) in the entire collection $C$, the feature functions in SDM are set as:
\begin{equation} \label{eq:sdm_t}
\resizebox{0.3 \textwidth}{!} 
{
 $f_T(q_i,D) =  \text{log} \left [\frac{ tf_{q_i,D} + \mu \frac{ cf_{q_i}}{|C|}}{|D| + \mu} \right ]$
}
\end{equation}

\begin{equation} \label{eq:sdm_o}
\resizebox{0.6 \textwidth}{!} 
{
$f_O(q_i,q_{i+1},D) =  \text{log} \left [\frac{tf_{\#1(q_i,q_{i+1}),D}  + \mu \frac{ cf_{\#1(q_i,q_{i+1)}}}{|C|}}{|D| + \mu} \right ]$

}
\end{equation}

\begin{equation} \label{eq:sdm_u}
\resizebox{0.6 \textwidth}{!} 
{
 $f_U(q_i,q_{i+1},D)=    \text{log} \left [\frac{tf_{\#uwN(q_i,q_{i+1}),D}  + \mu \frac{ cf_{\#uwN(q_i,q_{i+1)}}}{|C|}}{|D| + \mu} \right ]$
}
\end{equation}

where $\mu$ is the Dirichlet prior for smoothing, $\#1(q_i,q_{i+1})$ is a function that searches for exact matches of the phrase ``$q_i$ $ q_{i+1}$'' and $\#uwN(q_i,q_{i+1})$ is a function that searches for co-occurrences of $q_i$ and $q_{i+1}$ within a window of fixed-N terms (usually 8 terms) across document $D$. SDM has shown state-of-the-art performance in ad-hoc document retrieval when compared with several bigram dependence models and standard bag-of-words retrieval models, across short and long queries \cite{huston2014comparison}.

\subsection{MRF for Entity Retrieval}
The current state-of-the-art methods in ad-hoc entity retrieval from knowledge graphs are based on MRF \cite{zhiltsov2015fielded,nikolaev2016parameterized}. The Fielded Sequential Dependence Model (FSDM) \cite{zhiltsov2015fielded} extends SDM for structured document retrieval and it is applied to entity retrieval from knowledge graphs. In this context, entity documents are composed by fields representing metadata about the entity. Each entity document has five fields: names, attributes, categories, similar entity names and related entity names. FSDM builds individual language models for each field in the knowledge base. This corresponds to replacing SDM feature functions with those of the Mixture of Language Models  \cite{ogilvie2003combining}. The feature functions of FSDM are defined as:

\begin{equation} \label{eq:fsdm_t}
 \tilde{f}_T(q_i,D) =  \text{log} \sum_{j}^{F}  w_{j}^{T} \left [\frac{ tf_{q_i,D_j} + \mu_j \frac{ cf_{q_i,j}}{|C_j|}}{|D_j| + \mu_j} \right ]
\end{equation}

\begin{equation} \label{eq:fsdm_o}
\tilde{f}_O(q_i,q_{i+1},D) =  \text{log} \sum_{j}^{F}  w_{j}^{O}   \left [\frac{tf_{\#1(q_i,q_{i+1}),D_j}  + \mu_j \frac{ cf_{\#1(q_i,q_{i+1}),j}}{|C_j|}}{|D_j| + \mu_j} \right ]
\end{equation}

\begin{equation} \label{eq:fsdm_u}
 \tilde{f}_U(q_i,q_{i+1},D)=    \text{log}  \sum_{j}^{F}  w_{j}^{U}    \left [\frac{tf_{\#uwN(q_i,q_{i+1}),D_j}  + \mu_j \frac{ cf_{\#uwN(q_i,q_{i+1}),j}  }{|C_j|}}{|D_j| + \mu_j} \right ]
\end{equation}

where $\mu_j$ are the Dirichlet priors for each field and $w_j$ are the weights for each field and must be non-negative with constraint $\sum_{j}^{F} w_j= 1$. Coordinate Ascent was used in two stages to learn $w_j$ and $\lambda$ values \cite{zhiltsov2015fielded}.

The Parameterized Fielded Sequential Dependence Model (PFSDM) \cite{nikolaev2016parameterized} extends the FSDM by dynamically calculating the field weights $w_j$ to different query terms. Part-of-speech features are applied to capture the relevance of query terms to specific fields of entity documents. For instance, \textit{NNP} feature is positive if query terms are proper nouns, therefore the query terms should be mapped to the \textit{names} field. Therefore, the field weight contribution of a given query term $q_i$ and a query bigram $q_i$,$q_{i+1}$ in a field $j$ are a linear weighted combination of features:

\begin{equation}
w_{q_{i},j} = \sum_{k} \alpha_{j,k}^{U} \phi_k (q_i, j)
\end{equation}
\begin{equation}
w_{q_{i},q_{i+1},j} = \sum_k \alpha_{j,k}^{B} \phi_k (q_{i},q_{i+1}, j)
\end{equation}

where $\phi_k(q_i, j)$ is the $k$ feature function of a query unigram for the field $j$ and $\alpha_{j,k}^{U}$ is its respective weight. For bigrams, 
$\phi_k (q_{i},q_{i+1}, j)$ is the $k$ feature function of a query bigram for the field $j$ and $\alpha_{j,k}^{B}$ is its respective weight.
Consequently, PFSDM has $F * U + F * B + 3$ total parameters, where $F$ is the number of fields, $U$ is the number of field mapping features for unigrams, $B$ is the number of field mapping features for bigrams, plus the three $\lambda$ parameters. Their estimation is performed in a two stage optimization. First $\alpha$ parameters are learned separately for unigrams and then bigrams. This is achieved  by setting to zero the corresponding $\lambda$ parameters. In the second stage, the $\lambda$ parameters are learned. Coordinate Ascent is used in both stages.

 The ELR model exploits entity mentions in queries by defining a dependency between entity documents and entity links in the query \cite{hasibi2016exploiting}.

\section{Named Entity Disambiguation}
Given a mention in a document, Named Entity Disambiguation (NED) or Entity Linking aims to predict the entity in a reference knowledge base that the string refers to, or NIL if no such entity is available. Usually the reference knowledge base (KB) includes a set of documents, where each document describes one specific entity. Wikipedia is by far the most popular reference KB ~\cite{kulkarni2009collective}.

Previous research typically performs three steps to link an entity mention to a KB: 1) representation of the mention, i.e. extend the entity mention with relevant knowledge from the background document, 2) candidate generation, i.e. find all possible KB entries that the mention might refer to and their representation 3) disambiguation, by computing the similarity between the represented mention and the candidate entities. 

Entity Filtering, or targeted entity disambiguation, is a special case of NED in which there is only one candidate entity, i.e. the entity that is being monitored. There is an increasing interest in developing Entity Filtering methods for Social Media texts, considering its specificities and limitations~\cite{spina2011filter, munoz2012unsupervised}. These approaches focus on finding relevant keywords for positive and negative cases using co-occurrence, web and collection based features. Another line of work creates topic-centric entity extraction systems where entities belong to a certain topic and are used as evidence to disambiguate the short message given its topic~\cite{christoforaki2011searching}. Similarly, Hangya et al. \cite{hangya2013filtering} create features representing topic distributions over tweets using Latent Dirichlet Allocation (LDA). 

The majority of research work in NED is usually applied to disambiguate entities in reasonably long texts as news or blog posts. In recent years, there has been an increasing interest in developing NED methods for Social Media texts and its specificities and limitations \citep{cano2013making, derczynski2013microblog, liu2013entity, greenwood2012reputation}. A survey and evaluation of state-of-the-art NER and NED for Tweets concluded that current approaches do not perform robustly on ``ill-formed, terse, and linguistically compressed'' microblog texts \citep{derczynski2015analysis}. Some Twitter-specific methods reach F1 measures of over 80\%, but are still behind the state-of-the-art results obtained on well-formed news texts.

Social Media texts are too short to provide sufficient information to calculate context similarity accurately \citep{derczynski2013microblog,meij2012adding,greenwood2012reputation, liu2013entity, davis2012named}. In addition, most of state-of-the-art approaches leverage on neighboring entities in the documents but, once again, tweets are short and do not have more than one or two entities mentioned. Most of them \citep{shen2013linking,liu2013entity,davis2012named} extract information obtained from other tweets, and disambiguate entity mentions in these tweets collectively. The assumption is that Twitter users are content generators and tend to scatter their interests over many different messages they broadcast, which is not necessarily true \citep{kwak2010twitter}.

Entity Filtering has also been studied in the context of real-time classification. 
Davis et al.~\cite{davis2012named} propose a pipeline containing three stages. Clearly positive examples are exploited to create filtering rules comprising collocations, users and hashtags. The remaining examples are classified using a Expectation-Maximization (EM) model trained using the clearly positive examples.
Recently, Habib et al.~\cite{habib2016twitterneed} proposed an hybrid approach where authors first query Google to retrieve a set of possible candidate homepages and then enrich the candidate list with text from the Wikipedia. They extract a set of features for each candidate, namely, a language model and overlapping terms between tweet and document, as well as URL length and mention-URL string similarity. In addition, a prior probability of the mention corresponding to a certain entity on the YAGO~\cite{suchanek2007yago} knowledge base is also used.  

Recent work in NED or Entity Linking includes graph based algorithms for collective entity disambiguation, such as TagMe\cite{ferragina2010tagme}, Babelfy \cite{moro2014entity} and WAT \cite{piccinno2014tagme}. Word and entity embeddings have been also used for entity disambiguation \cite{he2013learning,fang2016entity,moreno2017combining}. More specifically,  Fang \cite{fang2016entity} and Moreno \cite{moreno2017combining} propose to learn an embedding space for both entities and words and then compute similarity features based on the combined representations.

\section{Sentiment Analysis}
In the last decade, the automatic processing of subjective and emotive text, commonly known as Sentiment Analysis, has triggered huge interest from the Text Mining research community~\cite{liu2012sentiment}. A typical task in Sentiment Analysis is text polarity classification and in the context of this work can be formalized as follows: given a text span that mentions a target entity, decide whether it conveys positive, negative or neutral sentiment towards the target. 

With the rise of Social Media, research on Sentiment Analysis shifted towards Twitter. New challenges have risen, including slang, misspelling, emoticons, poor grammatical structure~\cite{liu2012sentiment}. A number of competitions were organized, such as SemEval~\cite{rosenthal2015semeval}, leading to the creation of resources for research~\cite{mohammad2013nrc}.

There are two main approaches to sentiment polarity classification: lexicon-based - using a dictionary of terms and phrases with annotated polarity -- or supervised learning -- building a model of the differences in language associated with each polarity, based on training examples. 
In the supervised learning approach, a classifier is specifically trained for a particular type of text (e.g. tweets about politics). Consequently, it is possible to capture peculiarities of the language used in that context. As expected, this reduces the generality of the model, as it is biased towards a specific domain. Supervised learning approaches require training data. In Twitter, most of previous work obtained training data by assuming that emoticons represent the tweet polarity (positive, negative, neutral)~\cite{kouloumpis2011twitter}, or by using third party software, such as the Stanford Sentiment Analyzer~\cite{bamman2015contextualized}.

Lexicon-based approaches have shown to work effectively on conventional text~\cite{liu2010sentiment} but tend to be ill suited for Twitter data. With the purpose of overcoming this limitation, an algorithm that uses a human-coded lexicon specifically tailored to Social Media text was introduced ~\cite{thelwall2012sentiment}. SentiStrength has become a reference in recent years due to its relatively good performance and consistent performance on polarity classification of Social Media texts. Nevertheless, it is confined to a fixed set of words and it is context independent.

The recent interest in deep learning led to approaches that use deep learned word embeddings as features in a variety of Text Mining tasks \cite{bengio2013deep,mikolov2013distributed}. In Sentiment Analysis, recent work integrated polarity information of text into the word embedding by extending the probabilistic document model obtained from Latent Dirichlet Allocation \cite{maas2011learning}. While others learned task-specific embeddings from an existing embedding and sentences with annotated polarity \cite{labutov2013re}. Or learning polarity specific word embeddings from tweets collected using emoticons  \cite{sun2014radical} and directly incorporating the supervision from sentiment polarity in the loss functions of neural networks \cite{tang2014learning}.

\section{Word Embeddings}

The most popular and simple way to model and represent text data is the Vector Space Model \citep{salton1975vector}. A vector of features in a multi-dimensional feature space represents each lexical item (e.g. a word) in a document and each item is independent of other items in the document. This allows to compute geometric operations over vectors of lexical items using well established algebraic methods. However, the Vector Space Model faces some limitations. For instance, the same word can express different meanings in different contexts - the polysymy problem - or different words may be used to describe the same meaning - the synonymy problem. Since 2000, a variety of different methods (e.g. LDA \citep{blei2003latent}) and resources (e.g. DBpedia \citep{auer2007dbpedia}) have been developed to try to assign semantics, or meaning, to concepts and parts of text. 

Word embedding methods aim to represent words as real valued continuous vectors in a much lower dimensional space when compared to traditional bag-of-words models. Moreover, this low dimensional space is able to capture lexical and semantic properties of words. Co-occurrence statistics are the fundamental information that allows creating such representations. Two approaches exist for building word embeddings. One creates a low rank approximation of the word co-occurrence matrix, such as in the case of Latent Semantic Analysis \cite{deerwester1990indexing} and GloVe \cite{pennington2014glove}. The other approach consists in extracting internal representations from neural network models of text \cite{bengio2003neural,collobert2008unified,mikolov2013distributed}. Levy and Goldberg \cite{levy2014neural} showed that the two approaches are closely related.  

Although, word embedding research goes back several decades, it was the recent developments of Deep Learning and the word2vec framework \cite{mikolov2013distributed} that captured the attention of the NLP community. Moreover, Mikolov et al. \cite{mikolov2013linguistic} showed that embeddings trained using word2vec models (CBOW and Skip-gram) exhibit linear structure, allowing analogy questions of the form ``man:woman::king:??.'' and can boost performance of several text classification tasks.

In this context, the objective is to maximize the likelihood that words are predicted given their context. word2vec has two models for learning word embeddings, the skip-gram model (SG) and the continuous-bag-of-word model (CBOW). Here we focus on CBOW. More formally, every word is mapped to a unique vector represented by a column in a projection matrix $W \in \mathbb{R}^{d \times V}$ with $d$ as embedding dimension and $V$ as the total number of words in the vocabulary. Given a sequence of words $w_{-2}, w_{-1}, w_{t}, w_{1}, w_{2}, ..., w_{T}$, the objective is to maximize the average log probability:

\begin{equation}
\frac{1}{T}  \sum_{t=1}^{V}  \sum_{-c \leq j \leq  c, j \neq 0} \textnormal{log } P(w_t | w_{t+j})
\end{equation}

where $c$ is the size of the context window and $w_{t+j}$ is a word in the context window of the center word $w_t$. The context vector is obtained by averaging the embeddings of each word $w_{-c \leq j \leq  c, j \neq 0}$ and the prediction of the center word $w_t$ is performed using a softmax multiclass classifier over all vocabulary $V$:
\begin{equation}
P(w_t | w_{t+j}) = \frac{e^{y_{w_t}}}{\sum e^{y_{w_i}}}
\end{equation}
Each of $y_i$ is un-normalized log-probability for each output word $i$. After training, a low dimensionality embedding matrix $\textbf{E}$ encapsulating information about each word in the vocabulary and its surrounding contexts is learned, transforming a one-hot sparse representation of words into a compact real valued embedding vector of size $d \times 1$. This matrix can then be used as input to other learning algorithms tailored for specific tasks to further enhance performance.

For large vocabularies it is unfeasible to compute the partition function (normalizer) of softmax therefore Mikolov \cite{mikolov2013distributed} proposes to use the hierarchical softmax objective function or to approximate the partition function using a technique called negative sampling. Stochastic gradient descent is usually applied for training the softmax where the gradient is obtained via backpropagation.

There are several approaches to generating word embeddings. One can build models that \emph{explicitly} aim at generating word embeddings, such as Word2Vec or GloVe \cite{mikolov2013distributed,pennington2014glove}, or one can extract such embeddings as by-products of more general models, which implicitly compute such word embeddings in the process of solving other language tasks.

One of the issues of recent work in training word embeddings is the variability of experimental setups reported. For instance, in the paper describing GloVe \cite{pennington2014glove} the authors trained their model on five corpora of different sizes and built a vocabulary of 400K most frequent words. Mikolov et al. \cite{mikolov2013linguistic} trained with 82K vocabulary while Mikolov et al. \cite{mikolov2013distributed} was trained with 3M vocabulary. Recently, Arora et al. \cite{arora2015rand} proposed a generative model for learning embeddings that tries to explain some theoretical justification for nonlinear models (e.g. word2vec and GloVe) and some hyper parameter choices. The authors evaluated their model using 68K vocabulary. 

SemEval 2016-Task 4: Sentiment Analysis in Twitter organizers report that participants either used general purpose pre-trained word embeddings, or trained from Tweet 2016 dataset or ``from some sort of dataset'' \cite{nakov2016semeval}. However, participants neither report the size of vocabulary used neither the possible effect it might have on the task specific results.

Recently, Rodrigues et al. \cite{rodrigues2016lx} created and distributed the first general purpose embeddings for Portuguese. Word2vec gensim implementation was used and authors report results with different values for the parameters of the framework. Furthermore, authors used experts to translate well established word embeddings test sets for Portuguese language, which they also made publicly available and we use some of those in this work.

\section{Predicting Collective Attention}
Online Reputation Monitoring systems would be even more useful if they would be able to know in advance if social media users will talk a lot about the target entities or not. In recent years, a number of research works have studied the relationship and predictive behavior of user response to the publication of online media items, such as, commenting news articles, playing Youtube videos, sharing URLs or retweeting patterns \citep{ bandari2012pulse,yang2011patterns,tsagkias2009predicting,he2014predicting}. The first attempt to predict the volume of user comments for online news articles used both metadata from the news articles and linguistic features \citep{tsagkias2009predicting}. The prediction was divided in two binary classification problems: if an article would get any comments and if it would be high or low number of comments. Similarly, other studies found that shallow linguistic features (e.g. TF-IDF or sentiment) and named entities have good predictive power \citep{gottipati2012finding,louis2013makes}. 

Research work more in line with ours, tries to predict the popularity of news articles shares (url sharing) on Twitter based on content features \citep{bandari2012pulse}.  The authors considered the news source, the article's category, the article's author, the subjectivity of the language in the article, and number of named entities in the article as features. Recently, there was a large study of the life cycle of news articles in terms of distribution of visits, tweets and shares over time across different sections of the publisher \citep{castillo2014characterizing}. Their work was able to improve, for some content type, the prediction of web visits using data from social media after ten to twenty minutes of publication.

Other lines of work, focused on temporal patterns of user activities and have consistently identified broad classes of temporal patterns based on the presence of a clear peak of activity \citep{crane2008robust,lehmann2012dynamical, romero2011differences, yang2011patterns}. Classes differentiate by the specific amount and duration of activity before and after the peak. Crane and Sornette \citep{crane2008robust} define endogenous or exogenous origin of events based on being triggered by internal aspects of the social network or external, respectively. They find that hashtag popularity is mostly influenced by exogenous factors instead of epidemic spreading. Other work \citep{lehmann2012dynamical} extend these classes by creating distinct clusters of activity based on the distributions in different periods (before, during and after the peak) that can be interpreted based on semantics of hashtags. Consequently, the authors applied text mining techniques to semantically describe hashtag classes. Yang and Leskovec \citep{yang2011patterns} propose a new measure of time series similarity and clustering. The authors obtain six classes of temporal shapes of popularity of a given phrase (meme) associated with a recent event, as well as the ordering of media sources contribution to its popularity.

Recently, Tsytsarau et al. \citep{tsytsarau2014dynamics} studied the time series of news events and their relation to changes of sentiment time series expressed on related topics on social media. The authors proposed a novel framework using time series convolution between the importance of events and media response function, specific to media and event type. Their framework is able to predict time and duration of events as well as shape through time. 

\section{Political Data Science}
Content analysis of mass media has an established tradition in the social sciences, particularly in the study of effects of media messages,
encompassing topics as diverse as those addressed in seminal studies of newspaper editorials \citep{lasswell1952}, media agenda-setting
\citep{mccombs1972}, or the uses of political rhetoric \citep{moen1990}, among many others. By 1997, Riffe and Freitag \citep{riffe1997}, 
reported an increase in the use of content analysis in communication research and suggested that digital text and computerized means for its
extraction and analysis would reinforce such a trend. Their expectation has been fulfilled: the use of automated content analysis has by now surpassed the use of hand coding \citep{neuendorf2002}. The increase in the digital sources of text, on the one hand, and current advances in
computation power and design, on the other, are making this development both necessary and possible, while also raising awareness about the
inferential pitfalls involved \citep{hopkins2010, chall2}.

One avenue of research that has been explored in recent years concerns the use of social media to predict present and future political events, namely electoral results \cite{Bermingham, pred12, pred6, pred3, livne2011party, tumasjan2010predicting, Gayo-Avello2012, O'Connor2010, Chung2011}. Although there is no consensus about methods and their consistency \cite{Metaxas2011, Gayo-Avello2011}. Gayo-Avello \cite{pred_survey} summarizes the differences between studies conducted so far by stating that they vary about period and method of data collection, data cleansing and pre-processing techniques, prediction approach and performance evaluation. One particular challenge when using sentiment is how to aggregate opinions in a timely fashion that can be fed to the prediction method. Two main strategies have been used to predict elections: buzz, i.e., number of tweets mentioning a given candidate or party and the use of sentiment polarity. Different computational approaches have been explored to process sentiment in text, namely machine learning and linguistic based methods \cite{pang2008, Kouloumpis2011, nakov2013}. In practice, algorithms often combine both strategies. 

 Johnson et al. \cite{JSS} concluded that more than predicting elections, social media can be used to gauge sentiment about specific events, such as political news or speeches. Defending the same idea, Diakopoulos el al. \cite{Diakopoulos} studied the global sentiment variation based on Twitter messages of an Obama vs McCain political TV debate while it was still happening. Tumasjan et al. \cite{tumasjan2010predicting} used Twitter data to predict the 2009 Federal Election in Germany. They stated that ``the mere number of party mentions accurately reflects the election result''. Bermingham et al. \cite{Bermingham} correctly predicted the 2011 Irish General Elections also using Twitter data. Gayo-Avello et al. \cite{Gayo-Avello2011} also tested the share of volume as predictor in the 2010 US Senate special election in Massachusetts.

On the other hand, several other studies use sentiment as a polls result indicator. Connor et al. \cite{O'Connor2010} used a sentiment aggregate function to study the relationship between the sentiment extracted from Twitter messages and polls results. They defined the sentiment aggregate function as the ratio between the positive and negative messages referring an specific political target. 
They used the sentiment aggregate function as predictive feature in the regression model, achieving a correlation of 0.80 between the results and the poll results, capturing the important large-scale trends. Bermingham et al. \cite{Bermingham} also included in their regression model sentiment features. Bermingham et al. introduced two novel sentiment aggregate functions. For inter-party sentiment, they modified the share of volume function to represent the share of positive and negative volume. For intra-party sentiment , they used a log ratio between the number of positive and negative mentions of a given party. Moreover, they concluded that the inclusion of sentiment features augmented the effectiveness of their model. Gayo-Avello et al. \cite{Gayo-Avello2011} introduced a different aggregate function. In a two-party race, all negative messages on party $c2$ are interpreted as positive on party $c1$, and vice-versa.

In summary, suggestions for potentially independent or in other words predictive metrics appear in a wide variety of forms: the mention share that a party received within all party mentions during a given time-span \citep{Bermingham,sanders2013relating,skoric2012tweets,soler2012twitter,sang2012predicting,tumasjan2010predicting}, the mention share of political candidates  \citep{chen2012twitter,digrazia2013more,fink2013twitter,gaurav2013leveraging, skoric2012tweets}, the share of positive mentions a party received \citep{Bermingham,thapen2013towards}, the positive mention share of candidates \citep{O'Connor2010, shi2012predicting,fink2013twitter}, the share of users commenting on a candidate or party \citep{sang2012predicting}, the share of mentions for a candidate followed by a word indicative of electoral success or failure \citep{jensen2013psephological}, the relative increase of positive mentions of a candidate \citep{franch2013wisdom} or simply a collection of various potentially politically relevant words identified by their statistical relationship with polls or political actors in the past \citep{beauchamp2013predicting, contractor2013understanding,lampos2013user,marchetti2012learning}. 

 Suggestions for the dependent variable, metrics of political success, show a similar variety. They include the vote share that a party received on election day \citep{Bermingham,franch2013wisdom,sanders2013relating,skoric2012tweets,soler2012twitter}, the vote share of a party adjusted to include votes only for parties included in the analysis \citep{tumasjan2010predicting}, the vote share of candidates on election day \citep{digrazia2013more,fink2013twitter,gaurav2013leveraging,jensen2013psephological,skoric2012tweets}, campaign tracking polls \citep{beauchamp2013predicting,contractor2013understanding,fink2013twitter,lampos2013user,O'Connor2010,shi2012predicting,thapen2013towards}, politicians' job approval ratings \citep{marchetti2012learning,O'Connor2010}, and the number of seats in parliament that a party received after the election \citep{sang2012predicting}.

\chapter{Entity Retrieval for Online Reputation Monitoring}\label{ch:er1}

We start by presenting a formal definition of E-R queries and how can we model the E-R retrieval problem from a probabilistic perspective. We assume that a E-R query can be formulated as a sequence of individual sub-queries each targeting a specific entity or relationship. If we create specific representations for entities (e.g. context terms) as well as for pairs of entities, i.e. relationships then we can create a graph of probabilistic dependencies between sub-queries and entity/relationship representations. We show that these dependencies can be depicted in a probabilistic graphical model, i.e. a Bayesian network. Therefore, answering an E-R query can be reduced to a computation of factorized conditional probabilities over a graph of sub-queries and entity/relationship documents. 

However, it is not possible to compute these conditional probabilities directly from raw documents in a collection. Such as with traditional entity retrieval, documents serve as proxies to entities (and relationships) representations. It is necessary to fuse information spread across multiple documents. We propose two design patterns inspired from Model 1 and Model 2 of \citet{balog2006formal} to create entity/relationship centric and document centric representations.  

The first design pattern - Early Fusion - consists in aggregating context terms of entity and relationship occurrences to create two dedicated indexes, the \textit{entity} index and the \textit{relationship} index. Then it is possible to use any retrieval method to compute the relevance score of entity and relationship documents given the E-R sub-queries. The second design pattern - Late Fusion - can be applied on top of a standard document index alongside a set of entity occurrences in each document. First we compute the relevance score of documents given a E-R sub-query, then based on the entity occurrences of the top k results we compute individual entity or relationship scores. Once again any retrieval method can be used to score documents.

When combined with traditional retrieval methods (e.g. Language Models or BM25) these design patterns can be used to create unsupervised baselines for E-R retrieval. Finally, we follow a recent research line in entity retrieval  \citep{zhiltsov2015fielded,hasibi2016exploiting,nikolaev2016parameterized} which exploits term dependencies using the Markov Random Field (MRF) framework for retrieval\citep{metzler2005markov}. We introduce the Entity-Relationship Dependence Model (ERDM), a novel supervised Early Fusion-based model for E-R retrieval that creates a MRF to compute term dependencies of E-R queries and entity/relationship documents.  

\section{Entity-Relationship Retrieval}
 
E-R retrieval is a complex case of entity retrieval. E-R queries expect tuples of related entities as results instead of a single ranked list of entities as it happens with general entity queries. For instance, the E-R query ``Ethnic groups by country" is expecting a ranked list of tuples $<$\textit{ethnic group}, \textit{country}$>$ as results.  The goal is to search for multiple unknown entities and relationships connecting them.

\begin{table}[h]
\centering
\caption{E-R retrieval definitions.}
\label{tab:er_definitions}
\begin{tabular}{|p{1cm}|p{13cm}|}
\hline
$Q$ &  E-R query (e.g. ``congresswoman hits back at US president'').  \\ \hline
$Q^{E_i}$ &    Entity sub-query in $Q$  (e.g. ``congresswoman'').  \\ \hline
$Q^{R_{i-1,i}}$ &   Relationship sub-query in $Q$ (e.g. ``hits back at'').  \\ \hline
$D^{E_i}$& Term-based representation of an entity (e.g. <Frederica Wilson> = $\{$representative, congresswoman$\}$). We use the terminology \textit{representation} and \textit{document} interchangeably.  \\ \hline
$D^{R_{i-1,i}}$ &  Term-based representation of a relationship (e.g. <Frederica Wilson, Donald Trump> = $\{$hits,back$\}$). We use the terminology \textit{representation} and  \textit{document} interchangeably.  \\ \hline
$Q^{E}$ & The set of entity sub-queries in a E-R query (e.g. $\{$``congresswoman'',``US president'' $\}$). \\ \hline
$Q^{R}$ & The set of relationship sub-queries in a E-R query. \\ \hline
$D^{E}$ &  The set of entity documents to be retrieved by a E-R query. \\ \hline
$D^{R}$ &  The set of relationship documents to be retrieved by a E-R query. \\ \hline
$|Q|$ &   E-R query length corresponding to the number of entity and relationship sub-queries. \\ \hline
$T_E$ &   The entity tuple to be retrieved (e.g. <Frederica Wilson, Donald Trump>).  \\ \hline
\end{tabular}
\end{table}

In this section, we present a definition of E-R queries and a probabilistic formulation of the E-R retrieval problem from an Information Retrieval perspective. Table \ref{tab:er_definitions} presents several definitions that will be used throughout this chapter.

\subsection{E-R Queries} \label{erqueries}

E-R queries aim to obtain a ordered list of entity tuples $T_E=$ $<$$E_1, E_{2},..., E_n$$>$ as a result. Contrary to entity search queries where the expected result is a ranked list of single entities, results of E-R queries should contain two or more entities. For instance, the complex information need ``\textit{Silicon Valley companies founded by Harvard graduates}'' expects entity-pairs (2-tuples) $<$\textit{company}, \textit{founder}$>$ as results. In turn, ``\textit{European football clubs in which a Brazilian player won a trophy}" expects triples (3-tuples) $<$\textit{club}, \textit{player}, \textit{trophy}$>$ as results. 

Each pair of entities $E_{i-1}$, $E_{i}$ in an entity tuple is connected with a relationship $R(E_{i-1},E_i)$. A complex information need can be expressed in a relational format, which is decomposed into a set of sub-queries that specify types of entities $E$ and types of relationships $R(E_{i-1},E_{i})$ between entities. 

For each relationship sub-query there must be two sub-queries, one for each of the entities involved in the relationship. Thus a E-R query $Q$ that expects 2-tuples, is mapped into a triple of sub-queries $Q=\{Q^{E_1}$, $Q^{R_{1,2}}$, $Q^{E_{2}}\}$, where $Q^{E_1}$ and $Q^{E_{2}}$ are the entity attributes queried for $E_1$ and $E_{2}$ respectively, and $Q^{R_{1,2}}$ is a relationship attribute describing $R(E_i,E_{i+1})$. 

If we consider a E-R query as a chain of entity and relationship sub-queries $Q = \{Q^{E_1}$, $Q^{R_{1,2}}$, $Q^{E_{2}}$, ..., $Q^{E_{n-1}}$,$Q^{R_{n-1,n}}$, $Q^{E_n}  \}$ and we define the length of a E-R query $|Q|$ as the number of sub-queries, then the number of entity sub-queries must be $\frac{|Q|+1}{2}$ and the number of relationship sub-queries equal to $\frac{|Q|-1}{2}$. Consequently, the size of each entity tuple $T_E$ to be retrieved must be equal to the number of entity sub-queries. For instance, the E-R query ``soccer players who dated a top model'' with answers such as $<$\textit{Cristiano Ronaldo}, \textit{Irina Shayk}$>$) is represented as three sub-queries $Q^{E_1}=\{$\textit{soccer players}$\}$, $Q^{R_{1,2}}=\{$\textit{dated}$\}$, $Q^{E_{2}}=\{$\textit{top model}$\}$. 

Automatic mapping of terms from a E-R query $Q$ to sub-queries $Q^{E_i}$ or $Q^{R_{i-1,i}}$ is out of the scope of this work and can be seen as a problem of query understanding \cite{yahya2012natural,pound2012interpreting,sawant2013learning}. We assume that the information needs are decomposed into constituent entity and relationship sub-queries using Natural Language Processing techniques or by user input through an interface that enforces the structure $Q = \{Q^{E_1}$, $Q^{R_{1,2}}$, $Q^{E_{2}}$, ..., $Q^{E_{n-1}}$,$Q^{R_{n-1,n}}$, $Q^{E_n}  \}$.

\subsection{Modeling E-R Retrieval }

Our approach to E-R retrieval assumes that we have a raw document collection (e.g. news articles) and each document $D_j$ is associated with one or more entities $E_i$. In other words, documents contain mentions to one or more entities that can be related between them. Since our goal is to retrieve tuples of related entities given a E-R query that expresses entity attributes and relationship attributes, we need to create term-based representations for both entities and relationships. We denote a representation of an entity $E_i$ as $D^{E_i}$. 

In E-R retrieval we are interested in retrieving tuples of entities $T_E=$ $<$$E_1, E_{2},..., E_n$$>$ as a result. The number of entities in each tuple can be two, three or more depending on the structure of the particular E-R query. When a E-R query aims to get tuples of more than two entities, we assume it is possible to combine tuples of length two. For instance, we can associate two tuples of length two that share the same entity to retrieve a tuple of length three. Therefore we create representations of relationships as pairs of entities. We denote a representation of a relationship $R(E_{i-1}, E_i)$ as $D^{R_{i-1,i}}$. 

Considering the example query ``Which spiritual leader won the same award as a US vice president?'' it can be formulated in the relational format as $Q^{E_1}= \{$\textit{spiritual leader}$\}$, $Q^{R_{1,2}}=\{$\textit{won}$\}$, $Q^{E_{2}}=\{$\textit{award}$\}$, $Q^{R_{2,3}}=\{$\textit{won}$\}$, $Q^{E_{3}}=\{$\textit{US vice president}$\}$. Associating the tuples of length two $<$Dalai Lama, Nobel Peace Prize$>$ and $<$Nobel Peace Prize, Al Gore$>$ would result in the expected 3-tuple $<$Dalai Lama, Nobel Peace Prize, Al Gore$>$. 

For the sake of clarity we now consider an example E-R query with three sub-queries ($|Q|=3$). This query aims  to retrieve a tuple of length two, i.e. a pair of entities connected by a relationship. Based on the definition of a E-R query, each entity in the resulting tuple must be relevant to the corresponding entity sub-queries $Q^E$. Moreover, the relationship between the two entities must also be relevant to the relationship sub-queries $Q^R$. Instead of calculating a simple posterior $P(D|Q)$ as with traditional information retrieval, in E-R retrieval the objective is to rank tuples based on a joint posterior of multiple entity and relationship representations given a E-R query, such as $P(D^{E_{2}},D^{E_1},D^{R_{1,2}}|Q)$ when $|Q|=3$.  

E-R queries can be seen as chains of interleaved entity and relationship sub-queries. We take advantage of the chain rule to formulate the joint probability  $P(D^{E_2}, D^{E_{1}},D^{R_{1,2}},Q)$ as a product of conditional probabilities. Formally, we want to rank entity and relationship candidates in descending order of the joint posterior $P(D^{E_2}, D^{E_{1}},D^{R_{1,2}}|Q)$ as:

\begin{align}
\begin{split}\label{erprob}
  P(D^{E_2}, D^{E_{1}},D^{R_{1,2}}|Q)  & \stackrel{\text{$rank$}}{=}   \frac{P(D^{E_2}, D^{E_{1}},D^{R_{1,2}},Q)}{P(Q)}\\
	& \stackrel{\text{$rank$}}{=}   \frac{P(D^{E_2}| D^{E_{1}},D^{R_{1,2}},Q).P( D^{E_{1}}|D^{R_{1,2}},Q).P(D^{R_{1,2}}|Q).P(Q)}{P(Q)}\\
    & \stackrel{\text{$rank$}}{=}   P(D^{E_2}| D^{R_{1,2}},Q).P( D^{E_{1}}|D^{R_{1,2}},Q).P(D^{R_{1,2}}|Q)\\
    & \stackrel{\text{$rank$}}{\propto} P(D^{E_2}| D^{R_{1,2}},Q^{E_2}).P(D^{E_1}|D^{R_{1,2}},Q^{E_1}).P(D^{R_{1,2}}|Q^{R_{1,2}})\\
\end{split}\\
\end{align}

We consider conditional independence between entity representations within the joint posterior, i.e., the probability of a given entity representation $D^{E_i}$ being relevant given a E-R query is independent of knowing that entity $D^{E_{i+1}}$ is relevant as well. As an example, consider the query ``action movies starring a British actor''. Retrieving entity representations for ``action movies'' is independent of knowing that <Tom Hardy> is relevant to the sub-query ``British actor''. However, it is not independent of knowing the set of relevant relationships for sub-query ``starring''. If a given action movie is not in the set of relevant entity-pairs for ``starring'' it does not make sense to consider it as relevant. Consequently, $P(D^{E_2}| D^{E_{1}},D^{R_{1,2}},Q) = P(D^{E_2}| D^{R_{1,2}},Q)$.

Since E-R queries can be decomposed in constituent entity and relationship sub-queries, ranking candidate tuples using the joint posterior $P(D^{E_2}, D^{E_{1}},D^{R_{1,2}}|Q)$ is rank proportional to the product of conditional probabilities on the corresponding entity and relationship sub-queries $Q^{E_2}$, $Q^{E_1}$ and $Q^{R_{1,2}}$.

We now consider a longer E-R query aiming to retrieve a triple of connected entities. This query has three entity sub-queries and two relationship sub-queries, thus $|Q|=5$. As we previously explained, when there are more than one relationship sub-queries we need to join entity-pairs relevant to each relationship sub-query that have one entity in common. From a probabilistic point of view this can be seen as conditional dependence from the entity-pairs retrieved from the previous relationship sub-query, i.e. $P(D^{R_{2,3}}|D^{R_{1,2}},Q) \neq P(D^{R_{2,3}}|Q)$.  To rank entity and relationship candidates we need to calculate the following joint posterior:

\begin{align}
\begin{split}\label{erprob}
   P(D^{E_3},  D^{E_{2}},&D^{E_{1}}, D^{R_{2,3}},D^{R_{1,2}}|Q)\stackrel{\text{$rank$}}{=} P(D^{E_3}| D^{E_{2}},D^{E_{1}},D^{R_{2,3}},D^{R_{1,2}},Q).\\
   & P( D^{E_{3}}|D^{E_{2}},D^{R_{2,3}},D^{R_{1,2}},Q).P(D^{E_{1}}|D^{R_{2,3}},D^{R_{1,2}},Q). \\
   & P(D^{R_{2,3}}|D^{R_{1,2}},Q).P(D^{R_{1,2}}|Q)\\
&  \ \ \ \ \ \ \ \ \ \ \ \ \ \ \ \ \ \ \ \ \ \ \ \ \ \  \stackrel{\text{$rank$}}{=}   P(D^{E_3}| D^{R_{2,3}},Q).P( D^{E_{2}}|D^{R_{2,3}},D^{R_{1,2}},Q).\\
& P(D^{E_{1}}|D^{R_{1,2}},Q).P(D^{R_{2,3}}|D^{R_{1,2}},Q).P(D^{R_{1,2}}|Q)\\
&  \ \ \ \ \ \ \ \ \ \ \ \ \ \ \ \ \ \ \ \ \ \ \ \ \ \  \stackrel{\text{$rank$}}{\propto}   P(D^{E_3}| D^{R_{2,3}},Q^{E_3}).P( D^{E_{2}}|D^{R_{2,3}},D^{R_{1,2}},Q^{E_{2}}).\\
& P(D^{E_{1}}|D^{R_{1,2}},Q^{E_{1}}).P(D^{R_{2,3}}|D^{R_{1,2}},Q^{R_{2,3}}).P(D^{R_{1,2}}|Q^{R_{1,2}})\\
\end{split}\\
\end{align}
  
When compared to the previous example, the joint posterior for $|Q|=5$  shows that entity candidates for  $D^{E_{2}}$ are conditional dependent of both $D^{R_{2,3}}$ and $D^{R_{1,2}}$. In other words, entity candidates for $D^{E_{2}}$ must belong to entity-pairs candidates for both relationships representations that are connected with $E_2$, i.e. $D^{R_{2,3}}$ and $D^{R_{1,2}}$. 

We are now able to make a generalization of E-R retrieval as a factorization of conditional probabilities of a joint probability of entity representations $D^{E_i}$, relationship representations $D^{R_{i-1,i}}$, entity sub-queries $Q^{E_{i}}$ and relationship sub-queries $Q^{R_{i-1,i}}$. These set of random variables and their conditional dependencies can be easily represented in a probabilistic directed acyclic graph,i.e. a Bayesian network \citep{pearl1985bayesian}. 

In Bayesian networks, nodes represent random variables while edges represent conditional dependencies. Every other nodes that point to a given node are considered parents. Bayesian networks define the joint probability of a set of random variables as a factorization of the conditional probability of each random variable conditioned on its parents. Formally, $P(X_{1},\ldots ,X_{n})=\prod _{i=1}^{n}P(X_{i}|pa_{i})$, where $pa_{i}$ represents all parent nodes of $X_i$.

\begin{figure}
\centering
\begin{subfigure}{.4\linewidth}
    \centering
    \includegraphics[width=.8\textwidth]{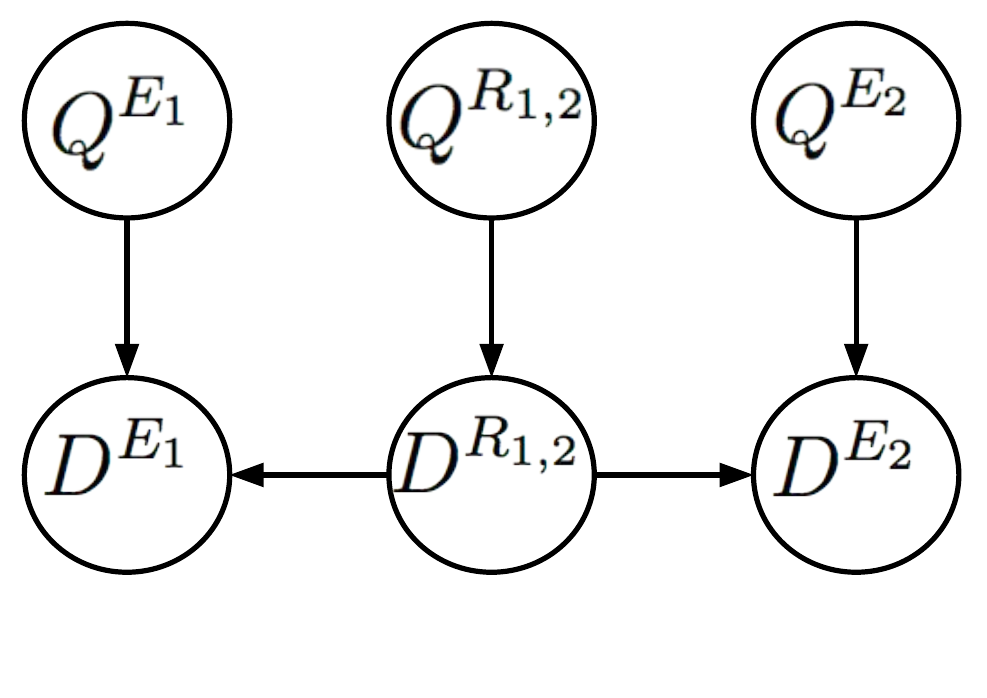}
    \caption{$|Q|=3$}
\end{subfigure}
    \hfill
\begin{subfigure}{.55\linewidth}
    \centering
	\includegraphics[width=1.0\textwidth]{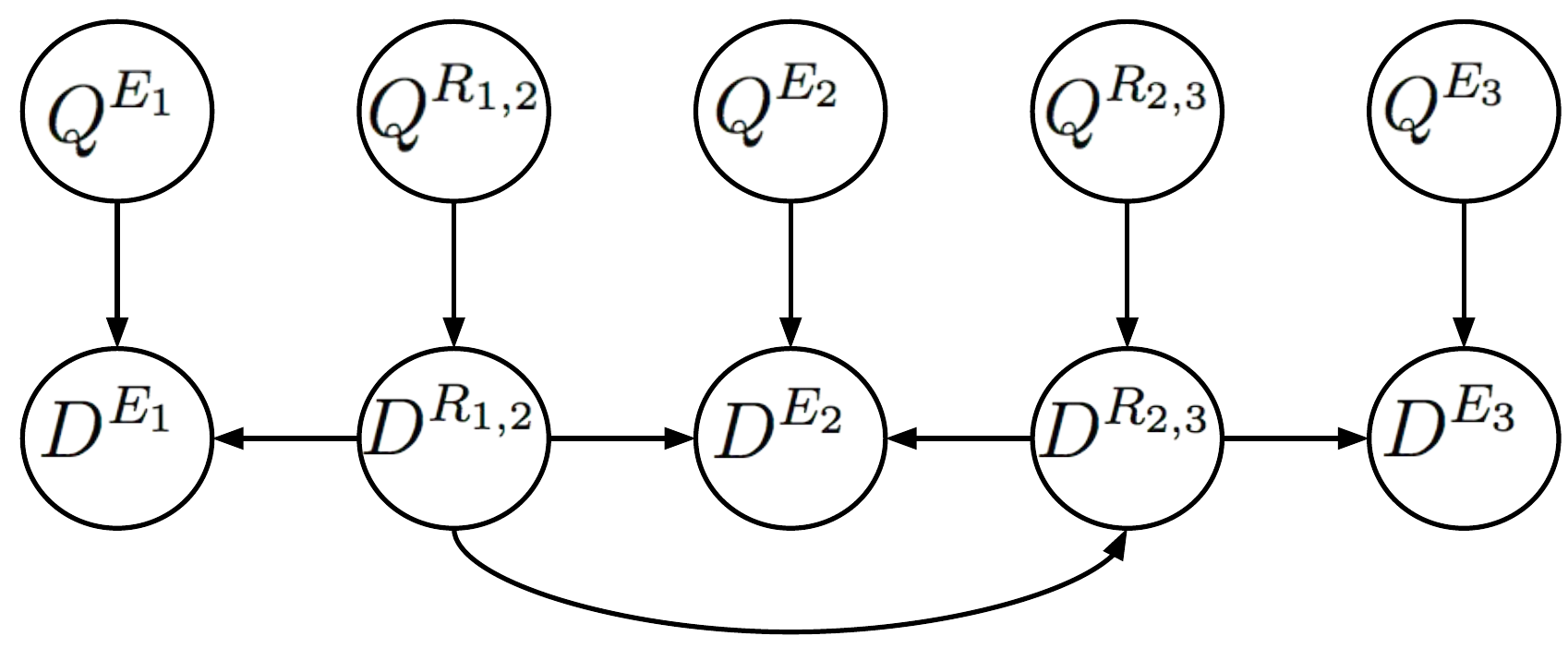}
    \caption{$|Q|=5$}
\end{subfigure}
   \hfill
   \bigskip
\begin{subfigure}{1.0\linewidth}
  \centering
	\includegraphics[width=0.8\textwidth]{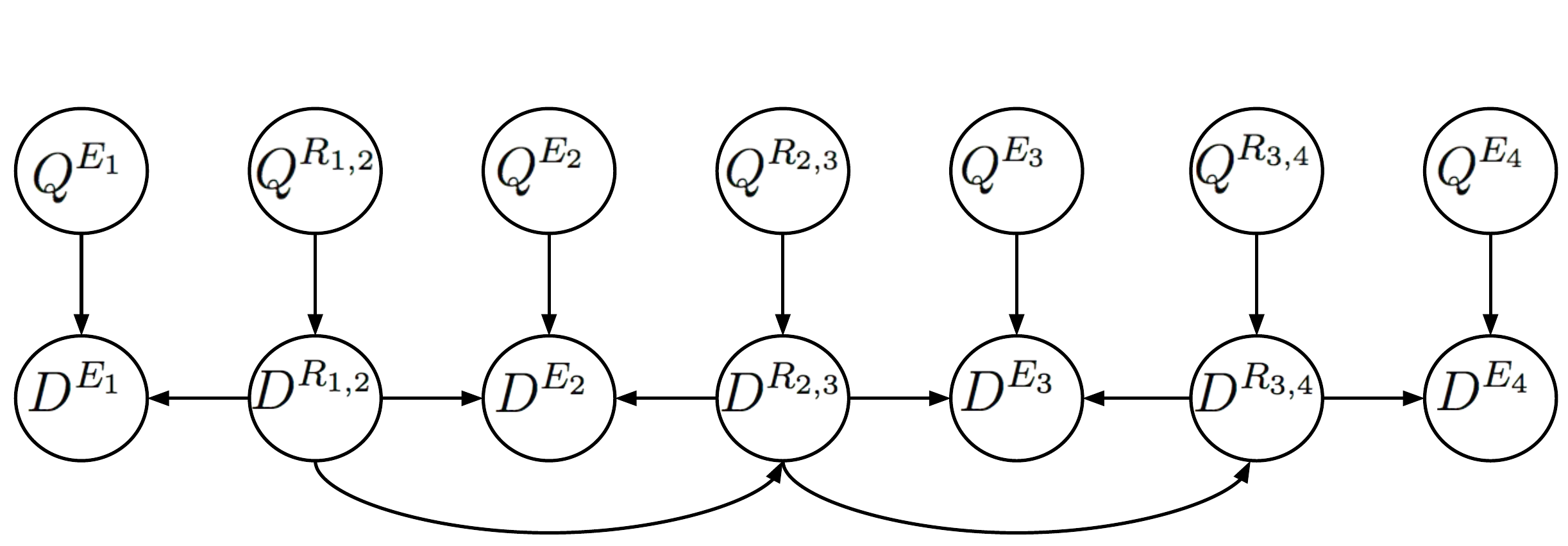}
  \caption{$|Q|=7$}
\end{subfigure} 
\caption{Bayesian networks for E-R Retrieval with queries of different lengths.}
\label{fig:bayesian_er}
\end{figure}

Figure \ref{fig:bayesian_er} depicts the representation of E-R retrieval for different query lengths $|Q|$ using Bayesian networks. We easily conclude that graphical representation contributes to establish a few guidelines for modeling E-R retrieval. First, each sub-query points to the respective document node. Second, relationship document nodes always point to the contiguous entity representations. Last, when there are more than one relationship sub-query, relationship documents also point to the subsequent relationship document.

Once we draw the graph structure for the number of sub-queries in $Q$ we are able to compute a product of conditional probabilities of each node given its parents. Adapting the general joint probability formulation of Bayesian networks to E-R retrieval we come up with the following generalization:

\begin{equation}
P(D^{E},D^{R}|Q) \stackrel{\text{$rank$}}{=}  \prod_{i=1}^{\frac{|Q|+1}{2}} P(D^{E_i}|D^{R_{i-1,i}},D^{R_{i,i+1}}, Q^{E_i})  \prod_{i=1}^{\frac{|Q|-1}{2}}   P(D^{R_{i,i+1}}|D^{R_{i-1,i}},   Q^{R_{i,i+1}})  
\end{equation}

We denote $D^{R}$ as the set of all candidate relationship documents in the graph and $D^{E}$ the set of all candidate entity documents in the graph. In Information Retrieval is often convenient to work in the log-space as it does not affect ranking and transforms the product of conditional probabilities in a summation, as follows:

\begin{align}
\begin{split}\label{eq:erlog}
P(D^{E},D^{R}|Q)   & \stackrel{\text{$rank$}}{=}  \text{log} \ P(D^{E},D^{R}|Q) \\
& \stackrel{\text{$rank$}}{=}  \sum_{i=1}^{\frac{|Q|+1}{2}}  \text{log} P(D^{E_i}|D^{R_{i-1,i}},D^{R_{i,i+1}}, Q^{E_i}) + \sum_{i=1}^{\frac{|Q|-1}{2}}    \text{log} P(D^{R_{i,i+1}}|D^{R_{i-1,i}},   Q^{R_{i,i+1}}) \\
\end{split}\\
\end{align}

We now present two design patterns to compute each conditional probability for every entity and relationship candidate documents.


\section{Design Patterns for Entity-Relationship Retrieval}

Traditional ad-hoc document retrieval approaches create direct term-based representations of raw documents. A retrieval model (e.g. Language Models) is then used to match the information need, expressed as a keyword query, against those representations. However, E-R retrieval requires collecting evidence for both entities and relationships that can be spread across multiple documents. It is not possible to create direct term-based representations. Raw documents serve as proxy to connect queries with entities and relationships.

Abstractly speaking, entity retrieval can be seen as a problem of object retrieval in which the search process is about fusing information about a given object, such as in the case of verticals (e.g. Google Finance). Recently, \citet{zhang2017design} presented two design patterns for fusion-based object retrieval. 

The first design pattern -- Early Fusion -- is an object-centric approach where a term-based representation of objects is created earlier in the retrieval process. First, it creates meta-documents by aggregating term counts across the documents associated with the objects. Later, it matches queries against these meta-documents using standard retrieval methods. 

The second design pattern - Late Fusion - is a document-centric approach where relevant documents to the query are retrieved first and then later in the retrieval process, it ranks objects associated with top documents. These design patterns represent a generalization of Balog's \textit{Model 1} and \textit{Model 2} for expertise retrieval \citep{balog2006formal}.

In essence, E-R retrieval is an extension, or a more complex case, of object-retrieval where besides ranking objects we need to rank tuples of objects that satisfy the relationship expressed in the E-R query. This requires creating representations of both entities and relationships by fusing information spread across multiple raw documents. We propose novel fusion-based  design patterns for E-R retrieval that are inspired from  the design patterns presented by \citet{zhang2017design} for single object-retrieval. We extend those design patterns to accommodate the specificities of E-R retrieval. We hypothesize that it should be possible to generalize the term dependence models to represent entity-relationships and achieve effective E-R retrieval without entity or relationship type restrictions (e.g. categories) as it happens with the Semantic Web based approaches.

\subsection{Early Fusion}
The Early Fusion strategy presented by \citet{zhang2017design} consists in creating a term-based representation for each object under retrieval, i.e., a meta-document containing all terms in the proximity of every object mention across a document collection.  As described in previous section, E-R queries can be formulated as a sequence of multiple entity queries $Q^{E}$ and relationship queries $Q^{R}$. In a Early Fusion approach, each of these queries should match against a previously created term-based representation. Since there are two types of queries, we propose to create two types of term-based representations, one for entities and other for relationships. 

Our Early Fusion design pattern is similar to \textit{Model 1} of \citet{balog2006formal}. It can be thought as creating two types of meta-documents $D^{E}$ and $D^{R}$. A meta-document $D^{E_i}$ is created by aggregating the context terms of the occurrences of $E_i$ across the raw document collection. On the other hand, for each each pair of entities $E_{i-1}$ and $E_i$  that co-occur close together across the raw document collection we aggregate context terms that describe the relationship to create a meta-document $D^{R_{i-1,i}}$

In our approach we focus on sentence level information about entities and relationships although the design pattern can be applied to more complex segmentations of text (e.g. dependency parsing). We rely on Entity Linking methods for disambiguating and assigning unique identifiers to entity mentions on raw documents $D$. We collect entity contexts across the raw document collection and index them in the \textit{entity index}. The same is done by collecting and indexing entity pair contexts in the \textit{relationship index}.  

We define the (pseudo) frequency of a term $t$ for an entity meta-document $D^{E_i}$ as follows:

\begin{equation} \label{tf_e}
f(t,D^{E_i}) =  \sum_{j=1}^{n} f(t,E_i,D_j) w(E_i,D_j)
\end{equation}

where $n$ is the total number of raw documents in the collection,  $f(t,E_i,D_j)$ is the term frequency in the context of the entity $E_i$ in a raw document $D_j$.  $w(E_i,D_j)$ is the entity-document association weight that corresponds to the weight of the document $D_j$ in the mentions of the entity $E_i$ across the raw document collection. Similarly, the term (pseudo) frequency of a term $t$ for a relationship meta-document $D^{R_{i-1,i}}$ is defined as follows:

\begin{equation} \label{tf_r}
f(t,D^{R_{i-1,i}}) =  \sum_{j=1}^{n} f(t,R_{i-1,i},D_j) w(R_{i-1,i},D_j)
\end{equation}

where $f(t,R_{i-1,i},D_j$ is the term frequency in the context of the pair of entity mentions corresponding to the relationship $R_{i-1,i}$ in a raw document $D_j$ and $w(R_{i-1,i},D_j)$ is the relationship-document association weight. In this work we use binary associations weights indicating the presence/absence of an entity mention in a raw document, as well as for a relationship. However, other weight methods can be used.

The relevance score for an entity tuple $T_E$ can then be calculated using the posterior $P(D^{E},D^{R}|Q)$ defined in previous section (equation \ref{eq:erlog}). We calculate the individual conditional probabilities  as a product of a retrieval score with an association weight. Formally we consider:

\begin{align}
\begin{split}\label{ef_eq}
 & \text{log}P(D^{E_i}|D^{R_{i-1,i}},D^{R_{i,i+1}}, Q^{E_i}) = score(D^{E_i}, Q^{E_i}) w(E_i, R_{i-1,i}, R_{i,i+1}) \\
	& \text{log} P(D^{R_{i,i+1}}|D^{R_{i-1,i}}, Q^{R_{i,i+1}}) = score(D^{R_{i,i+1}}, Q^{R_{i,i+1}}) w(R_{i,i+1}, R_{i-1,i}) \\
\end{split}\\
\end{align}

where $score(D^{R_{i,i+1}}, Q^{R_{i,i+1}})$ represents the retrieval score resulting of the match of the query terms of a relationship sub-query $Q^{R_{i,i+1}}$ and a relationship meta-document $D^{R_{i,i+1}}$. The same applies to the retrieval score $score(D^{E_i}, Q^{E_i})$ which corresponds to the result of the match of an entity sub-query $Q^{E_i}$ with a entity meta-document $D^{E_i}$. For computing both $score(D^{R_{i,i+1}}, Q^{R_{i,i+1}})$ and $score(D^{E_i}, Q^{E_i})$ any retrieval model can be used. Different scoring functions will be introduced below. 

We use a binary association weight for $w(E_i, R_{i-1,i}, R_{i,i+1})$ which represents the presence of a relevant entity $E_i$ to a sub-query $Q^{E_i}$ in its contiguous relationships in the Bayesian network, i.e.  $R_{i-1,i}$ and $R_{i,i+1}$ which must be relevant to the sub-queries $Q^{R_{i-1,i}}$ and $Q^{R_{i,i+1}}$. This entity-relationship association weight is the building block that guarantees that two entities relevant to sub-queries $Q^E$ that are also part of a relationship relevant to a sub-query $Q^R$ will be ranked higher than tuples where just one or none of the entities are relevant to the entity sub-queries $Q^E$. On the other hand, the entity-relationship association weight $w(R_{i,i+1}, R_{i-1,i})$ guarantees that consecutive relationships share one entity between them in order to create triples or 4-tuples of entities for longer E-R queries ($|Q|>3$).

The relevance score of an entity tuple $T_E$ given a query $Q$ is calculated by summing individual relationship and entity relevance scores for each $Q^{R_{i-1,i}}$ and $Q^{E_i}$ in $Q$. We define the score for a tuple $T_E$ given a query $Q$ as follows:

\begin{align}
\begin{split}\label{efd_eq}
  P(D^{E},D^{R}|Q) \stackrel{\text{$rank$}}{=} & \sum_{i=1}^{\frac{|Q|+1}{2}} score(D^{E_i}, Q^{E_i}) w(E_i, R_{i-1,i}, R_{i,i+1})+\\
	& \sum_{i=1}^{\frac{|Q|-1}{2}}  score(D^{R_{i,i+1}}, Q^{R_{i,i+1}}) w(R_{i,i+1}, R_{i-1,i})\\
\end{split}\\
\end{align}

Considering Dirichlet smoothing unigram Language Models (LM) the constituent retrieval scores can be computed as follows:
\begin{equation}
score_{LM}(D^{R_{i,i+1}}, Q^{R_{i,i+1}}) = \sum_{t \in D^{R_{i,i+1}} \cap Q^{R_{i,i+1}} } \text{log}  \left (\frac{ f(t,D^{R_{i,i+1}}) +  \frac{f(t,C^R)}{|C^R|}\mu^R}{|D^{R_{i,i+1}}| + \mu^R} \right ) 
\end{equation}

\begin{equation}
score_{LM}(D^{E_i}, Q^{E_i}) = \sum_{t \in D^{E_i}  \cap Q^{E_i}} \text{log} \left (\frac{f(t,D^{E_i})
 + \frac{ f(t,C^E)}{|C^E|}\mu^E }{|D^{E_i}| + \mu^E} \right )
\end{equation}

where $t$ is a term of a sub-query $Q^{E_i}$ or $Q^{R_{i,i+1}}$, $f(t,D^{E_i})$ and
$f(t,D^{R_{i,i+1}})$ are the (pseudo) frequencies defined in equations \ref{tf_e} and \ref{tf_r}. The collection frequencies $f(t,C^E)$, $f(t,C^R)$ represent the frequency of the term $t$ in either the \textit{entity index} $C^E$ or in the \textit{relationship index} $C^R$. $|D^{E_i}|$ and$|D^{R_{i,i+1}}|$ represent the total number of terms in a meta-document while $|C^R|$ and $|C^E|$ represent the total number of terms in a collection of meta-documents. Finally, $\mu^E$ and $\mu^R$ are the Dirichlet prior for smoothing which generally corresponds to the average document length in a collection.

\subsection{Association Weights}

Both Early Fusion and Late Fusion share three components: $w(R_{i,i+1},D_j)$, $w(E_i,D_j)$ and $w(E_i, R_{i,i+1})$. The first two represent document associations which determine the weight a given raw document contributes to the relevance score of a particular entity tuple $T_E$. The last one is the entity-relationship association which indicates the strength of the connection of a given entity $E_i$ within a relationship $R_{i,i+1}$. 

In our work we only consider binary association weights but other methods could be used. According to the binary method we define the weights as follows:

\begin{equation}
w(R_{i,i+1},D_j) = 1 \ \text{if}\ R(E_i,E_{i+1})\ \in\ D_j \ , 0 \ \text{otherwise}
\end{equation}
\begin{equation}
w(E_i,D_j) = 1 \ \text{if}\ E_i\ \in\ D_j \ , 0 \ \text{otherwise}
\end{equation}
\begin{equation}
w(E_i, R_{i-1,i},R_{i,i+1}) = 1 \ \text{if}\ E_i\ \in\ D^{R_{i-1,i}} and \  E_i\ \in\ D^{R_{i,i+1}} \ , 0 \ \text{otherwise}
\end{equation}
\begin{equation}
w(R_{i,i+1},R_{i-1,i}) = 1 \ \text{if}\ E_i\ \in\ D^{R_{i-1,i}} and \  E_i\ \in\ D^{R_{i,i+1}} \ , 0 \ \text{otherwise}
\end{equation}

 Under this approach the weight of a given association is independent of the number of times an entity or a relationship occurs in a document. A more general approach would be to assign real numbers to the association weights depending on the strength of the association \citep{balog2012expertise}. For instance, uniform weighting would be proportional to the inverse of the number of documents where a given entity or relationship occurs. Other option could be a TF-IDF approach.

\subsection{Early Fusion Example}

Let us consider an illustrative example of the Early Fusion design pattern for E-R retrieval using unigram Language Models and the E-R query $Q = \{$\textit{soccer players who dated a top model}$\}$. This query can be decomposed in three sub-queries, $Q^{E_i} = \{$ \textit{soccer players}$\}$, $Q^{E_{i+1}} = \{$ \textit{top model}$\}$ and $Q^{R_{i,i+1}} = \{$ \textit{dated} $\}$. The first two sub-queries target the \textit{entity index} and the last targets the \textit{relationship index}. Table \ref{entities_example} presents a toy \textit{entity index} with 3 entities as example for each of the two entity sub-queries, including the term frequency $f(t,D^{E_i})$ for each sub-query term.

\begin{table}[h]
\centering
\caption{Illustrative example of the \textit{entity index} in Early Fusion.}
\label{entities_example}
\begin{tabular}{|l|l|l|}
\hline
$E_i$          & $f(t,D^{E_i})$                                                           & $|D^{E_i}|$ \\ \hline
$<$Tom Brady$>$    & \begin{tabular}[c]{@{}l@{}}soccer:0\\ player:600\end{tabular} &  3000    \\ \hline
$<$Cristiano Ronaldo$>$        & \begin{tabular}[c]{@{}l@{}}soccer:800\\ player:800\end{tabular}                 &  5000    \\ \hline
$<$Lionel Messi$>$        & \begin{tabular}[c]{@{}l@{}}soccer:700\\ player:700\end{tabular}                 &  4000    \\ \hline
$<$Lu\'is Figo$>$ & \begin{tabular}[c]{@{}l@{}}soccer:200\\ player:200\end{tabular}                 &  800    \\ \hline
$<$Gisele Bundchen$>$    & \begin{tabular}[c]{@{}l@{}}top:400\\ model:400\end{tabular}           &   3000   \\ \hline
$<$Irina Shayik$>$    & \begin{tabular}[c]{@{}l@{}}top:300\\ model:300\end{tabular}                 &  2000    \\ \hline
$<$Helen Svedin$>$    & \begin{tabular}[c]{@{}l@{}}top:150\\ model:150\end{tabular}                 &  600    \\ \hline
... & ...                &  ...    \\ \hline
\end{tabular}
\end{table}

Considering the remaining variables required to calculate the $score_{LM}(D^{E_i}, Q^{E_i})$:

\begin{itemize}
\item [] $|C^E|$ = 100000
\item [] $|\mu^E|$ = 1500
\item [] $f(\textit{soccer},C^{E})$ = 3000
\item [] $f(\textit{player},C^{E})$ = 8000
\item [] $f(\textit{top},C^{E})$ = 8000
\item [] $f(\textit{model},C^{E})$ = 4000
\end{itemize}

We calculate the $score_{LM}(D^{E_i}, Q^{E_i})$ for the respective entities and sub-queries. For the first entity query -- ``soccer players'' -- the ranked list of relevant entities and the respective LM score would be the following:

\begin{enumerate}
\item $<$Lionel Messi$>$: -1.6947
\item $<$Cristiano Ronaldo$>$: -1.7351
\item $<$Lu\'is Figo$>$: -1.8291
\item $<$Tom Brady$>$: -2.7958
\end{enumerate}

For the second entity query -- ``top models'':

\begin{enumerate}
\item $<$Gisele Bundchen$>$: -1.6295
\item $<$Irina Shayik$>$: -1.7093
\item $<$Helen Svedin$>$: -1.9698
\end{enumerate}

Table \ref{rel_example} shows 3 relationships, i.e. entity pairs, relevant to the sub-query ``dated'' and the respective term frequency $f(t,D^{R_{i,i+1}})$.

\begin{table}[h]
\centering
\caption{Illustrative example of the \textit{relationship index} in Early Fusion.}
\label{rel_example}
\begin{tabular}{|l|l|l|}
\hline
$R_{i,i+1}$           & $f(t,D^{R_{i,i+1}})$        & $|D^{R_{i,i+1}}|$ \\ \hline
$<$Gisele Bundchen, Tom Brady$>$    & dated:500 &  800    \\ \hline
$<$Irina Shayik, Cristiano Ronaldo$>$    & dated:300 &  600    \\ \hline
$<$Helen Svedin, Lu\'is Figo$>$         & dated:100         &  200    \\ \hline
...    & ... &  ...    \\ \hline

\end{tabular}
\end{table}

Considering the remaining variables required to calculate the $score_{LM}(D^{R_{i,i+1}}, Q^{R_{i,i+1}})$:

\begin{itemize}
\item [] $|C^R|$ = 20000
\item [] $|\mu^R|$ = 500
\item [] $f(\textit{dated},C^{R})$ = 5000
\end{itemize}

We calculate the $score_{LM}(D^{R_{i,i+1}}, Q^{R_{i,i+1}})$for the respective relationship and the sub-query and we obtain the following ranked list:

\begin{enumerate}
\item $<$Gisele Bundchen, Tom Brady$>$:  -0.3180
\item $<$Irina Shayik, Cristiano Ronaldo$>$: -0.4130
\item $<$Helen Svedin, Lu\'is Figo$>$:    -0.4929
\end{enumerate}

We can now sum up individual scores for each sub-query and calculate the final score for the early fusion design pattern $score(T_{E}, Q)$ using the equation \ref{efd_eq}. The final ranked list of tuples is the following:

\begin{enumerate}
\item $<$Irina Shayik, Cristiano Ronaldo$>$: -3.8575
\item $<$Helen Svedin, Lu\'is Figo$>$:   -4.2919
\item $<$Gisele Bundchen, Tom Brady$>$:  -4.6977
\end{enumerate}

The entity tuple $<$Irina Shayik, Cristiano Ronaldo$>$ is the most relevant to the query ``soccer players who dated a top model''. Although $<$Gisele Bundchen, Tom Brady$>$ has higher individual scores in two sub-queries (``top model'' and ``dated'') it ranks last due to the poor relevance of Tom Brady to the sub-query ``soccer player''. The entity $<$Lionel Messi$>$ is the most relevant entity to the sub-query ``soccer player'' but it is not relevant to the relationship sub-query, therefore it is excluded from the final ranked list of entity tuples. 

\subsection{Late Fusion}

The Late Fusion design pattern presented by \citet{zhang2017design} is a document-centric strategy, i.e. first we query raw individual documents then we aggregate the associated objects with the relevant documents. Instead of creating term-based representations of entities and relationships (pairs of entities), in late fusion we use the raw documents as hidden variables, separating the E-R query from the relevant entity tuples to be retrieved. 

Our vision of ORM implies processing raw documents to detect entities occurrences and extract sentence level information that will be used in downstream Entity Retrieval and Text Mining tasks. Therefore, we are not interested in applying a Late Fusion strategy in this work. However, we believe it makes sense to present a theoretical formulation of a Late Fusion design pattern for E-R retrieval. We leave the practical experiments with Late Fusion for future work in the context of generic E-R retrieval.

The process of retrieving entity tuples using our late fusion strategy consists in processing each sub-query independently, as in the early fusion strategy, but in this case, we use a single index comprising a term based representation of the collection of raw documents. A retrieval model is used to calculate a relevance score of each individual raw document and a given sub-query. Once we have the relevant documents we use entity linking to extract the entities that are mentioned in each relevant raw document. Following this strategy we calculate aggregated counts of entity occurrences weighted by the individual relevance score of the individual raw documents. At the end, we join the results of each sub-query and calculate the overall relevance score of the entity tuples. 

Formally, we define the relevance score of an entity tuple $T_E$ given a query $Q$ as follows:

\begin{align}
\begin{split}\label{lf_eq}
  P(D^{E},D^{R}|Q) \stackrel{\text{$rank$}}{=} &  \sum_{i=1}^{\frac{|Q|+1}{2}} \sum_{j=1}^{n} score(D_j, Q^{E_i}) w(E_i,D_j) w(E_i,R_{i-1,i}, R_{i,i+1})+\\
	& \sum_{i=1}^{\frac{|Q|-1}{2}} \sum_{j=1}^{n} score(D_j, Q^{R_{i,i+1}}) w(R_{i,i+1},D_j) w(R_{i,i+1},R_{i-1,i})\\
\end{split}\\
\end{align}

where $score(D_j, Q^{R_{i,i+1}})$ represents the retrieval score resulting of the match of the query terms of a relationship sub-query $Q^{R_{i,i+1}}$ and a raw document $D_j$. The same applies to the retrieval score $score(D_j, Q^{E_i})$ which corresponds to the result of the match of an entity sub-query $Q^{E_i}$ with a raw document $D_j$. The weights $w(R_{i,i+1},D_j)$ and $w(E_i,D_j)$ represent association weights between relationships and raw documents, and entities and raw documents, respectively. We use binary association weights in this work but other weights can be used. We also use a binary association weight for $w(E_i,R_{i-1,i}, R_{i,i+1})$ and $w(R_{i,i+1},R_{i-1,i})$ which represent the entity-relationship association weights, similarly to what happens with the case of Early Fusion. 

For computing both $score(D_j, Q^{R_{i,i+1}})$ and $score(D_j, Q^{E_i})$ any retrieval model can be used. Considering BM25 the scores can be computed as follows:

\begin{equation}
score_{BM25}(D_j, Q^{R_{i,i+1}}) = \sum_{t \in D_j  \cap Q^{R_{i,i+1}} } \text{log} \frac{N - n(t) + 0.5}{n(t) + 0.5}   \ . \ \frac{ f(t,D_j)(K_1 + 1)} {  f(t,D_j) + K_1 (1 - b + b \frac{|D_j|}{avg(|D|)})} 
\end{equation}

\begin{equation}
score_{BM25}(D_j, Q^{E_i}) = \sum_{t \in D_j  \cap Q^{E_{i}}} \text{log} \frac{N - n(t) + 0.5}{n(t) + 0.5} \ . \   \frac{ f(t,D_j) (K_1 + 1)} {  f(t,D_j) + K_1 (1 - b + b \frac{|D_j|}{avg(|D|)})}
\end{equation}

where $t$ is a term of a sub-query $Q^{E_i}$ or $Q^{R_{i,i+1}}$ and $f(t,D_j)$
is the query term frequency in a raw document $D_j$. The inverse document frequency, $IDF(t)$, is computed as log$\frac{N - n(t) + 0.5}{n(t) + 0.5}$ with $N$ as the number of documents on the  collection and $n(t)$ the number of documents where the term occurs.$|D_j|$ is the total number of terms in a raw document $D_j$ and $avg(|D|)$ is the average document length. $K_1$ and $b$ are free parameters usually chosen as 1.2 and 0.75, in the absence of specific optimization.

\subsection{Late Fusion Example}

\begin{table}[h]
\centering
\caption{Illustrative example of the \textit{document index} in Late Fusion.}

\label{lateraw}
\begin{tabular}{|l|l|l|l|}
\hline
$D_j$ & $f(t,D_j)$                                                                    & $|D_j|$ & $E_i$ \\ \hline
docid-1   & \begin{tabular}[c]{@{}l@{}}soccer:10\\ player:10\end{tabular}                 &  200 & \begin{tabular}[c]{@{}l@{}}<Cristiano Ronaldo>\\ <Lionel Messi>\end{tabular}    \\ \hline
docid-2   & \begin{tabular}[c]{@{}l@{}}soccer:5\\ player:5\end{tabular}        &  150 & \begin{tabular}[c]{@{}l@{}}<Cristiano Ronaldo>\end{tabular}    \\ \hline
docid-3   & \begin{tabular}[c]{@{}l@{}}soccer:5\\ player:5\end{tabular}        &  100 & \begin{tabular}[c]{@{}l@{}}<Lu\'is Figo>\end{tabular}    \\ \hline
docid-4   & \begin{tabular}[c]{@{}l@{}}top:4\\ model:4\end{tabular}        &  150 & \begin{tabular}[c]{@{}l@{}}<Gisele Bundchen>\end{tabular}    \\ \hline
docid-5   & \begin{tabular}[c]{@{}l@{}}dated:5\end{tabular}        &  80 & \begin{tabular}[c]{@{}l@{}}<Gisele Bundchen>\\<Tom Brady>\end{tabular}    \\ \hline
docid-6   & \begin{tabular}[c]{@{}l@{}}top:6\\ model:6\end{tabular}        &  100 & \begin{tabular}[c]{@{}l@{}}<Irina Shayik>\end{tabular}    \\ \hline
docid-7   & \begin{tabular}[c]{@{}l@{}}model:4\\ dated:2 \\player:2\end{tabular}        &  100 & \begin{tabular}[c]{@{}l@{}}<Gisele Bundchen>\\<Adriana Lima>\\<Tom Brady>\end{tabular}    \\ \hline
docid-8   & \begin{tabular}[c]{@{}l@{}}\\dated:3\end{tabular}        &  120 & \begin{tabular}[c]{@{}l@{}}<Irina Shayik>\\<Cristiano Ronaldo>\end{tabular}    \\ \hline
docid-9   & \begin{tabular}[c]{@{}l@{}}top:2\\model:2\\ dated:2 \\soccer:2\\player:2\end{tabular}        &  150 & \begin{tabular}[c]{@{}l@{}}<Lu\'is Figo>\\<Helen Svedin>\end{tabular}    \\ \hline
...   & ...       &  ... & ...    \\ \hline

\end{tabular}
\end{table}

Considering the same toy example query introduced in the previous sub-section, we now have a single index, the \textit{document index}, as illustrated in  Table \ref{lateraw}. The remaining parameters required for calculating the $score_{BM25}(D_j, Q^{E_i})$ and $score_{BM25}(D_j, Q^{R_{i,i+1}})$ are the following:

\begin{itemize}
\item N=2000
\item n(soccer)=100
\item n(player)=130
\item n(dated)=60
\item n(top)=250
\item n(model)=80
\item avg($|D|$)=120
\end{itemize}

For the first entity sub-query, ``soccer players'', the relevant documents ranked by the $score_{BM25}(D_j, Q^{E_i})$ are the following:

\begin{enumerate}
\item docid-1 ($<$Cristiano Ronaldo$>$, $<$Lionel Messi$>$): 4.7606 

\item docid-3 ($<$Lu\'is Figo$>$): 4.6426

\item docid-2 ($<$Cristiano Ronaldo$>$): 4.3716

\item docid-9 ($<$Lu\'is Figo$>$, $<$Helen Svedin$>$): 3.2803 

\item docid-7 ($<$Gisele Bundchen$>$, $<$Adriana Lima$>$, $<$Tom Brady$>$): 1.8418

\end{enumerate}

For the second entity sub-query, ``top model'':

\begin{enumerate}
\item docid-6 ($<$Irina Shayik$>$): 3.1618
\item docid-4 ($<$Gisele Bundchen$>$): 2.7393 
\item docid-9 ($<$Lu\'is Figo$>$, $<$Helen Svedin$>$): 2.1694 
\item docid-7 ($<$Gisele Bundchen$>$, $<$Adriana Lima$>$, $<$Tom Brady$>$): 1.4714 

\end{enumerate}

For the relationship sub-query, ``dated'':

\begin{enumerate}
\item docid-5 (<Gisele Bundchen>, <Tom Brady>): 2.8081

\item docid-8 (<Irina Shayik>, <Cristiano Ronaldo>): 2.3668 

\item docid-7 ($<$Gisele Bundchen$>$, $<$Adriana Lima$>$, $<$Tom Brady$>$): 2.1728 

\item docid-9 ($<$Lu\'is Figo$>$, $<$Helen Svedin$>$): 1.9349 

\end{enumerate}

Since in Late Fusion there is no relationship meta-documents that could be used directly as entity tuples, we need to extract the candidate tuples from the raw documents retrieved using the relationship sub-query. When there are more than two entity associations in a relevant document we combine entities to create tuples. For instance, docid-7 has three entity associations therefore we extract three candidate tuples: $<$Gisele Bundchen, Tom Brady$>$, $<$Gisele Bundchen, Adriana Lima$>$ and $<$Adriana Lima, Tom Brady$>$. 

For each candidate tuple we sum up $scoreBM25(D_j, Q^{R_{i,i+1}}) w(R_{i,i+1},D_j)$ over every relevant document $D_j$ for the relationship sub-query that is associated with each entity tuple. The same applies to individual entities from the candidate tuples that are associated with relevant documents for each entity sub-query. For instance, for the entity sub-query ``soccer players'' we sum  $score(D_j, Q^{E_i}) w(E_i,D_j) w(E_i, R_{i,i+1})$ over the relevant documents that mentioned an entity that belongs to a candidate tuple.

When both entities of the candidate tuple are mentioned in relevant documents for both entity sub-queries, e.g. 
$<$Helen Svedin, Lu\'is Figo$>$, we assign each entity to the sub-query that maximizes the final score $score(T_{E}, Q)$, i.e., we use the scores of the entity sub-query ``soccer player'' for $<$Lu\'is Figo$>$ and the entity sub-query ``top model'' for $<$Helen Svedin$>$. The final ranked list of entity tuples is the following:

\begin{enumerate}
\item $<$Irina Shayik, Cristiano Ronaldo$>$: 14.0443
\item $<$Helen Svedin, Lu\'is Figo$>$:  12.5970 
\item $<$Gisele Bundchen, Tom Brady$>$:  10.9784
\item $<$Gisele Bundchen, Adriana Lima$>$:  9.3459
\item $<$Adriana Lima, Tom Brady$>$:  6.9245
\end{enumerate}

Once again <Lionel Messi> is excluded from the final ranked list of entity tuples because he is not associated with any document relevant to the relationship sub-query ``dated''. On the other hand, <Adriana Lima> is included in the final ranking although it is not true that she has dated either <Tom Brady> or <Gisele Bundchen>. In this example, the top three entity tuples are ranked in the same order as in the Early Fusion strategy example.

\subsection{Implementation}

In this section we proposed two design patterns for E-R retrieval: Early Fusion (EF) and Late Fusion (LF). Both can be seen as a flexible framework for ranking tuples of entities given a E-R query expressed as a sequence of entity and relationship sub-queries. 

This framework is flexible enough to allow using any retrieval method to compute individual retrieval scores between document and query nodes in a E-R graph structure. When using Language Models (LM) or BM25 as scoring functions, these design patterns can be used to create unsupervised baseline methods for E-R retrieval (e.g. EF-LM, EF-BM25, LF-LM, LF-BM25, etc.).

In the case of Early Fusion there is some overhead over traditional document search, since we need to create two E-R dedicated indexes that will store entity and relationship meta-documents. The \textit{entity} index is created by harvesting the context terms in the proximity of every occurrence of a given entity across the raw document collection. This process must be carried for every entity in the raw document collection. A similar process is applied to create the relationship index. For every two entities occurring close together in a raw document we extract the text between both occurrences as a term-based representation of the relationship between the two. Once again, this process must be carried for every pair of co-occurring entities in sentences across the raw document collection.  

Late Fusion requires less overhead and can be implemented on top of a web search engine with reduced effort. We only need to have a list of entity occurrences alongside each document. Therefore there is no need to create a separate index(es). On the other hand, it requires more processing on query time since we need to first rank raw documents for each sub-query and then aggregate entity occurrences at the top k documents retrieved. Moreover, it does not contain any proximity-based information on the entity occurrences, so two entities occurring very far in the text might be considered as relationship candidates. It might be prone to a higher false positive rate.   

One advantage of Early Fusing lies in its flexibility as we need to create two separate indexes for E-R retrieval it is possible to combine data from multiple sources in seamless way. For instance, one could use a well established knowledge base (e.g. DBpedia) as entity index and use a specific collection, such as a news collection or a social media stream, for harvesting relationships having a more transient nature.

Common to both design patterns is a challenge inherent to the problem of E-R retrieval: the size of the search space. Although the E-R problem is formulated as a sequence of independent sub-queries, the results of those sub-queries must be joined together. Consequently, we have a multi-dimensional search space in which we need to join results based on shared entities. 

This problem becomes particularly hard when sub-queries are short and contain very popular terms. Let us consider ``actor'' as $Q^{E_i}$, there will be many results to this sub-query, probably thousands. There is a high probability that will need to process thousands of sub-results before finding one entity that is also relevant to the relationship sub-query $Q^{R_{i-1,i}}$. If at the same time we have computational power constraints, we will probably apply a strategy of just considering top k results for each sub-query which can lead to reduced recall in the case of short sub-queries with popular terms.


\section{Entity-Relationship Dependence Model}

In this section we present the Entity-Relationship Dependence Model (ERDM), a novel supervised Early Fusion-based model for E-R retrieval. Recent approaches to entity retrieval \citep{zhiltsov2015fielded,nikolaev2016parameterized,hasibi2016exploiting} have demonstrated that using models based on Markov Random Field (MRF) framework for retrieval \citep{metzler2005markov} to incorporate term dependencies can improve entity search performance. This suggests that MRF could be used to model E-R query term dependencies among entities and relationships documents.

One of the advantages of the MRF framework for retrieval is its flexibility, as we only need to construct a graph $G$ representing dependencies to model, define a set of non-negative potential functions $\psi$ over the cliques of $G$ and to learn the parameter vector $\Lambda$ to score each document $D$ by its unique and unnormalized joint probability with $Q$ under the MRF \cite{metzler2005markov}.

The non-negative potential functions are defined using an exponential form $\psi(c;\Lambda) = \text{exp} [\lambda_c f(c)]$, where $\lambda_c$  is a feature weight, which is a free parameter in the model, associated with feature function $f(c)$. Learning to rank is then used to learn the feature weights that minimize the loss function. The model allows parameter and feature functions sharing across cliques of the same configuration, i.e. same size and type of nodes (e.g. 2-cliques of one query term node and one document node).

\subsection{Graph Structures}

The Entity-Relationship Dependence Model (ERDM) creates a MRF for modeling implicit dependencies between sub-query terms, entities and relationships. Each entity and each relationship are modeled as document nodes within the graph and edges reflect term dependencies. Contrary to traditional ad-hoc retrieval using MRF (e.g. SDM), where the objective is to compute the posterior of a single document given a query, the ERDM allows the computation of a joint posterior of multiple documents (entities and relationships) given a E-R query which consists also of multiple sub-queries.

\begin{figure}[h] 
\centering
\includegraphics[width=0.6\textwidth]{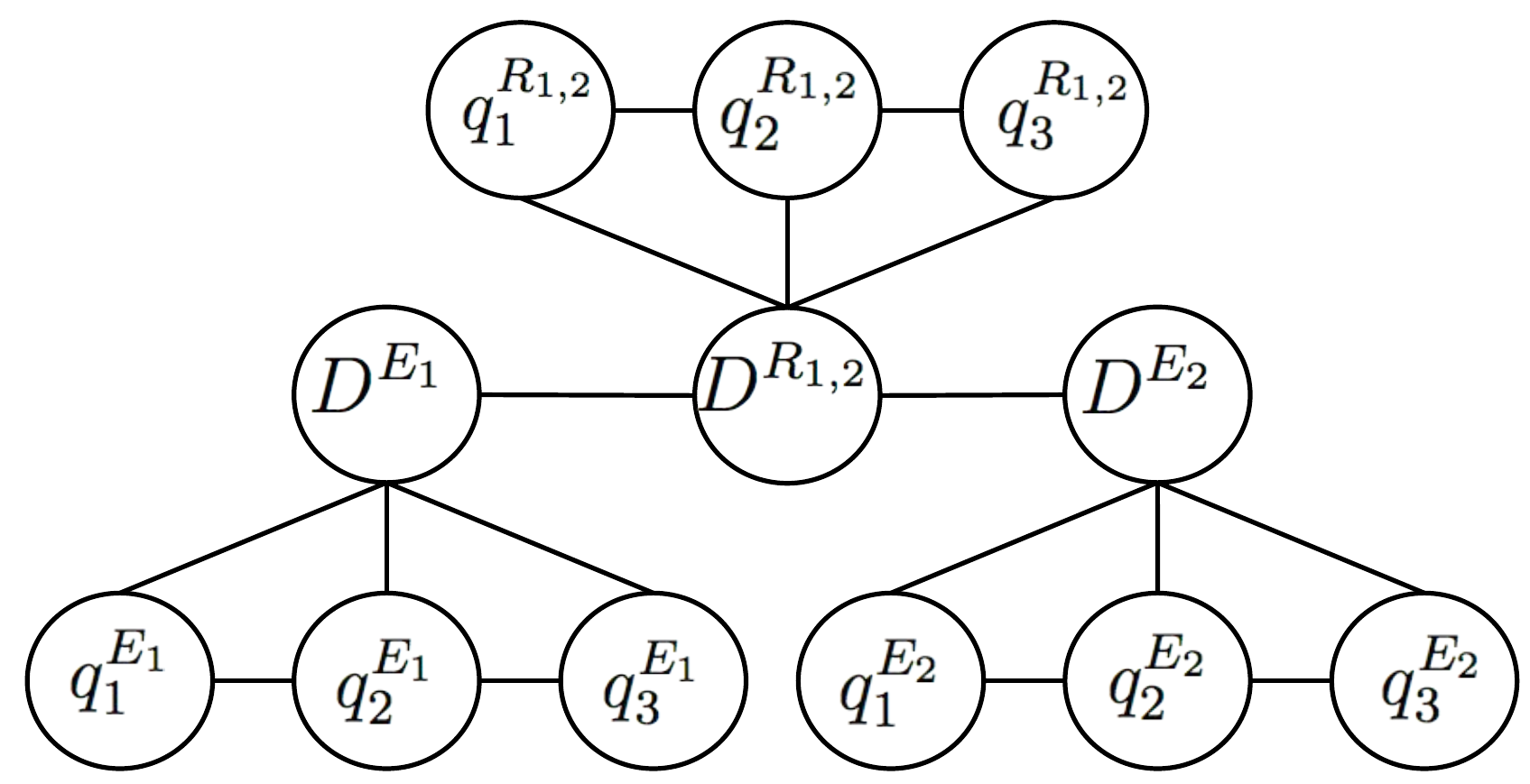}
\caption{Markov Random Field dependencies for E-R retrieval, $|Q|=3$.}\label{fig:erdm3}
\end{figure}

\begin{figure}[h] 
\centering
\includegraphics[width=0.9\textwidth]{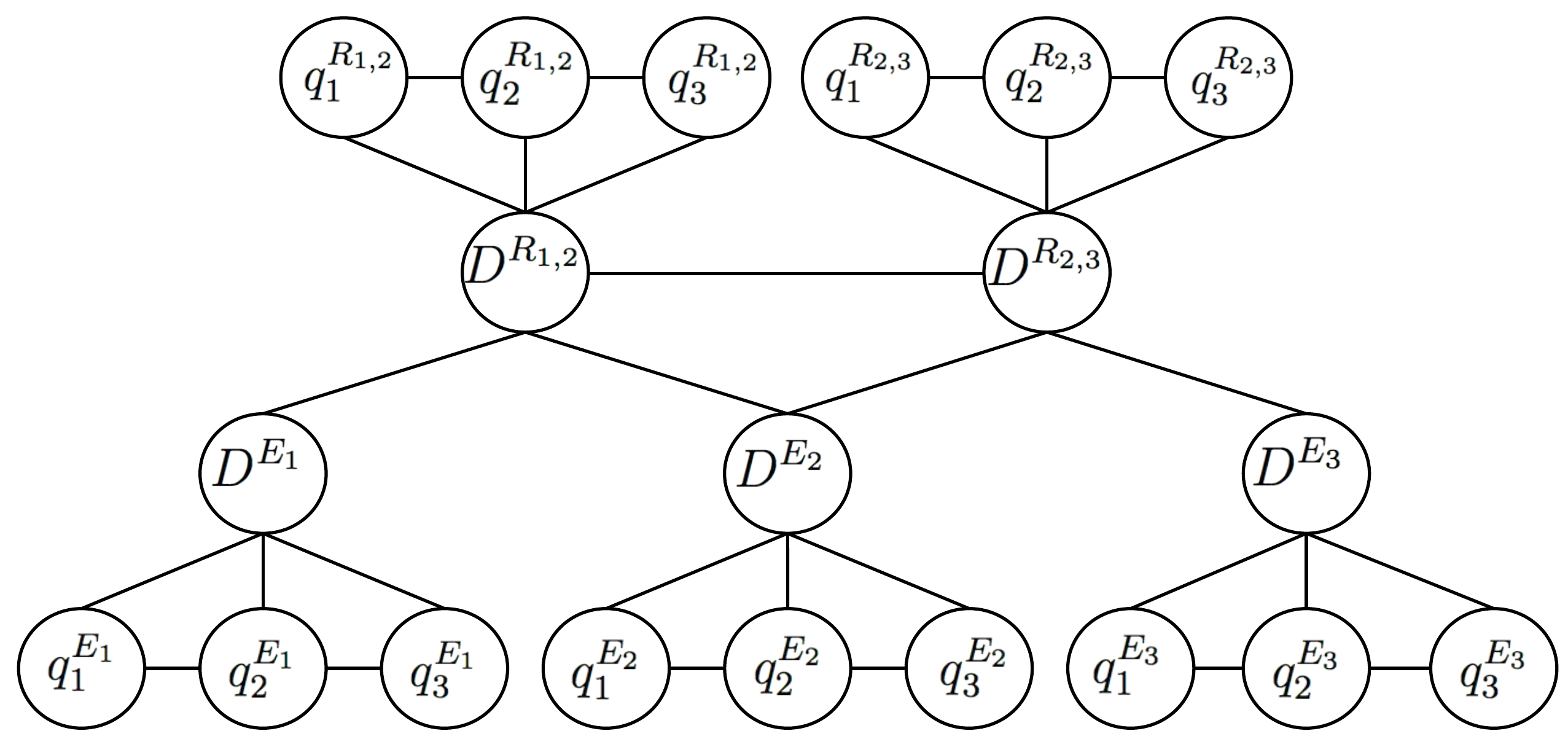}
\caption{Markov Random Field dependencies for E-R retrieval, $|Q|=5$.}\label{fig:erdm5}
\end{figure}

The graph structures of the ERDM for two E-R queries, one with $|Q|=3$ and other with $|Q|=3$ are depicted in Figure \ref{fig:erdm3} and Figure \ref{fig:erdm5}, respectively. Both graph structures contain two different types of query nodes and document nodes: entity query and relationship query nodes, $Q^E$ and $Q^R$, plus entity and relationship document nodes, $D^E$ and $D^R$. Within the MRF framework, $D^E$ and $D^R$ are considered ``documents'' but they are not actual real documents but rather objects representing an entity or a relationship between two entities. Unlike real documents, these objects do not have direct and explicit term-based representations. Usually, it is necessary to gather evidence across multiple real documents that mention the given object, in order to be able to match
them against keyword queries. Therefore, ERDM can be seen as Early Fusion-based retrieval model. The existence of two different types of documents implies two different indexes: the \textit{entity} index and the \textit{relationship} index.

The relationship-specific dependencies of ERDM are found in the 2-cliques formed by one entity document and one relationship document: $D^{E_{i-1}}$ - $D^{R_{i-1,i}}$, $D^{E_i}$- $D^{R_{i-1,i}}$  and for $|Q|=5$,  $D^{E_i}$- $D^{R_{i,i+1}}$ and $D^{R_{i-1,i}}$ - $D^{R_{i,i+1}}$. The graph structure does not need to assume any explicit dependence between entity documents given a relationship document. They have an implicit connection through the dependencies with the relationship document. The likelihood of observing an entity document $D^{E_i}$ given a relationship document $D^{R_{i-1,i}}$ is not affected by the observation of any other entity document. 

Explicit dependence between the two entity documents could be used to represent the direction of the relationship between the two entities. To support this dependence, relationship documents would need to account the following constraint: $R(E_{i-1},E_i) \neq R(E_{i},E_{i-1}), \text{ } \forall \text{ }  D^{R_{i-1,i}} \in C^R$, with $C^R$ representing the \textit{relationship} index. Then, we would compute an ordered feature function between entities in a relationship, similar to the ordered bigram feature function in SDM. In this work, we do not explicitly model asymmetric relationships. For instance, if a user searches for the relationship entity A ``criticized'' entity B but was in fact entity B who criticized entity A we assume that the entity tuple   <entity A, entity B> is still relevant for the information need expressed in the E-R query.

ERDM follows the SDM \citep{metzler2005markov} dependencies between query terms and documents due to its proved effectiveness in multiple contexts. Therefore, ERDM assumes a dependence between neighboring sub-query terms: 

\begin{equation}
P(q^{E_i}_{j}|D^{E_i},q^{E_i}_{j \neq l}) = P(q^{E_i}_{j}|D^{E_i},q^{E_i}_{j-1},q^{E_i}_{j+1})
\end{equation}

\begin{equation}
P(q^{R_{i-1,i}}_{j}|D^{R_{i-1,i}},q^{R_{i-1,i}}_{j \neq l},D^{E_i}) = P(q^{R_{i-1,i}}_{j}|D^{R_{i-1,i}},q^{R_{i-1,i}}_{j-1},q^{R_{i-1,i}}_{j+1})
\end{equation}

MRF for retrieval requires the definition of the sets of cliques (maximal or non-maximal) within the graph that one or more feature functions is to be applied to. The set of cliques in ERDM containing at least one document are the following:

\begin{itemize}
\item $T^{E}$ - set of 2-cliques containing an entity document node and exactly one term in a entity sub-query.

\item $O^{E}$ - set of 3-cliques containing an entity document node and two ordered terms in a entity sub-query.

\item $T^{R}$ - set of 2-cliques containing a relationship document node and exactly one term in a relationship sub-query.

\item $O^{R}$ - set of 3-cliques containing a relationship document node and two ordered terms in a relationship sub-query.

\item $S^{ER}$ - set of 2-cliques containing one entity document node and one relationship document node.

\item $S^{RER}$ - set of 3-cliques containing one entity document node and two consecutive relationship document nodes.

\end{itemize}

The joint probability mass function of the MRF is computed using the set of potential functions over the configurations of the maximal cliques in the graph \citep{metzler2005markov}. Non-negative potential functions are constructed from one or more real valued feature functions associated with the respective feature weights using an exponential form.

\subsection{Feature Functions}

ERDM has two types of feature functions: textual and non-textual. Textual feature functions measure the textual similarity between one or more sub-query terms and a document node. Non-textual feature functions measure compatibility between entity and relationship documents, i.e., if they share a given entity.

\begin{table}[h]
\centering
\caption{Clique sets and associated feature functions by type and input nodes.}
\label{tab:cliqfeats}
\begin{tabular}{|l|l|l|l|}
\hline
Clique Set & Feature Functions  & Type & Input Nodes \\ \hline
$T^{E}$   & $f^{E}_T$  & Textual &  $\{q^{E_i}_j,D^{E_i}\}$           \\ \hline
$O^{E}$ & $f^{E}_O$ and $f^{E}_U$ &Textual &   $\{q^{E_i}_j,q^{E_i}_{j+1},D^{E_i}\}$           \\ \hline
  $T^{R}$  & $f^{R}_T$  &Textual &   $\{q^{R_{i-1,i}}_j,D^{R_{i-1,i}}\}$           \\ \hline

$O^{R}$& $f^{R}_O$ and $f^{E}_U$   &Textual &   $\{q^{R_{i-1,i}}_j,q^{R_{i-1,i}}_{j+1} ,D^{R_{i-1,i}}\}$           \\ \hline
$S^{ER}$   & $f^{ER}_S$  &Non-textual &   $\{D^{E_i},D^{R_{i-1,i}}\}$           \\ \hline
$S^{RER}$   & $f^{RER}_S$  &Non-textual &   $\{D^{E_i},D^{R_{i-1,i}},D^{R_{i,i+1}}\}$           \\ \hline

\end{tabular}
\end{table}

Table \ref{tab:cliqfeats} presents an overview of the feature functions associated with clique sets and the type of input nodes. Although we could define a wide set of different feature functions, we decided to adapt SDM textual feature functions to ERDM clique configurations. Therefore we define unigram based feature functions $f^{E}_T$  and $f^{R}_T$ to 2-cliques containing a single sub-query term and a entity or relationship document node. 

For 3-cliques containing consecutive sub-query terms and a document node, we define two feature functions. One considers consecutive sub-query terms and matches ordered bigrams with entity or relationship documents. This feature function is denoted as $f^{E}_O$ and $f^{R}_O$, depending if the clique is $O^{E}$ or $O^{R}$. The second feature function matches bigrams with documents using an unordered window of 8 terms ($uw8$), i.e., it matches bigrams with documents if the two terms of the bigram occur with a maximum of 6 other terms between each other. This feature function is denoted as $f^{E}_U$ and $f^{R}_U$, depending if the clique is $O^{E}$ or $O^{R}$.

For each textual feature function we decided to use two variants: Dirichlet smoothing Language Models (LM) and BM25. We now present the summary of the textual feature functions used in this work.

\ \\
\textbf{LM}-T-E\\
\resizebox{0.6 \linewidth}{!} 
{
$f^{E}_{T,LM}(q^{E_i}_j,D^{E_i}) =  \text{log} \left (\frac{f(q^{E_i}_j,D^{E_i})
 + \frac{ f(q^{E_i}_j,C^E)}{|C^E|}\mu^E }{|D^{E_i}| + \mu^E} \right )$
 }\\

\ \\ \textbf{LM}-O-E\\

\resizebox{0.8 \linewidth}{!} 
{
$f^{E}_{O,LM}(q^{E_i}_j,q^{E_i}_{j+1},D^{E_i}) = \text{log} \left (\frac{f_{\#1}(q^{E_i}_j,q^{E_i}_{j+1},D^{E_i})
 + \frac{ f_{\#1}(q^{E_i}_j,q^{E_i}_{j+1},C^E)}{|C^E|}\mu^E }{|D^{E_i}| + \mu^E} \right )$
 }
 
\ \\ \textbf{LM}-U-E\\

\resizebox{0.8 \linewidth}{!} 
{
$f^{E}_{U,LM}(q^{E_i}_j,q^{E_i}_{j+1},D^{E_i}) =  \text{log} \left (\frac{f_{\#uw8}(q^{E_i}_j,q^{E_i}_{j+1},D^{E_i})
 + \frac{ f_{\#uw8}(q^{E_i}_j,q^{E_i}_{j+1},C^E)}{|C^E|}\mu^E }{|D^{E_i}| + \mu^E} \right )$
 }

\ \\ \textbf{LM}-T-R\\

\resizebox{0.7 \linewidth}{!} 
{
$f^{R}_{T,LM}(q^{R_{i-1,i}}_j,D^{R_{i-1,i}}) = \text{log}  \left (\frac{ f(q^{R_{i-1,i}}_j,D^{R_{i-1,i}}) +  \frac{f(q^{R_{i-1,i}}_j,C^R)}{|C^R|}\mu^R}{|D^{R_{i-1,i}}| + \mu^R} \right ) $
 }

\ \\ \textbf{LM}-O-R\\

\resizebox{1.0 \linewidth}{!} 
{
$f^{R}_{O,LM}(q^{R_{i-1,i}}_j,q^{R_{i-1,i}}_{j+1} ,D^{R_{i-1,i}}) = \text{log}  \left (\frac{ f_{\#1}(q^{R_{i-1,i}}_j,q^{R_{i-1,i}}_{j+1},D^{R_{i-1,i}}) +  \frac{f_{\#1}(q^{R_{i-1,i}}_j,q^{R_{i-1,i}}_{j+1},C^R)}{|C^R|}\mu^R}{|D^{R_{i-1,i}}| + \mu^R} \right ) $
 }
 
\ \\ \textbf{LM}-U-R\\

\resizebox{1.0 \linewidth}{!} 
{
$f^{R}_{U,LM}(q^{R_{i-1,i}}_j,q^{R_{i-1,i}}_{j+1} ,D^{R_{i-1,i}}) = \text{log}  \left (\frac{ f_{\#uw8}(q^{R_{i-1,i}}_j,q^{R_{i-1,i}}_{j+1},D^{R_{i-1,i}}) +  \frac{f_{\#uw8}(q^{R_{i-1,i}}_j,q^{R_{i-1,i}}_{j+1},C^R)}{|C^R|}\mu^R}{|D^{R_{i-1,i}}| + \mu^R} \right ) $
 }

\ \\
\ \\
Here, $f(q^{E_i}_j,D^{E_i})$ and $f(q^{R_{i-1,i}}_j,D^{R_{i-1,i}})$ represent the sub-query term frequencies in a entity document and relationship document, respectively.  The collection frequencies $f(q^{E_i}_j,C^E)$, $f(q^{R_{i-1,i}}_j,C^R)$ represent the frequency of sub-query term in either the \textit{entity} index $C^E$ or in the \textit{relationship} index $C^R$. The variants of these functions $f_{\#1}$ and $f_{\#uw8}$ represent ordered and unordered bigram matching frequency.  $|D^{E_i}|$ and$|D^{R_{i,i+1}}|$ represent the total number of terms in a meta-document while $|C^R|$ and $|C^E|$ represent the total number of terms in a collection of meta-documents. Finally, $\mu^E$ and $\mu^R$ are the Dirichlet prior for smoothing which generally corresponds to the average document length in a collection.

\ \\ \textbf{BM25}-T-E\\

\resizebox{0.85 \linewidth}{!} 
{
$f^{E}_{T,BM25}(q^{E_i}_j,D^{E_i}) =  \text{log} \frac{N^E - n(q^{E_i}_j) + 0.5}{n(q^{E_i}_j) + 0.5} \ . \   \frac{ f(q^{E_i}_j,D^{E_i}) (K_1 + 1)} {  f(q^{E_i}_j,D^{E_i}) + K_1 (1 - b + b \frac{|D^{E_i}|}{avg(|D^{E}|)})}$
 }

\ \\ \textbf{BM25}-O-E\\

\begin{align}
\begin{split}
f^{E}_{O,BM25}(q^{E_i}_j,q^{E_i}_{j+1},D^{E_i}) =&  \text{log} \frac{N^E - n_{\#1}(q^{E_i}_j,q^{E_i}_{j+1}) + 0.5}{n_{\#1}(q^{E_i}_j,q^{E_i}_{j+1}) + 0.5} \ \cdot\\
& \frac{ f_{\#1}(q^{E_i}_j,q^{E_i}_{j+1},D^{E_i}) (K_1 + 1)} {  f_{\#1}(q^{E_i}_j,q^{E_i}_{j+1},D^{E_i}) + K_1 (1 - b + b \frac{|D^{E_i}|}{avg(|D^{E}|)})}
 \end{split}\\
\end{align}

\ \\ \textbf{BM25}-U-E\\

\begin{align}
\begin{split}
f^{E}_{U,BM25}(q^{E_i}_j,D^{E_i})=&  \text{log} \frac{N^E - n_{\#uw8}(q^{E_i}_j,q^{E_i}_{j+1}) + 0.5}{n_{\#uw8}(q^{E_i}_j,q^{E_i}_{j+1}) + 0.5} \ \cdot\\
&\frac{ f_{\#uw8}(q^{E_i}_j,q^{E_i}_{j+1},D^{E_i}) (K_1 + 1)} {  f_{\#uw8}(q^{E_i}_j,q^{E_i}_{j+1},D^{E_i}) + K_1 (1 - b + b \frac{|D^{E_i}|}{avg(|D^{E}|)})}\\
\end{split}\\
\end{align}

\ \\ \textbf{BM25}-T-R\\

\begin{align}
\begin{split}
f^{R}_{T,BM25}(q^{R_{i-1,i}}_j,D^{R_{i-1,i}}) = & \text{log} \frac{N^R - n(q^{R_{i-1,i}}_j) + 0.5}{n(q^{R_{i-1,i}}_j) + 0.5} \ \cdot \\
&   \frac{ f(q^{R_{i-1,i}}_j,D^{R_{i-1,i}}) (K_1 + 1)} {  f(q^{R_{i-1,i}}_j,D^{R_{i-1,i}}) + K_1 (1 - b + b \frac{|D^{R_{i-1,i}}|}{avg(|D^{R}|)})}\\
\end{split}\\
\end{align}

\ \\ \textbf{BM25}-O-R\\

\begin{align}
\begin{split}
f^{R}_{O,BM25}(q^{R_{i-1,i}}_j,q^{R_{i-1,i}}_{j+1} ,D^{R_{i-1,i}}) = &  \text{log} \frac{N^R - n_{\#1}(q^{R_{i-1,i}}_j,q^{R_{i-1,i}}_{j+1}) + 0.5}{n_{\#1}(q^{R_{i-1,i}}_j,q^{R_{i-1,i}}_{j+1}) + 0.5} \cdot \\
& \frac{ f_{\#1}(q^{R_{i-1,i}}_j,q^{R_{i-1,i}}_{j+1} ,D^{R_{i-1,i}}) (K_1 + 1)} {  f_{\#1}(q^{R_{i-1,i}}_j,q^{R_{i-1,i}}_{j+1} ,D^{R_{i-1,i}}) + K_1 (1 - b + b \frac{|D^{R_{i-1,i}}|}{avg(|D^{R}|)})}\\
\end{split}\\
\end{align}

\ \\ \textbf{BM25}-U-R\\

\begin{align}
\begin{split}
f^{R}_{U,BM25}(q^{R_{i-1,i}}_j,q^{R_{i-1,i}}_{j+1} ,D^{R_{i-1,i}})=& \text{log} \frac{N^R - n_{\#uw8}(q^{R_{i-1,i}}_j,q^{R_{i-1,i}}_{j+1}) + 0.5}{n_{\#uw8}(q^{R_{i-1,i}}_j,q^{R_{i-1,i}}_{j+1}) + 0.5} \cdot \\
&  \frac{ f_{\#uw8}(q^{R_{i-1,i}}_j,q^{R_{i-1,i}}_{j+1} ,D^{R_{i-1,i}}) (K_1 + 1)} {  f_{\#uw8}(q^{R_{i-1,i}}_j,q^{R_{i-1,i}}_{j+1} ,D^{R_{i-1,i}}) + K_1 (1 - b + b \frac{|D^{R_{i-1,i}}|}{avg(|D^{R}|)})}\\
\end{split}\\
\end{align}

Here, $N^E$ and $N^R$ represent the total number of documents in the entity index and relationship index, respectively.  The document frequency of unigrams and bigrams is represented using $n()$,$n_{\#1}()$ and $n_{\#uw8}()$. $|D^{E_i}|$ and $|D^{R_{i-1,i}}|$ are the total number of terms in a entity or relationship document while $avg(|D^{E}|)$ and  $avg(|D^{R}|)$ are the average entity or relationship document length. $K_1$ and $b$ are free parameters usually chosen as 1.2 and 0.75, in the absence of specific optimization.

We define two non-textual features in ERDM. The first one, $f^{ER}_T$ is assigned to 2-cliques composed by one entity document and one relationship document and it is inspired in the feature function $f_E$ of Hasibi and Balog's ELR model \citep{hasibi2016exploiting}. It is defined as follows:

\begin{equation} \label{rsdm_der}
\resizebox{0.7 \textwidth}{!} 
{
 $f^{ER}_S(D^{E_i},D^{R_{i-1,i}}) =  \left [   (1 - \alpha) f(D^{E_i},D^{R_{i-1,i}}) + \alpha \frac{ n(E_i)} { N^R}    \right ]$
}
\end{equation}

where the linear interpolation implements the Jelinek-Mercer smoothing method with $\alpha \in [0,1]$ and $f(D^{E_i},D^{R_{i-1,i}}) = \{0,1\}$ which measures if the entity $E_i$ represented in $D^{E_i}$ belongs to the relationship $R(E_{i-1},E_i)$ represented in $D^{R_{i-1,i}}$. The background model employs the notion of entity popularity within the collection of relationship documents. $n(D^{E_i})$ represents the number of relationship documents $D^R$ that contain the entity $E_i$ and $N^R$ represents the total number of relationship documents in the \textit{relationship} index.

For E-R queries with more than one relationship sub-query, we draw an edge between consecutive relationship documents within the ERDM graph. This edge creates a 3-clique containing two relationship documents and one entity document. The feature function $f^{RER}_S$ measures if a given entity $E_i$ is shared between consecutive relationship documents within the graph. We opted to define a simple binary function:

\begin{equation}
f^{ER}_S(D^{E_i},D^{R_{i-1,i}}, D^{R_{i,i+1}}) = 1 \ if \  E_i \in D^{E_i} \cap D^{R_{i-1,i}} \cap D^{R_{i,i+1}}\ , \ 0 \ \text{otherwise}
\end{equation}

In summary, we described the set of feature functions associated with each clique configuration within the ERDM graph. We leave for future work the possibility of exploring other type of features to describe textual similarity and compatibility between different nodes in the ERDM graph, such as neural language models.

\subsection{Ranking}

We have defined the set of clique configurations and the real valued feature functions that constitute the non-negative potential functions over the cliques in the graph of ERDM. We can now formulate the calculation of the posterior $P(D^E,D^R|Q$ using the probability mass function of the MRF, as follows:

\begin{align}
\begin{split}\label{eq:3}
   P_{\Lambda}(D^E,D^R|Q)  {}  \stackrel{\text{$rank$}}{=} &  \sum_{c \in C(G)} \lambda_c f(c)\\
     \stackrel{\text{$rank$}}{=} & \lambda^{E}_T \sum_E \sum_{Q^{E_i}} f^{E}_T(q^{E_i}_j,D^{E_i}) +\\
& \lambda^{E}_O \sum_E \sum_{Q^{E_i}} f^{E}_O(q^{E_i}_j,q^{E_i}_{j+1},D^{E_i}) +\\
& \lambda^{E}_U \sum_E \sum_{Q^{E_i}} f^{E}_U(q^{E_i}_j,q^{E_i}_{j+1},D^{E_i}) +\\
& \lambda^{R}_T \sum_R \sum_{Q^{R_{i,j}}}  f^{R}_T(q^{R_{i-1,i}}_j,D^{R_{i-1,i}}) + \\
& \lambda^{R}_O \sum_R \sum_{Q^{R_{i,j}}} f^{R}_O(q^{R_{i-1,i}}_j,q^{R_{i-1,i}}_{j+1} ,D^{R_{i-1,i}}) +\\
& \lambda^{R}_U \sum_R \sum_{Q^{R_{i,j}}} f^{R}_U(q^{R_{i-1,i}}_j,q^{R_{i-1,i}}_{j+1} ,D^{R_{i-1,i}}) +\\
& \lambda^{ER}_S \sum_R \sum_{E} f^{ER}_S(D^{E_i},D^{R_{i-1,i}})+  \\
& \lambda^{RER}_S \sum_R \sum_{E} f^{ER}_S(D^{E_i},D^{R_{i-1,i}},D^{R_{i,i+1}})  \\
\end{split}\\
\end{align}

In essence, E-R retrieval using the ERDM corresponds to ranking candidate entity tuples using a linear weighted sum of the feature functions over the cliques in the graph. Therefore, we can apply any linear learning to rank algorithm to optimize the ranking with respect to the vector of feature weights $\Lambda$. Given a training set $\mathcal{T}$ composed by relevance judgments, a ranking of entity tuples $\mathcal{R}_{\Lambda}$ and an evaluation function $\mathcal{E}(\mathcal{R}_{\Lambda};\mathcal{T})$ that produces a real valued output, our objective is to find the values of the vector $\Lambda$ that maximizes $\mathcal{E}$. As explained in \citep{metzler2007linear}, we require $\mathcal{E}$ to only consider the ranking produced and not individual scores. This is the standard characteristic among information retrieval evaluation metrics (e.g. MAP or NDCG).

\subsection{Discussion}
In this section we introduced the Entity-Relationship Dependence Model (ERDM), a novel supervised Early Fusion-based model for E-R retrieval. Inspired by recent work in entity retrieval we believe that modeling term dependencies between sub-queries and entity/relationship documents can increase search performance. 

ERDM can be seen as an extension of the SDM model \citep{metzler2005markov} for ad-hoc document retrieval in a way that besides modeling query term dependencies we create graph structures that depict dependencies between entity and relationship documents. Consequently, instead of computing a single posterior $P(D|Q)$ we propose to use the MRF for retrieval for computing a joint posterior of multiple entity and relationship documents given a E-R query, $P(D^E,D^R|Q)$. 

Moreover, since ERDM is a supervised model, we believe that tuning weights of feature functions, besides optimizing search performance, can also help to explain the inter-dependencies between sub-query terms and the respective documents, but also how entity documents and relationship documents contribute to the overall relevance of entity tuples given a E-R query.

\section{Summary of the Contributions}
In this chapter we present several contributions to the problem of entity-relationship retrieval from a IR perspective:

\begin{itemize}
\item Generalization of the problem of entity-relationship search to cover entity types and relationships represented by any attribute and predicate, respectively, rather than a predefined set.
\item A general probabilistic model for E-R retrieval using Bayesian Networks.
\item Proposal of two design patterns that support retrieval approaches using the E-R model.
\item Proposal of a Entity-Relationship Dependence model that builds on the basic Sequential Dependence Model (SDM) to provide extensible entity-relationship representations and dependencies, suitable for complex, multi-relations queries. 
\end{itemize}

\chapter{Entity-Relationship Retrieval over a Web Corpus}\label{ch:er2}

We start this chapter by presenting a new semi-automatic method for generating E-R test collections together with a new E-R test collection, the RELink Query Collection comprising 600 E-R queries. We leverage web tabular data containing entities and relationships among them as they share the same row in a table. We exploit the Wikipedia Lists-of-lists-of-lists tree of articles containing lists of Entities in the form of tables. We developed a table parser that extracts tuples of entities from these tables together with associated metadata. This information is then provided to editors that create E-R queries fulfilled by the extracted tuples.

We then report a set of evaluations of the ERDM model using four different query sets. In order to leverage information about entities and relations in a corpus, it is necessary to create a representation of entity related information that is amenable to ER search. In our approach we focus on sentence level information about entities although the method can be applied to more complex segmentation of text. Our experiments are based on the ClueWeb-09-B data set with FACC1 text annotation that refer to  entities found in the text, including the variances of their surface forms. Each entity is designated by its unique ID and for each unique entity instance we created 'entity documents' comprising a collection of sentences that contain the entity. These context documents are indexed, comprising the \emph{entity index}. The same is done by creating \emph{entity pair documents} and the entity pair index.  These two indexes enable us to execute E-R queries using different retrieval models, including the ERDM that models the dependence between entities.

\section{RELink Query Collection \footnote{The material contained in this section was published in P. Saleiro, N. Milic-Frayling, E. M. Rodrigues, C. Soares, ``RELink: A Research Framework and Test Collection for Entity-Relationship Retrieval''\citep{saleiro-sigir}.}}

Improvements of entity-relationship (E-R) search techniques have been hampered by a lack of test collections, particularly for complex queries involving multiple entities and relationships. In this section we describe a method for generating E-R test queries to support comprehensive E-R search experiments. Queries and relevance judgments are created from content that exists in a tabular form where columns represent entity types and the table structure implies one or more relationships among the entities. Editorial work involves creating natural language queries based on relationships represented by the entries in the table. 

We have publicly released the RELink test collection comprising 600 queries and relevance judgments obtained from a sample of Wikipedia List-of-lists-of-lists tables. The latter comprise tuples of entities that are extracted from columns and labelled by corresponding entity types and relationships they represent. 

Improvement of methods for both extraction and search is hampered by a lack of query sets and relevance judgments, i.e., gold standards that could be used to compare effectiveness of different methods. In this section we introduce:
\begin{enumerate}
\item A low-effort semi-automatic method for acquiring instances of entities and entity relationships from tabular data. 
\item RELink Query Collection (QC) of 600 E-R queries with corresponding relevance judgments  
\end{enumerate}

Essential to our approach is the observation that tabular data typically includes entity types as columns and entity instances as rows. The table structure implies a relationship among table columns and enables us to create E-R queries that are answered by the entity tuples across columns.
Following this approach, we prepared and released the RELink QC comprising 600 E-R queries and relevance judgments based on a sample of Wikipedia \textit{List-of-lists-of-lists} tables. 

The query collection and the research framework are publicly available\footnote{\url{https://sigirelink.github.io/RELink/}}, enabling the community to expand the RELink Framework with additional document collections and alternative indexing and search methods. It is important to maintain and enhance the  RELink QC by providing updates to the existing entity types and creating new queries and relevant instances from additional tabular data.

\subsection{Tabular Data and Entity Relationships}
Information that satisfies complex E-R queries is likely to involve instances of entities and their relationships dispersed across Web documents. Sometimes such information is collected and published within a single document, such as a Wikipedia page. In such cases, traditional search engines can provide excellent search results without applying special E-R techniques or considering entity and relationship types. Indeed, the data collection, aggregation, and tabularization has been done by a Wikipedia editor. 

That also means that a tabular Wikipedia content, comprising various entities, can be considered as representing a specific information need, i.e., the need that motivated editors to create the page in the first place. Such content can, in fact, satisfy many different information needs. We focus on exploiting tabular data for exhaustive search for pre-specified E-R types. In order to specify E-R queries, we can use column headings as entity types. All the column entries are then relevance judgments for the entity query. Similarly, for a given pair of columns that correspond to distinct entities, we formulate the implied relationship. For example the pair $<$car, manufacturing plant$>$ could refer to ``is made in'' or ``is manufactured in'' relationships. The instances of entity pairs in the table then serve as evidence for the specific relationship. This can be generalized to more complex information needs that involve multiple entity types and relationships.

Automated creation of E-R queries from tabular content is an interesting research problem. For now we asked human editors to provide natural language and structured E-R queries for specific entity types. Once we collect sufficient amounts of data from human editors we will be able to automate the query creation process with machine learning techniques. For the RELink QC we compiled a set of 600 queries with E-R relevance judgments from Wikipedia lists about 9 topic areas.   

\subsection{Selection of Tables}\label{sectables}
Wikipedia contains a dynamic index ``\textit{The Lists of lists of lists}''\footnote{ \url{http://en.wikipedia.org/wiki/List_of_lists_of_lists}} which represents the root of a tree that spans curated lists of entities in various domains. We used a Wikipedia snapshot from October 2016 to traverse ``\textit{The Lists of lists of lists}'' tree starting from the root page and following every hyperlink of type ``\textit{List of}'' and their children. This resulted in a collection of 95,569 list pages. While most of the pages contain tabular data, only 18,903 include tables with consistent column and row structure. As in \cite{bhagavatula2013methods}, we restrict content extraction to \textit{wikitable} HTML class that typically denotes data tables in Wikipedia. We ignore other types of tables such as infoboxes. 

In this first instance, we focus on \textit{relational tables}, i.e., the tables that have a key column, referring to the \textit{main} entity in the table  \cite{lehmberg2016large}. For instance, the ''\textit{List of books about skepticism}'' contains a table ``\textit{Books}'' with columns ``Author'', ``Category'' and ``Title'', among others. In this case, the key column is ``Title'' which contains titles of books about skepticism. We require that any relationship specified for the entity types in the table must contain the  ``Title'' type, i.e., involve the ``Title'' column. 

In order to detect key columns we created a Table Parser that uses the set of heuristics adopted by Lehmberg et al. \cite{lehmberg2016large}, e.g., the ratio of unique cells in the column or text length. Once the key column is identified, the parser creates entity pairs consisting of the key column and one other column in the table. The content of the column cells then constitutes the set of relevant judgments for the relationship specified by the pair of entities. 

For the sake of simplicity we consider only those Wikipedia lists that contain a single relational table. Furthermore, our goal is to create queries that have verifiable entity and entity pair instances. Therefore, we selected only those relational tables for which the key column and at least one more column have cell content linked to Wikipedia articles.  

With these requirements, we collected 1795 tables. In the final step, we selected 600 tables by performing stratified sampling across semantic domains covered by Wikipedia lists. For each new table, we calcuated the Jaccard similarity scores between the title of the corresponding Wikipedia page and the titles of pages associated with tables already in the pool. By setting the maximum similarity threshold to 0.7 we obtained a set of 600 tables. 

\begin{figure}[t]
    \centering
        \centering
        \includegraphics[width= 0.8\textwidth]{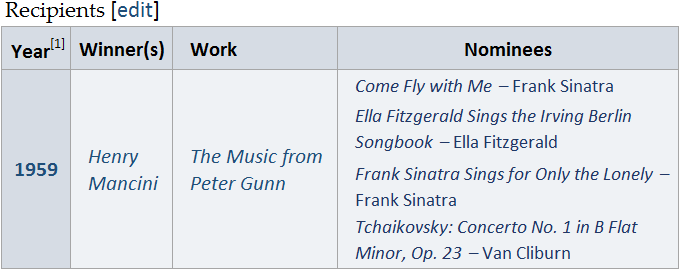}
        \caption{Example of Wikipedia table row.} \label{wiki1}
\end{figure}

The process of creating RELink queries involves two steps: (1) automatic selection of tables and columns within tables and (2) manual specification of information needs. For example,  in the table ``Grammy Award for Album of the Year'' the columns ``winner'', ``work'' were automatically selected to serve as entity types in the E-R query (Figure \ref{wiki1}). The relationship among these entities is suggested by the title and we let a human annotator to formulate the query.   

\begin{figure}[t]
    \centering
        \centering
        \includegraphics[width= 0.8\textwidth]{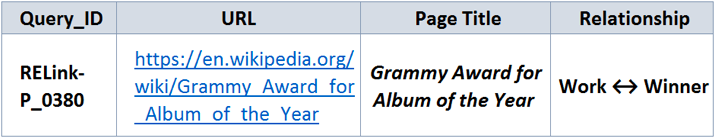}
        \caption{Example of metadata provided to editors.} \label{wiki2}
\end{figure}

The RELink query set was created by 6 annotators. We provided the annotators with access to the full table, metadata (e.g., table title or the first paragraph of the page) and entity pairs or triples to be used to specify the query (Figure \ref{wiki2}). For each entity pair or triple the annotators created a natural language information need and an E-R query in the relational format $Q = \{ Q^{E_{i-1}}, Q^{R_{i-1,i}}, Q^{E_i} \}$, as shown in Table \ref{qannot}. 
\subsection{Formulation of Queries}

The relational query format is introduced to support a variety of experiments with E-R queries. In essence, a complex information need is decomposed into a set of sub-queries that specify types of entities $E$ and types of relationships $R(E_{i-1},E_i)$ between entities. For each relationship query there is one query for each entity involved in the relationship. Thus a query $Q$ that expects a pair of entities for a given relationship, is mapped into three sub-queries $(Q^{E_{i-1}}$, $Q^{R_{i-1,i}}$, $Q^{E_i})$, where $Q^{E_{i-1}}$ and $Q^{E_i}$ are the entity types for $E_{i-1}$ and $E_i$ respectively, and $Q^{R_{i-1,i}}$ is a relationship type describing $R(E_{i-1},E_i)$. 

\begin{table}[t]
\caption{Examples of query annotations.}
\centering
\setlength\extrarowheight{3pt}
\begin{tabular}{|l|p{5cm}|p{4cm}|}
\cline{1-3}
ID  & NL Query & Relational Format   \\ \cline{1-3}
\small RELink\_P\_164     & \small  \textit{What are the regiments held by the Indian Army?}   &   \small  \{\textit{regiment}, \textit{held by}, \textit{Indian Army}\}         \\ \cline{1-3}
\small RELink\_T\_071 &  \small  \textit{In which seasons NHL players scored more than 50 goals and the team they represented?}  &  \small    \{\textit{NHL season}, \textit{scored more than 50 goals in}, \textit{NHL player}, \textit{played for}, \textit{NHL team}   \}             \\ \cline{1-3}
\end{tabular}
\label{qannot}
\end{table}

\begin{table}[t]
\caption{RELink collection statistics.}
\centering
\setlength\extrarowheight{1pt}
\begin{tabular}{l|l|l|l|}
\cline{2-4}
              & 2-entity  & 3-entity     & All   \\ \cline{1-4}  
\multicolumn{1}{|l|}{Total queries}      & 381  & 219 & \textbf{600} \\     \cline{1-4}
\multicolumn{1}{|l|}{Avg. queries length} & 56.5 & 83.8 & \textbf{66.5}   \\ \cline{1-4}
\multicolumn{1}{|l|}{Avg. $Q^E$ length} & 20.9 & 20.9 & \textbf{20.9}   \\ \cline{1-4}
\multicolumn{1}{|l|}{Avg. $Q^R$ length} & 11.8 & 12.6 & \textbf{12.3}   \\ \cline{1-4}
\multicolumn{1}{|l|}{\# uniq. entity attributes ($Q^E$)} & 679 & 592 & \textbf{1251}   \\ \cline{1-4}
\multicolumn{1}{|l|}{\# uniq. relationships ($Q^R$)} & 145 & 205 & \textbf{317}   \\ \cline{1-4}
\multicolumn{1}{|l|}{Avg. \# relevant judgments } & 67.9 & 41.8 & \textbf{58.5}   \\ \cline{1-4}

\end{tabular}
\label{stats}
\end{table}

\subsection{Collection Statistics}

RELink QC covers 9 thematic areas from the \textit{Lists-of-Lists-of-Lists} in Wikipedia: Mathematics and Logic, Religion and Belief Systems, Technology and Applied Sciences, Miscellaneous, People, Geography and Places, Natural and Physical Sciences, General Reference and Culture and the Arts. The most common thematic areas are Culture and the Arts with 70 queries and Geography and Places with 67 queries.

In Table \ref{stats} we show the characteristics of the natural language and relational queries. Among 600 E-R queries, 381 refer to entity pairs and 219 to entity triples. As expected, natural language descriptions of 3-entity queries are longer (on average 83.8 characters) compared to 2-entity queries (56.5 characters).

We further analyze the structure of relational queries and their components, i.e., entity queries $Q^E$ that specify the entity type and relationship queries  $Q^R$ that specify the relationship type.  Across 600 queries, there are 1251 unique entity types $Q^E$ (out of total 1419 occurrences). They are rather unique across queries: only 65 entity types occur in more than one E-R query and 44 occur in exactly 2 queries. The most commonly shared entity type is ``country'', present in 9 E-R queries.

In the case of relationships, there are 317 unique relationship types $Q^R$ (out of 817 occurrences) with a dominant type ``located in'' that occurs in 140 queries. This is not surprising since in many domains the key entity is tied to a location that is included in one of the columns. Nevertheless, there are only 44 relationship types $Q^R$ occurring more than once implying that RELink QC is a diverse set of queries, including 273 relationship types occurring only once. 

\section{Experimental Setup}

In this section we detail how we conducted our experiments in E-R retrieval. Since we only have access to test collections comprising general purpose E-R queries we decided to use a Web corpus as dataset, more precisely ClueWeb-09-B\footnote{\url{https://lemurproject.org/clueweb09/}}.The ClueWeb09 dataset was created to support research on information retrieval and related human language technologies and contains 1 billion web pages. The part B is a subset of the most popular 50 million English web pages, including the Wikipedia. Part B was created as a resource for research groups without processing power for processing the all ClueWeb09 collection. We used the ClueWeb-09-B Web collection with FACC1 text span annotations linked to Wikipedia entities to show how RELink can be used for E-R retrieval over Web content. We developed our prototype using Apache Lucene for indexing and search. We used a specific Python library (PyLucene) that allowed our customized implementation tailored for E-R retrieval.

\subsection{Data and Indexing}

\begin{figure}[h] 
\centering
\includegraphics[width=0.7\textwidth]{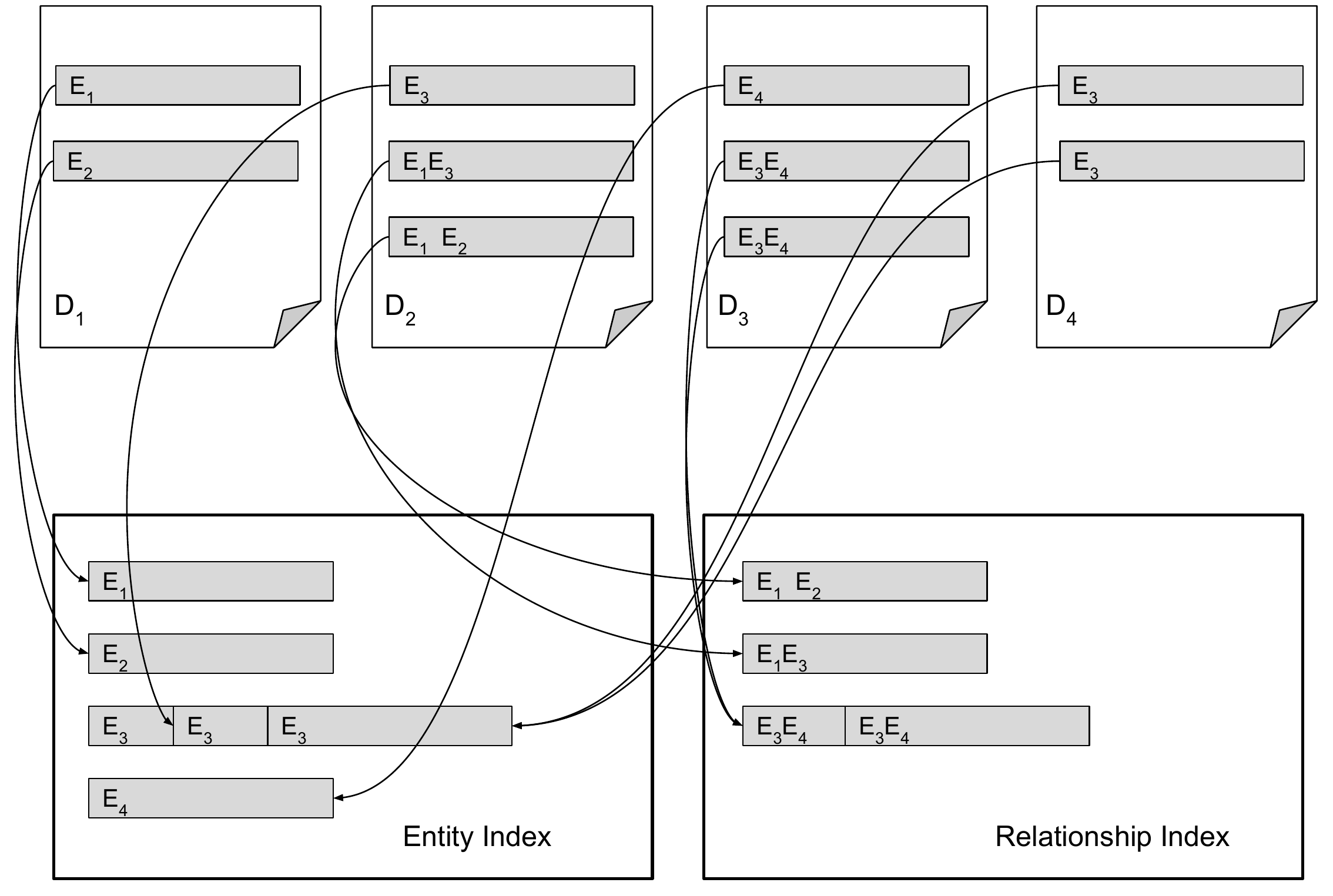}
\caption{Illustration of E-R indexing from a web corpus.}\label{ind}
\end{figure}

As a text corpus, we use ClueWeb-09-B combined with FACC1 text span annotations with links to Wikipedia entities (via Freebase). The entity linking precision and recall in FACC1 is estimated to be 80-85\% and 70-85\%, respectively \cite{gabrilovich2013facc1}. For our experiments we created two main indexes: one for entity extractions and one for entity pairs (relationships) extractions. We extract entity and pairs occurrences using an Open Information Extraction method like OLLIE \cite{schmitz2012open} over the annotated ClueWeb-09-B corpus as follows. For each entity annotation, we extract the sentence where it occurred as an entity context. For pairs of entities, we look for co-occurring entities in the same sentence and we extract the separating string, i.e., the context of the relationship connecting them. Figure \ref{ind} illustrates the indexing process adopted in this work. 

We obtained 476 million entity extractions and 418 million entity pairs extractions, as described in Table \ref{extractions}. In order to compute $|D^{E_i}|$ and $|D^{R_{i-1,i}}|$ we incrementally updated two auxiliary indices, containing the number of terms per entity and per entity pair, respectively. We ran our experiments using Apache Lucene and made use of GroupingSearch for grouping extractions by entity and entity pair at query time. To get the statistics for ordered and unordered bigrams we made use of SpanNearQuery.

\begin{table}[h]
\caption{ClueWeb09-B extractions statistics.}
\centering
\setlength\extrarowheight{1pt}
\begin{tabular}{l|l|l|l|}
\cline{2-4}
              & Total  & Unique     & Avg. doc. len.   \\ \cline{1-4}  
\multicolumn{1}{|l|}{Entities}     & 476,985,936 & 1,712,010  & 9977 \\     \cline{1-4}
\multicolumn{1}{|l|}{Entity pairs} & 418,079,378 & 71,660,094 & 138   \\ \cline{1-4}
\end{tabular}
\label{extractions}
\end{table}

\subsection{Retrieval Method and Parameter Tuning}

For experiments using ERDM we adopted a three stage retrieval method. First, queries $Q^{E_{i-1}}$,$Q^{E_i}$ are submitted against the entity index and $Q^{R_{i-1,i}}$ is submitted against the entity-pair index. Initial sets of top 20000 results grouped by entity or entity-pairs, respectively, are retrieved using Lucene's default search settings. Second, the feature functions of the specific retrieval model are calculated for each set, using an in-house implementation. This process is easily parallelized. The final ranking score for each entity-pair is then computed using the learned $\lambda$ weights. Evaluation scores are reported on the top 100 entity-pair results.

Parameter tuning for ERDM and baselines was directly optimized with respect to the Mean Average Precision (MAP). We make use of the RankLib's implementation of the coordinate ascent algorithm under the sum normalization and non-negativity constraints with 3 random restarts. Coordinate ascent is a commonly used optimization technique \cite{metzler2007linear} that iteratively optimizes a single parameter while holding all other parameters fixed.

Parameters are estimated using 5-fold cross validation for each of the 4 query sets separately. To be able to use the same train and test folds throughout all experiments, we first randomly create fixed train and test folds from the initial result set, for each query set. All reported evaluation metrics were macro-averaged over 5 folds.

We do not optimize the Dirichlet priors $\mu^E$ and $\mu^R$ in language models and set them equal to the traditional average document length, i.e., the average entity and entity pairs extractions length, respectively. The unordered window size $N$ for $f^{E}_U$ and $f^{R}_U$ is set to be 8, as suggested in \cite{metzler2005markov}. 

\subsection{Test Collections}
We ran experiments with a total of 548 E-R queries. We decided to just perform experiments using queries aiming 2-tuples of entities. We leave for future work the evaluation of queries aiming at triples. Besides RELink QC we used other 3 relationship-centric query sets, with pairs of Wikipedia entities as answers, i.e., relevance judgments. The query sets cover a wide range of domains as described in Table \ref{query_table}.  Query sets for entity-relationship retrieval are scarce. Generally entity retrieval query sets are not relationship-centric \cite{yahya2016relationship}.

\begin{table}[h]
\caption{Description of query sets used for evaluation.}
\centering
\setlength\extrarowheight{1pt}
\begin{tabular}{|l|l|p{7cm}|}
\cline{1-3}
\small Query Set  & Count & Domains   \\ \cline{1-3}
\small QALD-2     & 79     &   Geography and places, Politics and society, Culture and the Arts, Technology and science           \\ \cline{1-3}
 \small ERQ        & 28     &  \small Award, City, Club, Company, Film, Novel, Person, Player, Song, University                \\ \cline{1-3}
\small COMPLEX    & 60     & \small Cinema, Music, Books, Sports, Computing, Military conflicts                     \\ \cline{1-3}
\small RELink & 381    &  \small General Reference, Culture and the Arts, Geography and places, Mathematics and logic, Natural and physical Sciences, People, Religion and belief systems, Society and social sciences, Technology and applied science                    \\ \cline{1-3}
Total      & 548    &             \\ \cline{1-3}
\end{tabular}
\label{query_table}
\end{table}

One exception is the QALD-2 query set used in the DBpedia-entity collection \cite{balog2013test}. It contains a subset of relational queries, e.g.``\textit{Who designed the Brooklyn Bridge?}''. Most of relational queries in QALD-2 have a fixed relevant entity, e.g., ``\textit{Brooklyn Bridge}'' and can be easily transformed from single entity relevance judgments into pairs. From the 79 relational queries in QALD-2, we identified 6 with no fixed relevant entity in the query (e.g. ``\textit{Give me the capitals of all countries in Africa.}''). In these cases, for provided single entity relevance judgment we needed to annotate the missing entity manually to create a pair. For instance, given a capital city in Africa we identified the corresponding African country.

In addition, we used two benchmarks created in previous work using Semantic-Web-based approaches: ERQ \cite{li2012entity} and COMPLEX \cite{yahya2016relationship}. Neither ERQ nor COMPLEX provide complete relevance judgments and consequently, we manually evaluated each answer in our experiments. 
ERQ consists of 28 queries that were adapted from INEX17 and OWN28 \cite{li2012entity}. However, 22 of the queries have a given fixed entity in the query (e.g. \textit{``Find Eagles songs''}). Only 6 queries are asking for pairs of unknown entities, such as ``\textit{Find films starring Robert De Niro and please tell directors of these films.}''. 

COMPLEX queries were created with a semi-automatic approach \cite{yahya2016relationship}. It contains 70 queries from which we removed 10 that expect 3-tuples of entities. This query set consists of pure relationship-centric queries for unknown pairs of entities, such as ``\textit{Currency of the country whose president is James Mancham} ``\textit{Kings of the city which led the Peloponnesian League.}''  and ``\textit{Who starred in a movie directed by Hal Ashby?}''.

We used four different retrieval metrics, Mean Average Precision at 100 results (MAP), precision at 10 (P@10), mean reciprocal rank (MRR) and normalized discounted cumulative gain at 20 (NDCG@20). 


\section{Results and Analysis}

We start by performing a simple experiment for comparing Early Fusion and ERDM using both Language Models (LM) and BM25 as retrieval functions. Since we are only interested in comparing relative performance we opted to scale down our experimental setup. Instead of computing the term frequency for every extraction for a given entity or relationship we cap to 200 the number for each group of documents retrieved in the first passage. We tried several different values and for values below 200 extraction the performance reduced significantly. For 200, while the performance reduces it is not dramatic. This setup reduces the experimental runtime and since we had limited resources this proved to be useful. 

Table \ref{eferdm} depicts the results for this comparative evaluation. We decided to only use the three test collections specifically tailored for relationship retrieval. As we can see the results are very similar between EF and ERDM for both LM and BM25 variants. In the three test collections ERDM presents slightly better performance than the corresponding EF variant (e.g. BM25). However when performing statistical significance tests we obtained p-values above 0.05 when comparing EF and ERDM. This is very interesting as it shows that for general purpose E-R evaluation the overhead of computing sequential dependencies does not carry significant improvements.

\begin{table}[h]
\centering
\caption{Early Fusion and ERDM comparison using LM and BM25.}
\label{eferdm}
\begin{tabular}{|l|l|l|l|l|}
\hline
          & \multicolumn{4}{c|}{\textbf{ERQ}}     \\ \hline
          & MAP  & P@10    & MRR    & NDCG@20 \\ \hline
EF-LM     & \textbf{0.251}    & 0.15    & 0.3408 & 0.3508  \\ \hline
EF-BM25   & 0.1939   & 0.1423  & 0.1783 & 0.2861  \\ \hline
ERDM-LM   & \textbf{0.2611}   & 0.1615  & 0.3151 & 0.3589  \\ \hline
ERDM-BM25 & 0.2106   & 0.1462  & 0.2839 & 0.3257  \\ \hline
          & \multicolumn{4}{c|}{\textbf{COMPLEX}} \\ \hline
          & MAP  & P@10    & MRR    & NDCG@20 \\ \hline
EF-LM     & 0.1703   & 0.0596  & 0.1839 & 0.2141  \\ \hline
EF-BM25   & \textbf{0.1855}   & 0.0719  & 0.1907 & 0.2454  \\ \hline
ERDM-LM   & 0.1719   & 0.0789  & 0.2466 & 0.2492  \\ \hline
ERDM-BM25 & \textbf{0.1955}   & 0.0772  & 0.2257 & 0.248   \\ \hline
          & \multicolumn{4}{c|}{\textbf{RELink(381 queries)}}  \\ \hline
          & MAP  & P@10    & MRR    & NDCG@20 \\ \hline
EF-LM     & 0.0186   & 0.0063  & 0.0192 & 0.0249  \\ \hline
EF-BM25   & \textbf{0.0203}   & 0.0071  & 0.0227 & 0.0259  \\ \hline
ERDM-LM   & 0.0213   & 0.0058  & 0.0273 & 0.0255  \\ \hline
ERDM-BM25 & \textbf{0.0213}   & 0.0061  & 0.0265 & 0.0275  \\ \hline
\end{tabular}
\end{table}

On the other hand, we detect sensitivity to the retrieval function used. In ERQ, both ERDM-LM and EF-LM outperform BM25 but the opposite happens for COMPLEX and RELink. This sensitivity means that we cannot generalize the assumption that one of the retrieval functions is more adequate for E-R retrieval. 

Another important observation has to do with the overall lower results on the RELink test collection in comparison with ERQ and COMPLEX. Contrary to our expectations ClueWeb-09B has very low coverage of entity tuples relevant to the RELink test collection. 

We now present the results of comparing ERDM with three baselines using sequential dependence to evaluate the impact of modeling dependencies between query terms. The first baseline method, BaseEE, consists in submitting two queries against the entity index: $Q^{E_{i-1}} + Q^{R_{i-1,i}}$ and $Q^{R_{i-1,i}} + Q^{E_i}$. Entity-pairs are created by cross product of the two entity results set retrieved by each query. For each method we compute the Sequential Dependence Model(SDM) \citep{metzler2005markov} scores.

The second baseline method, BaseE, consists in submitting again a single query $Q$ towards the entity index used in ERDM. Entity-pairs are created by cross product of the entity results set with itself. The third baseline method, BaseR, consists in submitting a single query $Q$ towards an entity-pair index. This index is created using the full sentence for each entity-pair co-occurrence in ClueWeb-09-B, instead of just the separating string as in ERDM. This approach aims to capture any entity context that might be present in a sentence. ERDM relies on the entity index for that purpose.

In this evaluation we decided to not cap the number of extractions to compute term frequencies inside each group of results returned from the first passage with Lucene GroupingSearch. Due to the low coverage of ClueWeb for the entire RELink collection, we decided to just perform the evaluation using the top 100 queries with highest number of relevance judgments in our indexes. We also include results for the adapted QALD-2 test collection.

\begin{table}[h]
\centering
\caption{Results of ERDM compared with three baselines.}
\label{res}
\begin{tabular}{|l|l|l|l|l|}
\hline
          & \multicolumn{4}{c|}{\textbf{QALD-2}}           \\ \hline
          & MAP    & P@10     & MRR     & NDCG@20 \\ \hline
BaseEE & 0.0087 & 0.0027   & 0.0093  & 0.0055  \\ \hline
BaseE  & 0.0306 & 0.004684 & 0.0324  & 0.0363  \\ \hline
BaseR  & 0.0872 & 0.01678  & 0.0922  & 0.0904  \\ \hline
ERDM      & \textbf{0.1520} & 0.0405   & 0.1780  & 0.1661  \\ \hline
          & \multicolumn{4}{c|}{\textbf{ERQ}}              \\ \hline
          & MAP    & P@10     & MRR     & NDCG@20 \\ \hline
BaseEE & 0.0085 & 0.004    & 0.00730 & 0.0030  \\ \hline
BaseE  & 0.0469 & 0.01086  & 0.0489  & 0.038   \\ \hline
BaseR  & 0.1041 & 0.05086  & 0.1089  & 0.1104  \\ \hline
ERDM      & \textbf{0.3107} & 0.1903   & 0.37613 & 0.3175  \\ \hline
          & \multicolumn{4}{c|}{\textbf{COMPLEX}}          \\ \hline
          & MAP    & P@10     & MRR     & NDCG@20 \\ \hline
BaseEE & 0.0035 & 0        & 0.00430 & 0       \\ \hline
BaseE  & 0.0264 & 0.005    & 0.03182 & 0.1223  \\ \hline
BaseR  & 0.0585 & 0.01836  & 0.0748  & 0.0778  \\ \hline
ERDM      & \textbf{0.2879} & 0.1417   & 0.32959 & 0.3323  \\ \hline
          & \multicolumn{4}{c|}{\textbf{RELink(100 queries)}}           \\ \hline
          & MAP    & P@10     & MRR     & NDCG@20 \\ \hline
BaseEE & 0.03   & 0.01     & 0.0407  & 0.02946 \\ \hline
BaseE  & 0.0395 & 0.019    & 0.0679  & 0.03948 \\ \hline
BaseR  & 0.0451 & 0.021    & 0.0663  & 0.07258 \\ \hline
ERDM      & \textbf{0.1249} & 0.048    & 0.1726  & 0.1426  \\ \hline
\end{tabular}
\end{table}

Table \ref{res} presents the results of our experiments on each query set. We start by comparing the three baselines among each other. As follows from Table \ref{res}, BaseR baseline outperforms BaseEE and BaseE on all query sets, while BaseEE is the worst performing baseline. The BaseR retrieval is the only relationship-centric approach from the three baselines, as its document collection comprises entity-pairs that co-occurred in ClueWeb-09-B corpus. BaseEE and BaseE retrieve entity pairs that are created in a post-processing step which reduces the probability of retrieving relevant results. This results shows the need for a relationship-centric document collection when aiming to answer entity-relationship queries.

ERDM significantly outperform all baselines on all query sets. We performed statistical significance testing of MAP using ERDM against each baseline obtaining p-values below 0.05 on all the query sets. This results show that our Early Fusion approach using two indexes (one for entities and other for relationships) is adequate and promising. We believe this approach can become a reference for future research in E-R retrieval from an IR-centric perspective.

Nevertheless, based on the absolute results obtained on each evaluation metric and for each query set we can conclude that E-R retrieval is still very far from being a solved problem. There is room to explore new feature functions and retrieval approaches. This is a very difficult problem and the methods we proposed are still far from optimal performance. Queries such as ``\textit{Find world war II flying aces and their services}'' or ``Which mountain is the highest after Annnapurna?'' are examples of queries with zero relevant judgments returned. 

On the other hand, ERDM exhibits interesting performance in some queries with high complexity, such as ``Computer scientists who are professors at the university where Frederick Terman was a professor.'' We speculate about some aspects that might influence performance. 

One aspect has to do with the lack of query relaxation in our experimental setup. The relevant entity tuples might be in our indexes but if the query terms used to search for entity tuples do not match the query terms harvested from ClueWeb-09B it is not possible to retrieve those relevant judgments. Query relaxation approaches should be tried in future work. More specifically, with the recent advances in word embeddings it is possible to expand queries with alternative query terms that are in the indexes.

On the other hand, we adopted a very simple approach for extracting entities and relationships. The use of dependency parsing and more complex methods of relation extraction would allow to filter out noisy terms. We also leave this for future work. Moreover, to further assess the influence of the extraction method we propose to use selective text passages containing the target entity pairs and the query terms associated as well. Then different extraction methods could be tried and straightforward evaluation of their impact.

\begin{figure*}[h]
    \centering
    \begin{subfigure}[t]{0.5\textwidth}
        \centering
        \includegraphics[width= 1.0\linewidth]{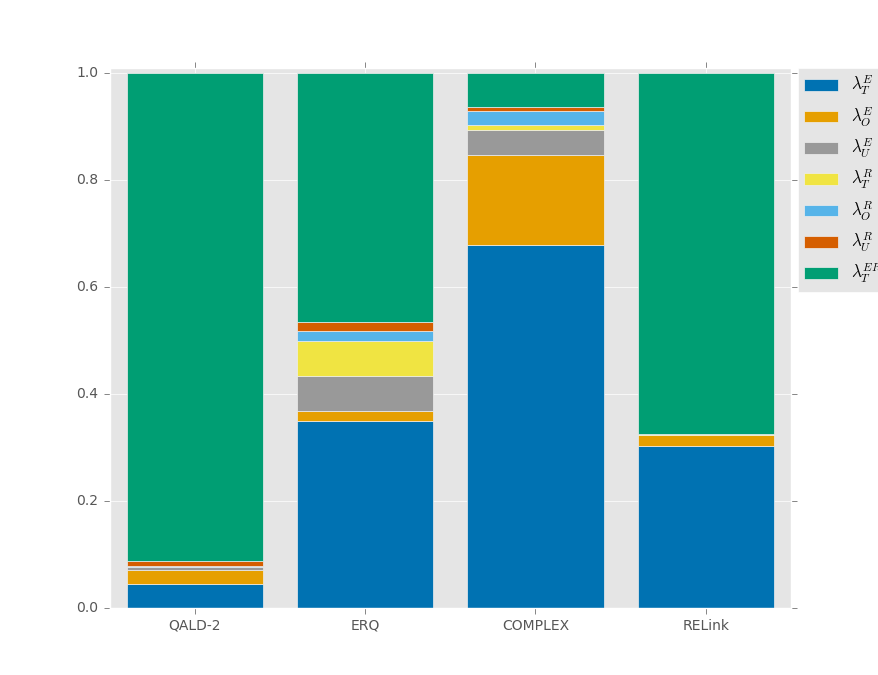}
        \caption{}
    \end{subfigure}%
    ~ 
    \begin{subfigure}[t]{0.5\textwidth}
        \centering
        \includegraphics[width= 1.0 \linewidth]{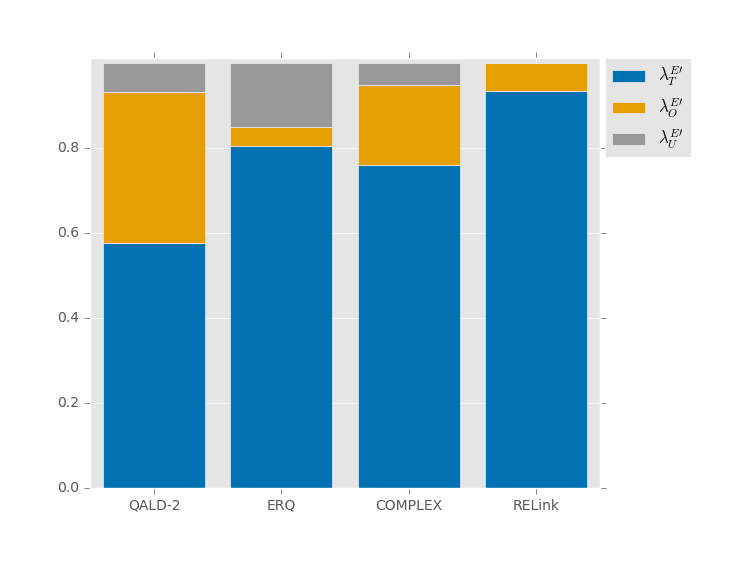}
        \caption{}
    \end{subfigure}
    ~
    \begin{subfigure}[t]{0.5\textwidth}
    \centering
    \includegraphics[width = 1.0\linewidth]{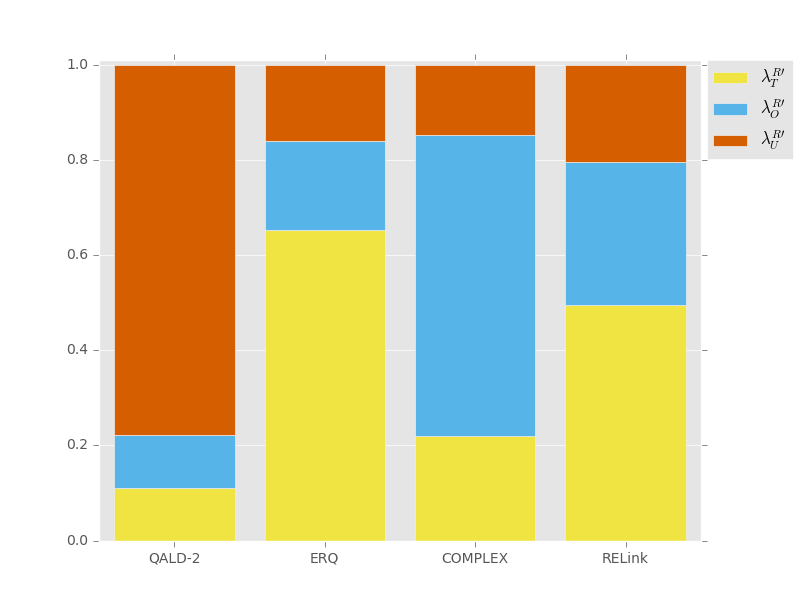}
    
          \caption{}
    \end{subfigure}
    \caption{Values of $\lambda$ for ERDM: (a) all $\lambda$, (b) $\lambda^{E'}$, (c) $\lambda^{R'}$. (b) and (c) were obtained using sum normalization. }
    \label{lambdas}
\end{figure*}

To understand how much importance is attributed to the different types of clique sets, we plot the values of the lambda parameters:  $\lambda^E$ parameters represent the feature importance of the set of functions targeting the dependence between entity query terms and the entity documents in overall ranking score for entity-pairs; $\lambda^R$ represent the importance of the feature functions of the relationship type queries and finally, the value for $\lambda^{ER}$ which is assigned to the feature function that evaluates if each entity retrieved from both entity type queries belongs to the entity-pair retrieved from the relationship type query. 

We plot the feature weights learned on each query set, as depicted in Figure \ref{lambdas}. We see that $\lambda^{ER}$  and $\lambda^{E}_{T}$ (weight for the unigram language model in the entity type queries) dominate the ranking function. We further evaluated the relative weights for each one of the three SDM-like functions using a sum normalization of the three weights for both entity documents and entity-pair documents. We observe that $\lambda^{E}_{T}$ dominates on every query set, however the same does not happen with $\lambda^{R}_{T}$. For relationship type queries the bigram features have higher values for COMPLEX and RELink.

\section{Summary of the Contributions}
In this chapter we presented the following contributions to the E-R retrieval research area:

\begin{enumerate}

\item Indexing method that supports generalization of entity types and entity-relationships to any attribute and predicate, respectively
\item A semi-automatic method for generating E-R test collections, which resulted in the RELink Query Collection comprising 600 E-R queries.
\item Results of experiments at scale, with a comprehensive set of queries and corpora.
\end{enumerate}

\chapter{Entity Filtering and Financial Sentiment Analysis}
\label{ch:text}

In this chapter we present the work developed to tackle two fundamental Text Mining problems in ORM: Entity Filtering and Sentiment Analysis. We start by describing our participation at the Filtering task of RepLab 2013 \citep{amigo2013overview}. We developed a supervised method to classify tweets as relevant or non-relevant to given target entity. This method obtained the first place at the competition. Entity Filtering can be seen as target based Named Entity Disambiguation (NED). Given a target entity under study, we need to develop a binary classifier to filter out tweets that are not talking about the target entity. This task is fundamental in ORM as downstream tasks such as Sentiment Analysis or entity-centric predictions would produce misleading results if noisy signals were used.  

Sentiment Analysis has been widely studied over the last decade. It is a research area with several ramifications as it is dependent on the type of texts and the objective of the analysis. We decided to focus our efforts in a not so well explored sub-area of Sentiment Analysis. SemEval 2017 Task 5 focused on fine-grained sentiment analysis of financial news and microblogs. As one of the use cases of ORM is to track the online reputation of companies and try to assess its impact on the stock market we decided it was a specific task within Sentiment Analysis in which we could make a contribution. We obtained the fourth place in the Microblogs sub-task using one of the evaluation metrics. The task consisted in predicting a real continuous variable from -1.0 to +1.0 representing the polarity and intensity of sentiment concerning companies/stocks mentioned in short texts. We modeled it as a regression analysis problem.

\section{Entity Filtering\footnote{Most of the material contained in this section was published in P.Saleiro,  E. M. Rodrigues, C. Soares, E. Oliveira, ``TexRep: A Text Mining Framework for Online Reputation Monitoring''  \citep{saleiro-ngc}}}
The relationship between people and public entities has changed with the rise of social media. Online users of social networks, blogs and micro-blogs are able to directly express and spread opinions about public entities, such as politicians, artists, companies or products. Online Reputation Monitoring (ORM) aims to automatically process online information about public entities. Some of the common tasks within ORM consist in collecting, processing and aggregating social network messages to extract opinion trends about such entities. 

Twitter, one of the most used online social networks, provides a search system that allows users to query for tweets containing a set of keywords. ORM systems often use Twitter as a source of information when monitoring a given entity. However, search results are not necessarily relevant to that entity because keywords can be ambiguous. For instance, a tweet containing the word ``columbia'' can be related with several entities, such as a federal state, a city or a university. Furthermore, tweets are short which results in a reduced context for entity disambiguation. When monitoring the reputation of a given entity on Twitter, it is first necessary to guarantee that all tweets are relevant to that entity. Consequently, other processing tasks, such as sentiment analysis will benefit from filtering out noise in the data stream.

In this work, we tackle the aforementioned problem by applying a supervised learning approach. Given a set of entities $E = \{e_1, e_2, ..., e_i, ...\}$, a stream of texts $S = \{s_1, s_2, ..., s_i, ...\}$ (e.g. tweets), we are interested in monitoring the mentions of an entity $e_i$ on the stream $S$, i.e. the discrete function $f_m(e_i, S)$. We cast the prediction of $f_m$ as a supervised learning classification problem, in which we want to infer the target variable $\hat{f_m}(e_i, S) \in \{0,1\}$ 

We implemented a large set of features that can be generated to describe the relationship between an entity representation and a text mention. We use metadata (e.g. entity names, category) provided in the user configurations, text represented with TF-IDF, similarity between texts and Wikipedia, Freebase entities disambiguation, feature selection of terms based on frequency and feature matrix transformation using SVD. The learning algorithms from scikit-learn Python library that were tested for Entity Filtering include Naive Bayes, SVM, Random Forests, Logistic Regression and MultiLayer Perceptron.

\subsection{Task Overview}

RepLab 2013 \cite{amigo2013overview} focused on monitoring the online reputation of entities on Twitter. The Filtering task consisted in determining which tweets are relevant to each entity. The corpus consists of a collection of tweets obtained by querying the Twitter Search API with 61 entity names during the period from the June 2012 until the  December 2012. The corpus contains tweets both in English and Spanish. The balance between both languages varies for each entity. Tweets were manually annotated as ``Related'' or ``Unrelated'' to the respective target entity.

The data provided to participants consists in tweets and a list of 61 entities. For each tweet in the corpus we have the target entity id, the language of the tweet, the timestamp and the tweet id. The content of each URL in the tweets is also provided. Due to Twitter's terms of service, the participants were responsible to download the tweets using the respective id. The data related with entities contain the query used to collect the tweets (e.g. ``BMW''), the official name of the entity (e.g. ``Bayerische Motoren Werke AG''), the category of the entity (e.g. ``automotive''), the content of its homepage and both Wikipedia articles in English and Spanish.

\subsection{Pre-processing}

The Entity Filtering module includes methods to normalize texts by removing all punctuation, converting text to lower case, removing accents and converting non-ASCII characters to their ASCII equivalent. Lists of stop words for several languages are also available, which are used to filter out non relevant words. We rely on the Natural Language Toolkit (NLTK) to provide those lists. 

Contrary to other types of online texts (e.g. news or blog posts) tweets contain informal and non-standard language including emoticons, spelling errors, wrong letter casing, unusual punctuation and abbreviations.  Therefore, when dealing with tweets, the Entity Filtering module uses a tokenizer~\cite{Laboreiro} optimized for segmenting words in tweets. After tokenization we extract user mentions and URLS and hashtags textual content.

\subsection{Features}

Many different types of features can be used to optimize relevance classification, including language models, keyword similarities between tweets and entities as well as external resources projections. We implemented a large number of those. We assume that future users of our framework for ORM will provide entity-specific data (e.g. homepage/Wikipedia content) prior to training and configuring the Entity Filtering module.

\begin{description}

\item[\textbf{Language Model}:] text is encapsulated in a single feature to avoid high dimensionality issues when adding other features. A TF-IDF representation of unigrams, bigrams and trigrams for training a text classifier which calculates the probability of a text being related to the expected entity. The output probabilities of the classifier are used as a feature.

\item[\textbf{Keyword similarity}:] similarity scores between metadata and the texts, obtained by calculating the ratio of the number of common terms in the texts and the terms of query and entity name. Similarities at character level are also available in order to include possible spelling errors in the text. 

\item[\textbf{Web similarity}:] similarity between the text and the normalized content of the entity's homepage and normalized Wikipedia articles are also available. The similarity value is the number of common terms multiplied by logarithm of the number of terms in tweet.

\item[\textbf{Freebase}:] For each keyword of the entity's query that exists in the text, two bigrams are created, containing the keyword and the previous/subsequent word. These bi-grams are submitted to the Freebase Search API and the list of retrieved entities are compared with the id of the target entity on Freebase. A Freebase score is computed by using the inverse position of the target entity in the list of results retrieved. If the target entity is the first result, the score is 1, if it is the second, the score is 0.5, and so on. If the target entity is not in the results list, the score is zero. The feature corresponds to the maximum score of the extracted bigrams of each text.

\item[\textbf{Category classifier}:] a sentence category classifier is created using the Wikipedia articles of each entity.  Each sentence  of the Wikipedia articles is annotated with the category of the corresponding entity. TF-IDF for unigrams, bigrams and trigrams are calculated and a multi-class classifier (SVM) is trained to classify each text. The feature is the probability of the text being relevant to its target class.

\end{description}

\subsection{Experimental Setup}

The dataset used for the competition consists of a collection of tweets both in English and Spanish, possibly relevant to 61 entities from four domains: automotive, banking, universities and music.The dataset consists of a collection of tweets obtained by querying the Twitter Search API with 61 entity names during the period from the June 2012 until the  December 2012. The balance between both languages varies for each entity. The complementary data about each target entity is the following:
\begin{itemize}
\item query used to collect the tweets (e.g. ``BMW'')
\item official name of the entity (e.g. ``Bayerische Motoren Werke AG'')
\item category of the entity (e.g. ``automotive'')
\item content of entity homepage
\item Wikipedia article both in English and Spanish
\end{itemize}

Tweets were manually annotated as ``Related'' or ``Unrelated'' to the respective target entity. The dataset is divided in training, test and development (Table \ref{tab:data}). The training set consists in a total of 45,671 tweets from which we were able to download 43,582. Approximately 75\% of tweets in the training set are labeled as ``Related''. We split the training dataset into a development set and a validation set, containing 80\% and 20\% of the original, respectively. We adopted a randomly stratified split approach per entity, i.e., we group tweets of each target entity and randomly split them preserving the balance of ``Related''/``Unrelated'' tweets. The test dataset consists of 90,356 tweets from which we were able to download 88,934.

\begin{table}
\centering
\begin{tabular}{l l l l}
\hline\noalign{\smallskip}
  Dataset & Related & Unrelated & Total \\ 
\noalign{\smallskip}\hline\noalign{\smallskip}
  Training & 33,193  & 10,389 & 43,582 \\ 
  Development & 26,534  & 8,307 & 34,841 \\ 
  Validation & 6,659 & 2,082 & 8,741 \\ 
  Test & 75,470  & 21,378 & 96,848   \\
  \noalign{\smallskip}\hline
\end{tabular}
\caption{RepLab 2013 Filtering Task dataset description.}\label{tab:data}
\end{table}

We used the development set for trying new features and test algorithms. We divided the development set in 10 folds generated with the randomly stratified approach. We used the validation set to validate the results obtained in the development set. The purpose of this validation step is to evaluate how well the Entity Filtering classifier generalizes from its training data to the validation data and thus estimate how well it will generalize to the test set. It allows us to spot overfitting. After validation, we trained the classifier using all of the data in the training dataset and evaluated in the test set.

\subsection{Results}\label{sec:Results}

We created different classifier runs using different learners, features and we also created entity specific models as explained in Table~\ref{tab:submissions}, \cite{saleiro2013popstar}. We applied selection of features based on frequency and transformation of content representation using SVD. The learners tested include Naive Bayes (NB), SVM, Random Forests (RF), Logistic Regression (LR) and MultiLayer Perceptron (MLP). 
The evaluation measures used are accuracy and the official metric of the competition, F-measure which is the harmonic mean of Reliability and Sensitivity~\cite{fmeasure}. We present results for the top 4 models regarding the F-measure. We replicated the best system at RepLab 2013 in the run 1.

\begin{table}[h]
\centering
\begin{tabular}{ l  l  l  l  }
  \hline\noalign{\smallskip}
  Run & Learner & Features &  No. of models\\
  \noalign{\smallskip}\hline\noalign{\smallskip}
  1 & SVM & All  & global \\ 
  2 & RF & All   & global \\ 
  3 & RF & All   & per entity \\

  \noalign{\smallskip}\hline
\end{tabular}
\caption{Entity filtering versions description.}\label{tab:submissions}
\end{table}

Table~\ref{tab:results} shows the results of top performing runs and the official baseline of the competition. This baseline classifies each tweet with the label of the most similar tweet of target entity in the training set using Jaccard similarity coefficient. The baseline results were obtained using 99.5\% of the test set.

\begin{table}[h]
\begin{tabular}{llllll}
\hline\noalign{\smallskip}
  Run & Acc. (Val. Set) & Acc. & R & S & F-measure \\ 
\noalign{\smallskip}\hline\noalign{\smallskip}
  1 & 0.944 & 0.906 &  0.759 & 0.428  & 0.470 \\ 
  2 & 0.945 & 0.908 &  0.729 & 0.451  & 0.488 \\ 
  3 & 0.948 & 0.902 &  0.589 & 0.444  & 0.448 \\ 
  Official Baseline & - & 0.8714  & 0.4902 & 0.3199 & 0.3255 \\
  Best RepLab & - & 0.908 &  0.729 & 0.451  & 0.488 \\
\noalign{\smallskip}\hline
\end{tabular}
\caption{Official results for each version plus our validation set accuracy.}\label{tab:results}
\end{table}

Based on the results achieved we are able to conclude that the models of our classifier are able to generalize successfully. Results obtained in the validation set are similar to those obtained in the test set. During development, solutions based on one model per entity were consistently outperformed by solutions based on global models. We also noticed during development that language specific models (English and Spanish) did not exhibit improvements in global accuracy, therefore we opted to use language as a feature. Results show that the best model uses the Random Forests classifier with 500 estimators for training a global model. Though, the Language Modeling feature encapsulates text by using a specific model trained just with TF-IDF of n-grams of tweets.

\begin{figure}[h]
\centering
    \includegraphics[width=0.8\linewidth]{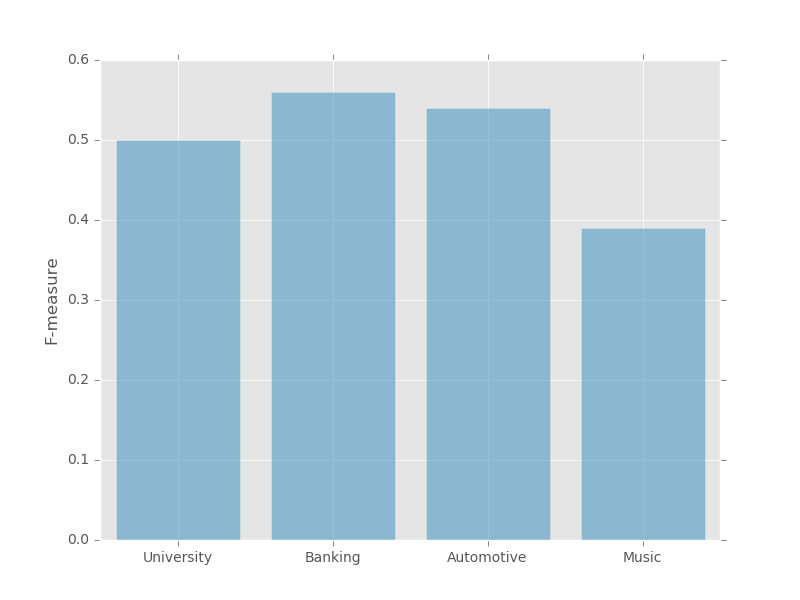}
\caption{Results grouped by entity's category using Run 2.}
\label{fig:categs}
\end{figure}

We performed a ``break down'' analysis for each one of the four categories of RepLab 2013 using Run 2 model, as depicted in Figure \ref{fig:categs}. We observe that University, Banking and Automotive categories exhibit similar average F-measure results, all above 0.50. In contrast, results for Music shows it is a rather difficult category of entities to disambiguate (achieving F-measure of 0.39). In fact, some of the entity names of this category contain very ambiguous tokens, such as ``Alicia Keys'', ``U2'', ``The Wanted'' or ``The Script''.  

The main goal of this task was to classify tweets as relevant or not to a given target entity. We have explored several types of features, namely similarity between keywords, language models and we have also explored external resources such as Freebase and Wikipedia. Results show that it is possible to achieve an Accuracy over 0.90 and an F-measure of 0.48 in a test set containing more than 90000 tweets of 61 entities. In future work, we expect to include the possibility of using entity-specific embedding to learn a joint embedding space of entities and words, similar to \cite{moreno2017combining}.


\section{Financial Sentiment Analysis\footnote{The material contained in this section was published in P. Saleiro, E. M. Rodrigues, C. Soares, E. Oliveira, ``FEUP at SemEval-2017 Task 5: Predicting Sentiment Polarity and Intensity with Financial Word Embeddings'' \citep{saleiro-semeval}}}

Sentiment Analysis on financial texts has received increased attention in recent years~\citep{nardo2016}. Nevertheless, there are some challenges yet to overcome~\citep{smailovic2014stream}. Financial texts, such as microblogs or newswire, usually contain highly technical and specific vocabulary or jargon, making the development of specific lexical and machine learning approaches necessary. Most of the research in Sentiment Analysis in the financial domain has focused in analyzing subjective text, labeled with explicitly expressed sentiment.

However, it is also common to express financial sentiment in an implicit way. Business news stories often refer to events that  might indicate a positive or negative impact, such as in the news title ``company X will cut 1000 jobs''. Economic indicators, such as unemployment and change over time such as drop or increase can also provide clues on the implicit sentiment~\citep{musat2010impact}. Contrary to explicit expressions (subjective utterances), factual text types often contain objective statements that convey a desirable or undesirable fact~\citep{liu2012sentiment}.

Recent work proposes to consider all types of implicit sentiment expressions ~\citep{van2015fine}.
The authors created a fine grained sentiment annotation procedure to identify polar expressions (implicit and explicit expressions of positive and negative sentiment). A target (company of interest) is identified in each polar expression to identify the sentiment expressions that are relevant. The annotation procedure also collected information about the polarity and the intensity of the sentiment expressed towards the target. However, there is still no automatic approach, either lexical-based or machine learning based, that tries to model this annotation scheme.

In this work, we propose to tackle the aforementioned problem by taking advantage of unsupervised learning of word embeddings in financial tweets and financial news headlines to construct a domain-specific syntactic and semantic representation of words. We combine bag-of-embeddings with traditional approaches, such as pre-processing techniques, bag-of-words and financial lexical-based features to train a regressor for sentiment polarity and intensity. We study how different regression algorithms perform using all features in two different sub-tasks at SemEval-2017 Task 5: microblogs and news headlines mentioning companies/stocks. Moreover, we compare how different combinations of features perform in both sub-tasks. The system source code and word embeddings developed for the competition are publicly available.%
\footnote{\url{https://github.com/saleiro/Financial-Sentiment-Analysis}}

\subsection{Task Overview}

The task 5 of SemEval 2017~\citep{semeval2017task5} consisted of fine-grained sentiment analysis of financial short texts and it was divided in two  sub-tasks based on the type of text. Sub-task 5.1 -- Microblogs -- consisted of stocktwits and tweets focusing on stock market events and assessments from investors and traders. Companies/stocks were identified using stock symbols, the so called cashtags, e.g.``\$AMZN'' for the company Amazon.com, Inc. Sub-task 5.2 -- News Headlines -- consisted of sentences extracted from Yahoo Finance and other financial news sources on the Internet. In this case, companies/stocks were identified using their canonical name and were previously annotated by the task organizers.

\begin{table*}[h]
\centering
\begin{tabular}{|l|l|p{5cm}|l|}
\hline
  Sub-task & Company & Text Span & Sentiment Score\\
  \hline
  5.1 - Microblogs & JPMorgan &``its time to sell banks" & -0.763 \\
  5.2 - Headlines & Glencore &``Glencore's annual results beat forecasts" & +0.900 \\
  \hline
\end{tabular}
\caption{Training set examples for both sub-tasks.} \label{examp}
\end{table*}

The goal of both sub-tasks was the following: predict the sentiment polarity and intensity for each of the companies/stocks mentioned in a short text instance (microblog message or news sentence). The sentiment score is a real continuous variable in the range of -1.0 (very negative/bearish) to +1.0 (very positive/bullish), with 0.0 designating neutral sentiment. Table~\ref{examp} presents two examples from the training set. Task organizers provided 1700 microblog messages for training and 800 messages for testing in sub-task 5.1, while in sub-task 5.2, 1142 news sentences were provided for training and 491 for testing. Submissions were evaluated using the cosine similarity~\citep{semeval2017task5}. 

\subsection{Financial Word Embeddings}
\citeauthor{mikolov2013efficient} \cite{mikolov2013efficient} created word2vec, a computationally efficient method to learn distributed representation of words, where each word is represented by a distribution of weights (embeddings) across a fixed set of dimensions. Furthermore,~\citeauthor{mikolov2013distributed} \cite{mikolov2013distributed} showed that this representation is able to encode syntactic and semantic similarities in the embedding space. 

The training objective of the skip-gram model, defined by ~\citeauthor{mikolov2013distributed} \cite{mikolov2013distributed}, is to learn the target word representation (embeddings) that maximize the prediction of its surrounding words in a context window. Given the $w_t$ word in a vocabulary the objective is to maximize the average log probability: 

\begin{equation}
\frac{1}{T}  \sum_{t=1}^{T}  \sum_{-c \leq j \leq  c, j \neq 0} \textnormal{log } P(w_{t+j} | w_t)
\end{equation}

where $c$ is the size of the context window, $T$ is the total number of words in the vocabulary and $w_{t+j}$ is a word in the context window of $w_t$. After training, a low dimensionality embedding matrix $\textbf{E}$ encapsulates information about each word in the vocabulary and its use (surrounding contexts).

We used word2vec to learn word embeddings in the context of financial texts using unlabeled tweets and news headlines mentioning companies/stocks from S\&P 500. Tweets were collected using the Twitter streaming API with cashtags of stocks titles serving as request parameters. Yahoo Finance API was used for requesting financial news feeds by querying the canonical name of companies/stocks. The datasets comprise a total of 1.7M tweets and 626K news titles.

We learned separate word embeddings for tweets and news headlines using the skip-gram model. We tried several configurations of word2vec hyperparameters. The setup resulting in the best performance in both sub-tasks was skip-gram with 50 dimensions, removing words occurring less than 5 times, using a context window of 5 words and 25 negative samples per positive example. 

Even though the text collections for training embeddings were relatively small, the resulting embedding space exhibited the ability to capture semantic word similarities in the financial context. We performed simple algebraic operations to capture semantic relations between words, as described in ~\citeauthor{mikolov2013linguistic} \cite{mikolov2013linguistic}.  For instance, the skip-gram model trained on tweets shows that vector (``bearish'') - vector(``loss'') + vector(``gain'') results in vector (``bullish'') as most similar word representation. 

\subsection{Approach} \label{app}
In this section we describe the implementation details of the proposed approach.

\subsubsection{Pre-Processing}
A set of pre-processing operations are applied to every microblog message and news sentence in the training/test sets of sub-tasks 5.1 and 5.2, as well as in the external collections for training word embeddings:

\begin{itemize}
\item \textbf{Character encoding and stopwords}: every message and headline was encoded in UTF-8. Standard english stopword removal is also applied.

\item \textbf{Company/stock and cash obfuscation}: both cashtags and canonical company names strings were replaced by the string \textit{\_company\_}. Dollar or Euro signs followed by numbers were replaced by the string \textit{\_cash\_amount\_}.

\item \textbf{Mapping numbers and signs}: numbers were mapped to strings using bins (0-10, 10-20, 20-50, 50-100, $>$100). Minus and plus signs were coverted to \textit{minus} and \textit{plus}, ``B'' and ``M'' to \textit{billions} and \textit{millions}, respectively. The \% symbol was converted to \textit{percent}. Question and exclamation marks were also converted to strings.

\item \textbf{Tokenization, punctuation, lowercasing}: tokenization was performed using Twokenizer~\citep{gimpel2011part}, the remaining punctuation was removed and all characters were converted to lowercase.
\end{itemize}

\subsubsection{Features}

We combined three different group of features: bag-of-words, lexical-based features and bag-of-embeddings. 

\begin{itemize}
\item \textbf{Bag-of-words}: we apply standard bag-of-words as features. We tried unigrams, bi-grams and tri-grams with unigrams proving to obtain higher cosine similarity in both sub-tasks.

\item \textbf{Sentiment lexicon features}: we incorporate knowledge from manually curated sentiment lexicons for generic Sentiment Analysis as well as lexicons tailored for the financial domain. The Laughran-Mcdonald financial sentiment dictionary~\cite{bodnaruk2015using} has several types of word classes: positive, negative, constraining, litigious, uncertain and modal. For each word class we create a binary feature for the match with a word in a microblog/headline and a polarity score feature (positive - negative normalized by the text span length).  As a general-purpose sentiment lexicon we use MPQA~\citep{wilson2005recognizing} and created binary features for positive, negative and neutral words, as well as, the polarity score feature.

\item \textbf{Bag-of-Embeddings}: we create bag-of-embeddings by taking the average of word vectors for each word in a text span. We used the corresponding embedding matrix trained on external Twitter and Yahoo Finance collections for sub-task 5.1 and sub-task 5.2, respectively. 

\end{itemize}

\subsection{Experimental Setup}

In order to avoid overfitting we created a validation set from the original training datasets provided by the organizers. We used a 80\%-20\% split and sampled the validation set using the same distribution as the original training set.
We sorted the examples in the training set by the target variable values and skipped every 5 examples. Results are evaluated using Cosine similarity~\citep{semeval2017task5} and Mean Average Error (MAE). The former gives more importance to differences in the polarity of the predicted sentiment while the latter is concerned with how well the system predicts the intensity of the sentiment.


We opted to model both sub-tasks as single regression problems. Three different regressors were applied: Random Forests (RF), Support Vector Machines (SVM) and MultiLayer Perceptron (MLP). Parameter tuning was carried using 10 fold cross validation on the training sets.

\subsection{Results and Analysis}

In this section we present the experimental results obtained in both sub-tasks. We provide comparison of different learning algorithms using all features, as well as, a comparison of different subsets of features, to understand the information contained in each of them and also how they complement each other.

\subsubsection{Task 5.1 - Microblogs}
\label{sec:length}

Table~\ref{res1} presents the results obtained using all features in both validation set and test sets.  Results in the test set are worse than in the validation set with the exception of MLP. The official score obtained in sub-task 5.1 was 0.6948 using Random Forests (RF), which is the regressor that achieves higher cosine similarity and lower MAE in both training and validation set.

\begin{table}[h]
\centering
\begin{tabular}{|l|l|l|l|}
\hline
  Regressor & Set & Cosine & MAE\\
  \hline
  RF & Val & 0.7960 & 0.1483 \\
  RF & \textbf{Test} & \textbf{0.6948} & \textbf{0.1886} \\
  SVR & Val & 0.7147 & 0.1944 \\
  SVR & \textbf{Test} & 0.6227 & 0.2526 \\
  MLP & Val & 0.6720 & 0.2370 \\
  MLP & \textbf{Test} & 0.6789 & 0.2132 \\
  \hline
\end{tabular}
\caption{Microblog results with all features on validation and test sets.} \label{res1}
\end{table}

We compared the results obtained with different subsets of features using the best regressor, RF, as depicted in Table~\ref{res2}. Interestingly, bag-of-words (BoW) and bag-of-embeddings (BoE) complement each other, obtaining better cosine similarity than the system using all features. Financial word embeddings (BoE) capture relevant information regarding the target variables. As a single group of features it achieves a cosine similarity of 0.6118 and MAE of 0.2322. It is also able to boost the overall performance of BoW with gains of more than 0.06 in cosine similarity and reducing MAE more than 0.03. 

The individual group of features with best performance is Bag-of-words while the worst is a system trained using Lex (only lexical-based features). While Lex alone exhibits poor performance, having some value but marginal, when combined with another group of features, it improves the results of the latter, as in the case of BoE + Lex and BoW + Lex.

\begin{table}[h]
\centering
\begin{tabular}{|l|l|l|}
\hline
  Features &  Cosine & MAE\\
  \hline
  Lex &  0.3156 & 0.3712 \\
  BoE &  0.6118 & 0.2322 \\
  BoW &  0.6386 &0.2175 \\
  BoE + Lex &  0.6454 & 0.2210 \\
  Bow + Lex &  0.6618 & 0.2019 \\
  Bow + BoE &  \textbf{0.7023} & \textbf{0.1902} \\
  All &  0.6948 & 0.1886 \\

  \hline
\end{tabular}
\caption{Features performance breakdown on test set using RF.} \label{res2}
\end{table}

\subsubsection{Task 5.2 - News Headlines}

Results obtained in news headlines are very different from the ones of the previous sub-task, proving that predicting sentiment polarity and intensity in news headlines is a completely different problem compared to microblogs.
Table~\ref{res3} shows that MLP obtains the best results in the test set using both metrics while SVR obtains the best performance in the validation set. The best regressor of sub-task 5.1, RF is outperformed by both SVR and MLP. The official result obtained at sub-task 5.2 was a cosine similarity of 0.68 using MLP. 

\begin{table}[h]
\centering
\begin{tabular}{|l|l|l|l|}
\hline
  Regressor & Set & Cosine & MAE\\
  \hline
  RF & Val & 0.5316 &0.2539 \\
  RF & \textbf{Test} & 0.6562 & 0.2258\\
  SVR & Val & 0.6397 &0.2422 \\
  SVR & \textbf{Test} &  0.6621 & 0.2424 \\
  MLP & Val &  0.6176 &0.2398 \\
  MLP & \textbf{Test} & \textbf{0.6800} & \textbf{0.2271} \\

  \hline
\end{tabular}
\caption{News Headlines results with all features on validation and test sets.} \label{res3}
\end{table}

Table~\ref{res4} shows the results of the different groups of features in sub-task 5.2 for MLP regressor. The most evident observation is that word embeddings are not effective in this scenario. On the other hand, lexical based features have significantly better performance in news headlines than in microblogs. Despite this, the best results are obtained using all features.

\begin{table}[h]
\centering
\begin{tabular}{|l|l|l|}
\hline
  Features &  Cosine & MAE\\
  \hline
  BoE &  0.0383 & 0.3537 \\
  Lex &  0.5538 & 0.2788 \\
  BoW &  0.6420 &0.2364 \\
  BoE + Lex &  0.5495 & 0.2830 \\
  BoW + Lex &  \textbf{0.6733} & \textbf{0.2269} \\
  BoW + BoE &   0.6417 &  0.2389 \\
  All & 0.6800 & 0.2271 \\

  \hline
\end{tabular}
\caption{Features performance breakdown on test set using MLP.} \label{res4}
\end{table}

\subsubsection{Analysis}

Financial word embeddings were able to encapsulate valuable information in sub-task 5.1 - Microblogs but not so much in the case of sub-task 5.2 - News Headlines. We hypothesize that as we had access to a much smaller dataset ($\sim 600$K) for training financial word embeddings for news headlines, this resulted in reduced ability to capture semantic similarities in the financial domain. Other related works in Sentiment Analysis usually take advantage of a much larger dataset for training word embeddings \citep{deriu2016swisscheese}.

On the other hand, lexical features showed poor performance in microblog texts but seem to be very useful using news headlines. The fact that microblogs have poor grammar, slang and informal language reveals that financial lexicons created using well written and formal financial reports, result better in news headlines rather than in microblog texts. 

After inspecting microblog texts and headlines in which our models showed poor performance we believe it would be important to also encapsulate syntactic and semantic dependencies in our models. For instance, our model predicted a sentiment score of -0.467 for the microblog message ``was right to reject the offer'' while the true value is 0.076. Similar examples include ``Glencore shares in record crash as profit fears grow'' and ``I would rather be a buyer at these levels then trying to sell'', in which our models has absolute errors around 0.5. Other type of errors have to do with intensity of the sentiment in which our model correctly predicts the polarity but still has a large error.

\subsection{Concluding Remarks}

Work reported here reported is concerned with the problem of predicting sentiment polarity and intensity of financial short texts. Previous work showed that sentiment is often depicted in an implicit way in this domain. We created financial-specific continuous word representations in order to obtain domain specific syntactic and semantic relations between words. We combined traditional bag-of-words and lexical-based features with bag-of-embeddings to train a regressor of both sentiment polarity and intensity. Results show that different combination of features attained different performances on each sub-task. Future work will consist on collecting larger external datasets for training financial word embeddings of both microblogs and news headlines. We also have planned to perform the regression analysis using Deep Neural Networks.

\section{Summary of the Contributions}
In this chapter we present some contributions to two fundamental Text Mining problems in ORM.

\begin{itemize}
\item  A supervised learning approach for Entity Filtering on tweets, achieving state-of-the-art performance using a relatively small training set. 

\item Created and made available word embeddings trained from financial texts.

\item A supervised learning approach for fine-grained sentiment analysis of financial texts.
\end{itemize}

\chapter{Text-based Entity-centric Prediction}\label{ch:predict}

In this chapter we explore the predictive power of entity-centric information in online news and social media in the context of ORM. We address two different predictive tasks. The first is concerned with predicting entity popularity on Twitter based on signals extracted from the news cycle. We aim to study different sets of signals extracted from online news mentioning specific entities that could influence or at least are correlated with future popularity of those entities on Twitter. We know that entity popularity on social media can be influenced by several factors but we are only interested in exploring the interplay between online news and social media for entities that are frequently mentioned on the news cycle such as politicians or footballers. This could be particularly interesting for anticipating public relations damage control once a polemic news article is published. Or even for editorial purposes to maximize buzz on social media.   

The second predictive task consists in using entity-centric sentiment polarity extracted from tweets to predict political polls. There has been several research work trying to assess the predictive power of social media to predict the outcome of political opinion surveys or elections. However, each study proposes its own method of aggregating polarity scores over time, however, there is not a consensus on which sentiment aggregate function is the most adequate for this problem. We propose to use and contrast several sentiment aggregate functions reported in the literature, by assessing their predictive power on a specific case comprising data collected during the Portuguese bailout (2011-2013).

\section{Exploring Online News for Reputation Monitoring on Twitter \footnote{The material contained in this section was published in P. Saleiro and C. Soares, ``Learning from the News: Predicting Entity Popularity on Twitter'' \citep{saleiro-ida}}}

Online publication of news articles has become a standard behavior of news outlets, while the public joined the movement either using desktop or mobile terminals. The resulting setup consists of a cooperative dialog between news outlets and the public at large. Latest events are covered and commented by both parties in a continuous basis through the social media, such as Twitter. When sharing or commenting news on social media, users tend to mention the most predominant entities mentioned in the news story. Therefore, entities, such as public figures, organizations, companies or geographic locations, can act as latent connections between online news and social media. 

Online Reputation Monitoring (ORM) focuses on continuously tracking what is being said about entities on social media and online news. Automatic collection and processing of comments and opinions on social media is now crucial to understand the reputation of individuals and organizations and therefore to manage their public relations. However, ORM systems would be even more useful if they would be able to know in advance if social media users will talk a lot about the target entities or not. 

We hypothesize that for entities that are frequently mentioned on the news (e.g. politicians) it is possible to establish a predictive link between online news and popularity on social media. We cast the problem as a supervised learning classification approach: to decide whether popularity will be high or low based on features extracted from the news cycle. We define four set of features: signal, textual, sentiment and semantic. We aim to respond to the following research questions: 

\begin{itemize}
\setlength\itemsep{-3pt}
\item Is online news a valuable source of information to effectively predict entity popularity on Twitter? 
\item Do online news carry different predictive power based on the nature of the entity under study? 
\item How do different thresholds for defining high and low popularity affect the effectiveness of our approach? 
\item Does the performance remain stable for different prediction times? 
\item What is the most important feature set for predicting entity popularity on Twitter based on the news cycle? 
\item Do individual sets of features exhibit different importance for different entities? 
\end{itemize}
\vspace{3pt}

\subsection{Approach}\label{sec:appr}
The starting point of our hypothesis is that for entities that are frequently mentioned on the news (e.g. politicians) it is possible to predict popularity on social media using signals extracted from the news cycle. The first step towards a solution requires the definition of entity popularity on social media.

\subsubsection{Entity Popularity}
There are different ways of expressing the notion of popularity on social media. For example, the classical way of defining it is through the number of followers of a Twitter account or the number of likes in a Facebook page. Another notion of popularity, associated with entities, consists on the number of retweets or replies on Twitter and post likes and comments on Facebook.
We define entity popularity based on named entity mentions in social media messages. Mentions consist of specific surface forms of an entity name. For example, ``Cristiano Ronaldo'' might be mentioned also using just ``Ronaldo'' or ``\#CR7''.  

Given an set of entities $E = \{e_1, e_2, ..., e_i, ...\}$, a daily stream of social media messages $S = \{s_1, s_2, ..., s_i, ...\}$ and a daily stream of online news articles $N = \{n_1, n_2, ..., n_i, ...\}$ we are interested in monitoring the mentions of an entity $e_i$ on the social media stream $S_t$, i.e. the discrete function $f_m(e_i, S_t)$. Let $T$ be a daily time frame $T = [t_p, t_{p+h}]$, where the time $t_p$ is the time of prediction and $t_{p+h}$ is the prediction horizon time. We want to learn a target popularity function $f_p$ on social media stream $S$ as a function of the given entity $e_i$, the online news stream $N$ and the time frame $T$:
$$f_p(e_i, N, T) = \sum_{t=t_p}^{t=t_{p+h}}  f_m(e_i, S_t)$$
which corresponds to integrating $f_m(e_i,S)$ over $T$.  

Given a day $d_i$, a time of prediction $t_p$, we extract features from the news stream $N$ until $tp$ and predict $f_p$ until the prediction horizon $tp+h$. We measure popularity on a daily basis, and consequently, we adopted $t_{p+h}$ as 23:59:59 everyday. For example, if $t_p$ equals to 8 a.m, we extract features from $N$ until 07:59:59 and predict $f_p$ in the interval 08:00 - 23:59:59 on day $d_i$. In the case of $t_p$ equals to midnight, we extract features from $N$ on the 24 hours of previous day $d_{i-1}$ to predict $f_p$ for the 24 hours of $d_i$.

We cast the prediction of $f_p(e_i, N, T)$ as a supervised learning classification problem, in which we want to infer the target variable $\hat{f_p}(e_i, N, T) \in \{0,1\}$ defined as:
$$   
 \hat{f_p}=
    \begin{cases}
      0 (\textit{low}), & \text{if}\ P(f_p(e_i, N, T) \leq \delta) = k \\
      1 (\textit{high}), & \text{if}\ P(f_p(e_i, N, T) > \delta) = 1 - k \\
    \end{cases}
$$
where $\delta$ is the inverse of cumulative distribution function at $k$ of $f_p(e_i, N, T)$ as measured in the training set, a similar approach to Tsagkias et al. \cite{tsagkias2009predicting}. For instance, $k=0.5$ corresponds to the median of $f_p(e_i, N, T)$ in the training set and higher values of $k$ mean that $f_p(e_i, N, T)$ has to be higher than $k$ examples on the training set to consider $\hat{f_p}=1$, resulting in a reduced number of training examples of the positive class \textit{high}.

\subsubsection{News Features}

Previous work has focused on the influence of characteristics of the social media stream $S$ in the adoption and popularity of memes and hashtags \cite{romero2011differences}. In contrast, the main goal of this work is to investigate the predictive power of the online news stream $N$. Therefore we extract four types of features from $N$ which we label: (i) \textit{signal}, (ii) \textit{textual}, (iii) \textit{sentiment} and (iv) \textit{semantic}, as depicted in Table \ref{table:feats}. 
One important issue is how can we filter relevant news items to $e_i$. There is no consensus on how to link a news stream $N$ with a social media stream $S$. Some works use URLs from $N$, shared on $S$, to filter simultaneously relevant news articles and social media messages \cite{bandari2012pulse}. As our work is entity oriented, we select news articles with mentions of $e_i$ as our relevant $N$.  

\textit{Signal Features -} This type of features depict the ``signal'' of the news cycle mentioning $e_i$ and we include a set of counting variables as features, focusing on the total number of news mentioning $e_i$ in specific time intervals, mentions on news titles, the average length of news articles, the different number of news outlets that published news mentioning $e_i$ as well as, features specific to the day of the week to capture any seasonal trend on the popularity. The idea is to capture the dynamics of news events, for instance, if $e_i$ has a sudden peak of mentions on $N$, a relevant event might have happened which may influence $f_p$.

\textit{Textual features -} To collect textual features we build a daily profile of the news cycle by aggregating all titles of online news articles mentioning $e_i$ for the daily time frame $[0, t_p]$ in $d_i$. We select the top 10,000 most frequent terms (unigrams and bi-grams) in the training set and create a document-term matrix $R$. Two distinct methods were applied to capture textual features. 

The first method is to apply TF-IDF weighting to $R$. We employ Singular Value Decomposition (SVD) to capture similarity between terms and reduce dimensionality. It computes a low-dimensional linear approximation $\sigma$. The final set of features for training and testing is the  TF-IDF weighted term-document matrix $R$ combined with $\sigma R$ which produces 10 real valued latent features. When testing, the system uses the same 10,000 terms from the training data and calculates TF-IDF using the IDF from the training data, as well as, $\sigma$ for applying SVD on test data.

The second method consists in applying Latent Dirichlet allocation (LDA) to generate a topic model of 10 topics (features). The system learns a topic-document distribution $\theta$ and a word distribution over topics $\varphi$ using the training data for a given entity $e_i$. When testing, the system extracts the word distribution of the news title vector $r$ on a test day $d'_i$. Then, by using $\varphi$ learned on training data, it calculates the probability of $r$ belonging to one of the 10 topics learned before. 
The objective of extracting this set of features is to create a characterization of the news stream that mentions $e_i$, namely, which are the most salient terms and phrases on each day $d_i$ as well as the latent topics associated with $e_i$. By learning our classifier we hope to obtain correlations between certain terms and topics and $f_p$.

\textit{Sentiment features -} We include several types of word level sentiment features. The assumption here is that subjective words on the news will result in more reactions on social media, as exposed in \cite{dos2015breaking}.  Once again we extract features from the titles of news mentioning $e_i$ for the daily time frame $[0, t_p]$. We use a sentiment lexicon as \textit{SentiWordNet} to extract subjective terms from the titles daily profile and label them as positive, neutral or negative polarity. We compute count features for number of positive, negative, neutral terms as well as difference and ratio of positive and negatives terms. Similar to textual features we create a TFIDF weighted term-document matrix $R$ using the subjective terms from the title and apply SVD to compute 10 real valued sentiment latent features.

\textit{Semantic features -} We use the number of different named entities recognized in $N$ on day $d_i$  until $tp$, as well as, the number of distinct news category tags extracted from the news feeds metadata. These tags, common in news articles, consist of author annotated terms and phrases that describe a sort of semantic hierarchy of news categories, topics and news stories (e.g. ``european debt crisis''). We create  a TF-IDF weighted entity-document and TF-IDF tag-document matrices and applied SVD to each of them to reduce dimensionality to 10. The idea is to capture interesting entity co-occurrences as well as, news stories that are less transient in time and might be able to trigger popularity on Twitter.

\begin{table}[t]
\centering
\caption{Summary of the four type of features we consider.}
\hspace*{-0.2cm}
\begin{tabular}{l l l}
\setlength{\tabcolsep}{4pt}

\\
\hline
\textbf{Number} &\textbf{Feature} & \textbf{Description}\\
\hline
\multicolumn{3}{l}{\texttt{Signal}} \\
1 & \textit{news} &   number of news mentions of $e_i$ in $[0, t_{p}]$ in $d_i$ \\
2 & \textit{news} $d_{i-1}$ &   number of news mentions of $e_i$ in $[0, t_{p}]$ in $d_{i-1}$\\
3 & \textit{news total} $d_{i-1}$ & number of news mentions of $e_i$ in $[0, 24[$ in $d_{i-1}$\\
4 & \textit{news titles} & number of title mentions in news of $e_i$ in $[0, t_{p}]$ in $d_i$ \\
5 & \textit{avg content} & average content length of news of $e_i$ in $[0, t_{p}]$ in $d_i$\\
6 & \textit{sources} & number of different news sources of $e_i$ in $[0, t_{p}]$ in $d_i$\\
7 & \textit{weekday} &   day of week\\
8 & \textit{is weekend} &   true if weekend, false otherwise\\
\hline
\multicolumn{3}{l}{\texttt{Textual}} \\
9-18 & \textit{tfidf titles} &   TF-IDF of news titles $[0, t_{p}]$  in $d_i$\\
19-28 & \textit{LDA titles} &   LDA-10 of news titles $[0, t_{p}]$  in $d_i$\\
\hline
\multicolumn{3}{l}{\texttt{Sentiment}} \\
29 & \textit{pos} &   number of positive words in news titles $[0, t_{p}]$ in $d_i$\\
30 & \textit{neg} &   number of negative words in news titles $[0, t_{p}]$ in $d_i$\\
31 & \textit{neu} &   number of neutral words in news titles $[0, t_{p}]$ in $d_i$\\
32 & \textit{ratio} &   $positive/negative$\\
33 & \textit{diff} &   $positive - negative$\\
34 & \textit{subjectivity} & $(positive + negative + neutral)/ \sum words$\\
35-44 & \textit{tfidf subj} & TF-IDF of subjective words (pos, neg and neu)\\
\hline
\multicolumn{3}{l}{\texttt{Semantic}} \\
45 & \textit{entities} &   number of entities in news $[0, t_{p}]$  in $d_i$\\
46 & \textit{tags} &   number of tags in news $[0, t_{p}]$  in $d_i$\\
47-56 & \textit{tfidf entities} &   TF-IDF of entities in news $[0, t_{p}]$  in $d_i$\\
57-66 & \textit{tfidf tags} &   TF-IDF of news tags $[0, t_{p}]$  in $d_i$\\
\hline
\end{tabular}
\label{table:feats}
\end{table}

\subsubsection{Learning Framework}

Let $x$ be the feature vector extracted from the online news stream $N$ on day $d_i$ until $tp$. We want to learn the probability $P(\hat{f_p}=1|X=x)$. This can be done using the inner product between $x$ and a weighting parameter vector $w \in \mathbb{R}$, $\textbf{w}^\top \textbf{x}$. 

Using logistic regression and for binary classification one can unify the definition of $p(\hat{f_p}=1|x)$ and $p(\hat{f_p}=0|x)$ with

$$ p(\hat{f_p}|x) = \frac{1}{1 + e^{-\hat{f_p} w^\top x}}  $$

Given a set of $z$ instance-label pairs ($x_i$,$\hat{f_p}_i$), with $i = 1,...,z$ and $\hat{f_p}_i \in \left\{0,1\right\}$ we solve the binary class L2 penalized logistic regression optimization problem, where $C > 0$  

$$\underset{w}{min\,} \frac{1}{2} w^\top w + C \sum_{i=1}^n \log(1 + e^{- \hat{f_p}_i w^\top x_i})$$

We apply this approach following an entity specific basis, i.e. we train an individual model for each entity. Given a set of entities $E$ to which we want to apply our approach and a training set of example days $D = \{d_1, d_2, ..., d_i, ...\}$, we extract a feature vector $x_i$ for each entity $e_i$ on each training day $d_i$. Therefore, we are able to learn a model of $w$ for each $e_i$. The assumption is that popularity on social media $f_p$ is dependent of the entity $e_i$ and consequently we extract entity specific features from the news stream $N$. For instance, the top 10,000 words of the news titles mentioning $e_i$ are not the same for $e_j$. 

\subsection{Experimental Setup}
\begin{figure}[h]
\centering
    \includegraphics[width=0.95\textwidth]{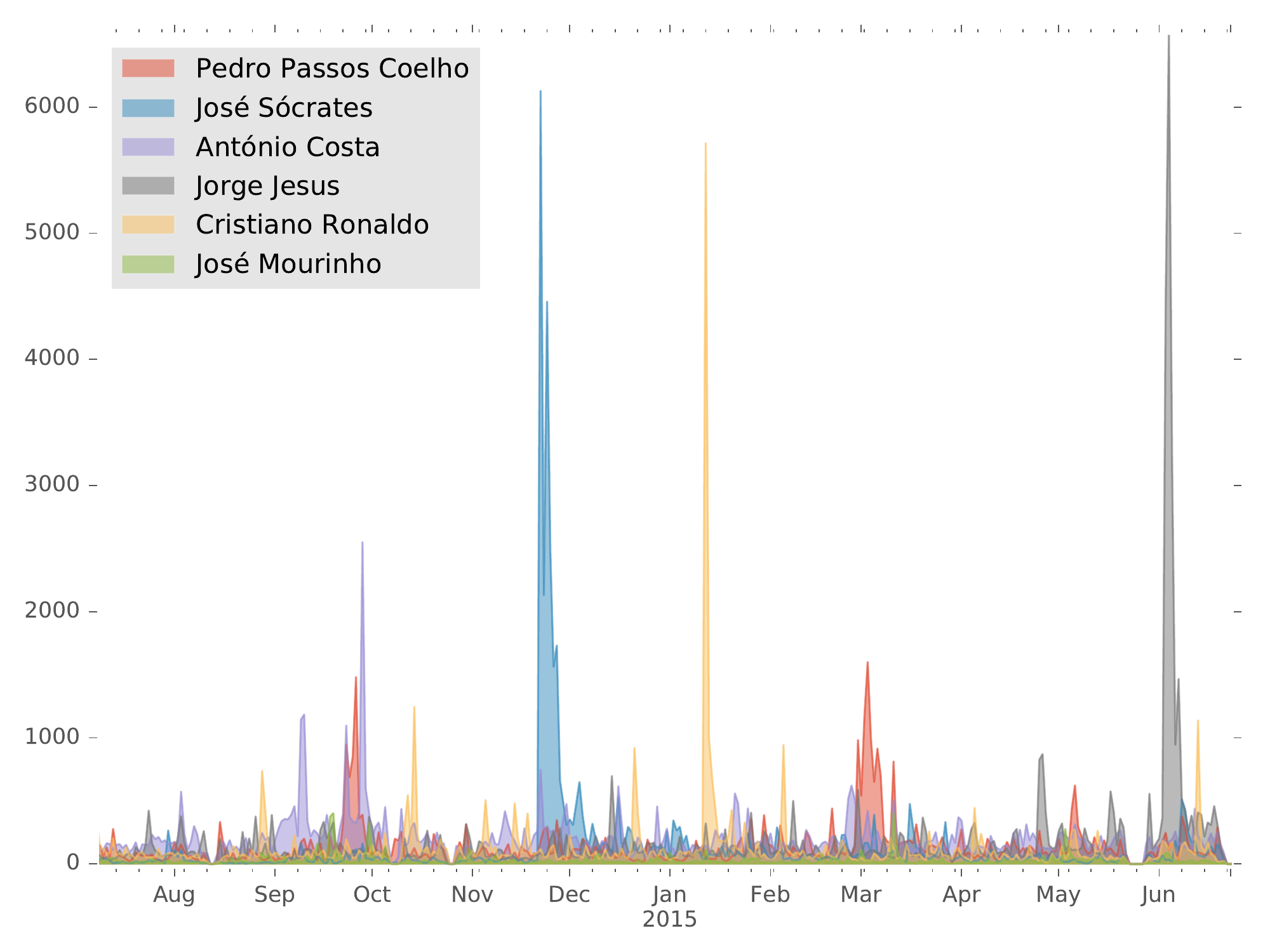}
\caption{Daily popularity on Twitter of entities under study.}
\label{fig:architecture}
\end{figure}

This work uses Portuguese news feeds and tweets collected from January 1, 2013 to January 1, 2016, consisting of over 150 million tweets and 5 million online news articles\footnote{Dataset is available for research purposes. Access requests via e-mail.}. To collect and process raw Twitter data, we use a crawler, which recognizes and disambiguates named entities on Twitter \cite{boanjak2012twitterecho}.
News data is provided by a Portuguese online news aggregator\footnote{http://www.sapo.pt}. This service handles online news from over 60 Portuguese news outlets and it is able to recognize entities mentioned on the news.

We choose the two most common news categories: politics and football and select the 3 entities with highest number of mentions on the news for both categories. The politicians are two former Prime-ministers, Jos{\'e} S{\'o}crates and Pedro Passos Coelho and the incumbent, Ant{\'o}nio Costa. The football entities are two coaches, Jorge Jesus and Jos{\'e} Mourinho, and the most famous Portuguese football player, Cristiano Ronaldo. 

Figure \ref{fig:architecture} depicts the behavior of daily popularity of the six entities on the selected community stream of Twitter users for  each day from July 2014 until July 2015. As expected, it is easily observable that in some days the popularity on Twitter exhibits bursty patterns. For instance, when Jos{\'e} S{\'o}crates was arrested in November 21st 2014 or when Cristiano Ronaldo won the FIFA Ballon d'Or in January 12th 2015.

\begin{figure}[t]
\centering
    \includegraphics[width=0.7\textwidth]{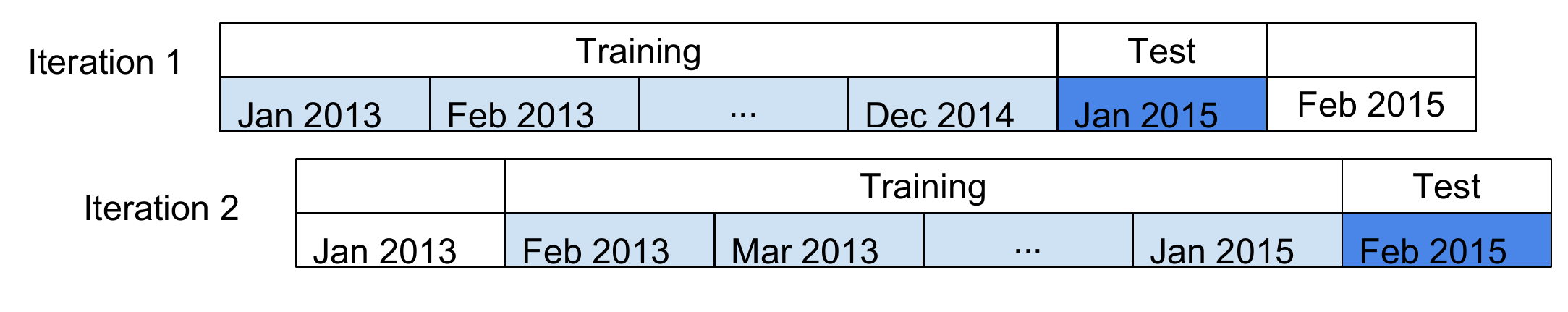}
\caption{Training and testing sliding window - first 2 iterations.}
\label{fig:sliding}
\end{figure}

We defined the years of 2013 and 2014 as training set and the whole year of 2015 as test set. We applied a monthly sliding window setting in which we start by predicting entity popularity for every day of January 2015 (i.e. the test set) using a model trained on the previous 24 months, 730 days (i.e. the training set). Then, we use February 2015 as the test set, using a new model trained on the previous 24 months. Then March and so on, as depicted in Figure \ref{fig:sliding}. We perform this evaluation process, rolling the training and test set until December 2015, resulting in 365 days under evaluation.

The process is applied for each one of the six entities, for different time of predictions $t_p$ and for different values of the decision boundary $k$. We test $tp = {0, 4, 8, 12, 16, 20}$ and $k = {0.5, 0.65, 0.8}$. Therefore, we report results in Section \ref{sec:results} for 18 different experimental settings, for each one of the six entities. The goal is to understand how useful the news cycle is for predicting entity popularity on Twitter for different entities, at different hours of the 24 hours cycle and with different thresholds for considering popularity as \textit{high} or \textit{low}.

\subsection{Results and Discussion}\label{sec:results}
Results are depicted in Table \ref{table:res}. We report F1 on positive class since in online reputation monitoring is more valuable to be able to predict \textit{high} popularity than \textit{low}. Nevertheless, we also calculated overall Accuracy results, which were  better than the F1 reported here. Consequently, this means that our system is fairly capable of predicting \textit{low} popularity.
We organize this section based on the research questions we presented in the beginning of this section.
\begin{table}[t]
\centering
\caption{F1 score of popularity \textit{high} as function of $t_p$ and $k$ equal to 0.5, 0.65 and 0.8 respectively.}
\setlength{\tabcolsep}{10pt}
\begin{tabular}{l c c c c c c }
\\
\hline
 \textbf{Entity  \textbackslash       $t_p$(hour)}&\textbf{0} &\textbf{4} & \textbf{8} & \textbf{12} & \textbf{16}& \textbf{20}\\
\hline
 \multicolumn{7}{l}{\textbf{$k=0.50$}} \\
Ant{\'o}nio Costa	& 0,76	& 0,67	& 0,74	& 0,77	& 0,75	& 0,72 \\
Jos{\'e} S{\'o}crates & 	0,77 & 	0,66 & 	0,73 & 	0,75 & 	0,75 & 	0,75 \\
Pedro Passos Coelho &	0,72 & 	0,63 & 	0,70 & 	0,70	 & 0,74	 & 0,71\\
Cristiano Ronaldo	& 0,35	& 0,41 &	0,45 &	0,37	& 0,35& 0,32 \\
Jorge Jesus & 0,73	 & 0,68	 & 0,69	 & 0,68 & 	0,69 & 	0,70 \\
Jos{\'e} Mourinho & 	0,62 & 	0,46 & 	0,51 & 	0,56	 & 0,55	 & 0,45 \\
\hline
\multicolumn{7}{l}{\textbf{$k=0.65$}} \\
Ant{\'o}nio Costa	 & 0,61 & 	0,60 & 	0,66	 & 0,64	 & 0,60 & 	0,60 \\
Jos{\'e} S{\'o}crates	 & 0,63 & 	0,57	 & 0,62 & 	0,66 & 	0,64	 & 0,62 \\
Pedro Passos Coelho	 & 0,58	 & 0,57 & 	0,65	 & 0,67 & 	0,67 & 	0,65 \\
Cristiano Ronaldo & 	0,29 & 	0,35	 & 0,42 & 	0,41 & 	0,36 & 	0,30 \\
Jorge Jesus	 & 0,63	 & 0,61	 & 0,63 & 	0,59	 & 0,62	 & 0,64 \\
Jos{\'e} Mourinho	 & 0,56	 & 0,39	 & 0,48 & 	0,56	 & 0,47	 & 0,38 \\
\hline
\multicolumn{7}{l}{\textbf{$k=0.80$}} \\
Ant{\'o}nio Costa	 & 0,48	 & 0,51 & 	0,55	 & 0,53	 & 0,44 & 	0,49 \\
Jos{\'e} S{\'o}crates	 & 0,48	 & 0,42	 & 0,47	 & 0,53	 & 0,47 & 	0,35 \\
Pedro Passos Coelho & 	0,47	 & 0,46 & 	0,56	 & 0,56 & 	0,52	 & 0,54 \\
Cristiano Ronaldo & 	0,14 & 	0,29	 & 0,31 & 	0,26 & 	0,20	 & 0,21 \\
Jorge Jesus	 & 0,50	 & 0,48 & 	0,51	 & 0,48 & 	0,57	 & 0,56 \\
Jos{\'e} Mourinho & 	0,32 & 	0,32 & 	0,36	 & 0,41	 & 0,41 & 	0,36 \\
\hline
\end{tabular}
\label{table:res}
\end{table}

\begin{itemize}
\item [] Is online news valuable as source of information to effectively predict entity popularity on Twitter? 

\item [] Do online news carry different predictive power based on the nature of the entity under study?

\end{itemize}

Results show that performance varies with each target entity $e_i$. In general, results are better in the case of predicting popularity of politicians. In the case of football public figures, Jorge Jesus exhibits similar results with the three politicians but Jos{\'e} Mourinho and especially Cristiano Ronaldo represent the worst results in our setting. For instance, when Cristiano Ronaldo scores three goals in a match, the burst on popularity is almost immediate and not possible to predict in advance.  

Further analysis showed that online news failed to be informative of popularity in the case of live events covered by other media, such as TV. Interviews and debates on one hand, and live football games on the other, consist of events with unpredictable effects on popularity. Cristiano Ronaldo can be considered a special case in our experiments. He is by far the most famous entity in our experiments and in addition, he is also an active Twitter user with more than 40M followers. This work focus on assessing the predictive power of online news and its limitations. We assume that for Cristiano Ronaldo, endogenous features from the Twitter itself would be necessary to obtain better results. 

\begin{itemize}
\item [] How do different thresholds for defining high and low popularity affect the effectiveness of our approach?
\end{itemize}
Our system exhibits top performance with $k=0.5$, which corresponds to balanced training sets, with the same number of $high$ and $low$ popularity examples on each training set. 
Political entities exhibit F1 scores above 0.70 with $k=0.5$. On the other hand, as we increase $k$, performance deteriorates. We observe that for $k=0.8$, the system predicts a very high number of false positives. It is very difficult to predict extreme values of popularity on social media before they happen. We plan to tackle this problem in the future by also including features about the target variable in the current and previous hours, i.e., time-series auto-regressive components.

\begin{figure}[t]
\centering
    \includegraphics[width=0.9\textwidth]{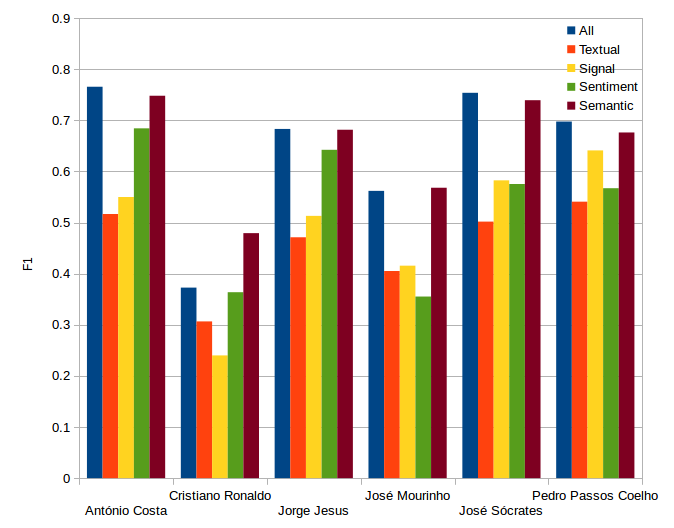}
\caption{Individual feature type F1 score for $t_p=12$ at $k=0.5$.}
\label{fig:feats}
\end{figure}

\begin{itemize}
\item [] Does performance remain stable for different time of predictions?
\end{itemize}
Results show that time of prediction affects the performance of the system, specially for the political entities. In their case, F1 is higher when time of prediction is noon and 4 p.m. which is an evidence that in politics, most of the news events that trigger popularity on social media are broadcast by news outlets in the morning. It is very interesting to compare results for midnight and 4 a.m./8 a.m. The former use the news articles from the previous day, as explained in Section \ref{sec:appr}, while the latter use news articles from the first 4/8 hours of the day under prediction. In some examples, Twitter popularity was triggered by events depicted on the news from the previous day and not from the current day.

\begin{itemize}
\item [] What is the most important feature set for predicting entity popularity on Twitter based on the news cycle?
\item [] Do individual set of features exhibit different importance for different entities? 
\end{itemize}

Figure \ref{fig:feats} tries to answer these two questions. The first observation is that the combination of all groups of features does not lead to substantial improvements. Semantic features alone achieve almost the same F1 score as the combination of all features. However in the case of Mourinho and Ronaldo, the combination of all features lead to worse F1 results than the semantic set alone. 

Sentiment features are the second most important for all entities except Jos{\'e} Mourinho. Signal and Textual features are less important and this was somehow a surprise. Signal features represent the surface behavior of news articles, such as the volume of news mentions of $e_i$ before $t_p$ and we were expecting an higher importance. Regarding Textual features, we believe that news articles often refer to terms and phrases that explain past events in order to contextualize a news article. 

In future work, we consider alternative approaches for predicting future popularity of entities that do not occur everyday on the news, but do have social media public accounts, such as musicians or actors. In opposition, entities that occur often on the news, such as economics ministers and the like, but do not often occur in the social media pose also a different problem.

\section{Predicting Political Polls using Twitter\\ Sentiment\footnote{The material contained in this section was published in P. Saleiro, L. Gomes, C. Soares, ``Sentiment Aggregate Functions for Political Opinion Polling using Microblog Streams'' \citep{saleiro-c3s2e}}}
Surveys and polls using the telephone are widely used to provide information of what people think about parties or political entities \cite{JSS}. Surveys randomly select the electorate sample, avoiding selection bias, and are designed to collect the perception of a population regarding some subject, such as in politics or marketing. However this method is expensive and time consuming \citep{JSS}. Furthermore, over the years it is becoming more difficult to contact people and persuade them to participate in these surveys \citep{Kohut}.

On the other hand, the rise of social media, namely Twitter and Facebook, has changed the way people interact with news. This way, people are able to react and comment any news in real time \citep{Bermingham}. One challenge that several research works have been trying to solve is to understand how opinions expressed on social media, and their sentiment, can be a leading indicator of public opinion. 
However, at the same time there might exist simultaneously positive, negative and neutral opinions regarding the same subject. Thus, we need to obtain a value that reflects the general image of each political target in social media, for a given time period. To that end, we use sentiment aggregate functions. In summary, a sentiment aggregate function calculates a global value based on the number of positive, negative, and neutral mentions of each political target, in a given period. We conducted an exhaustive study and collected and implemented several sentiment aggregate functions from the state of the art \citep{Bermingham, pred12, pred6, pred3, livne2011party, tumasjan2010predicting, Gayo-Avello2012, O'Connor2010, Chung2011}.

Thus, the main objective of our work is to study and define a methodology capable of successfully estimating the poll results, based on opinions expressed on social media, represented by sentiment aggregators. We applied this problem to the Portuguese bailout case study, using Tweets from a sample of the Portuguese Tweetosphere and Portuguese polls as gold standard. Given the monthly periodicity of polls, we needed to aggregate the data by month. This approach allows each aggregate value to represent the monthly sentiment for each political party. Due to the absence of a general sentiment aggregate function suitable for different case studies, we decided to include all aggregate functions as features of the regression model. Therefore the learning algorithm is able to adapt to the most informative aggregate functions through time.

\subsection{Methodology}\label{sec:methodology}

To collect and process raw Twitter data, we use an online reputation monitoring platform \cite{saleiro2015popmine} which can be extended by researchers interested in tracking political opinion on the web. It collects tweets from a predefined sample of users, applies named entity disambiguation \cite{saleiro2013popstar} and generates indicators of both frequency of mention and polarity (positivity/negativity) \cite{filgueiras2013popstar} of mentions of entities over time. In our case, tweets are collected from the stream of 100 thousand different users, representing a sample of the Portuguese community on Twitter. This sample was obtained by expanding a manually annotated seed set of 1000 users using heuristics such as, as language of posts, language of followers posts or geo-location \cite{boanjak2012twitterecho}.

The platform automatically classifies each tweet according to its sentiment polarity. If a message expresses a positive, negative or neutral opinion regarding an entity (e.g. politicians), it is classified as positive, negative or neutral mention, respectively. The sentiment classifier uses a corpus of 1500 annotated tweets as training set and it has achieved an accuracy over 80\% using 10-fold cross validation. These 1500 tweets were manually annotated by 3 political science students. 

Mentions of entities and respective polarity are aggregated by counting positive, negative, neutral and total mentions for each entity in a given period. Sentiment aggregate functions use these cumulative numbers as input to generate a new value for each specific time period. Since we want to use sentiment aggregate functions as features of a regression model to produce an estimate of the political opinion, we decided to use traditional poll results as gold standard.

\subsubsection{Sentiment Aggregate Functions}
Let $M_{e_i}$ be a mention on Twitter of an entity $e_i$, then $M_{e_i}^{+}$, $M_{e_i}^{*}$  and $M_{e_i}^{-}$ are positive, neutral and negative classified mentions of entity $e_i$ on Twitter. Therefore, given a time frame $T$ (e.g. a month), sentiment aggregate functions applied to the aggregated data between polls are the following:
 

\begin{itemize}

\item $entity buzz$: $\sum_T M_{e_i}$, the sum of the number of mentions (buzz) of a given entity in the time frame $T$.

\item $entity positives$: $\sum_T M_{e_i}^{+}$, sum of the positively classified mentions of a given entity in the time frame $T$.

\item $entity neutrals$:  $\sum_T M_{e_i}^{*}$,  the sum of the neutral classified mentions of a given entity in a time frame $T$.

\item $entity negatives$: $\sum_T M_{e_i}^{-}$, the sum of the negatively classified mentions of a given entity in a time frame $T$.

\item $entity subjectivity$: $\frac{\sum_T M_{e_i}^{+} + M_{e_i}^{-}}{\sum_T M_{e_i}}$, the ratio of positive and negative classified mentions of entity $e_i$ over its buzz in a time frame $T$.

\item $entity polarity$:  $ \frac{\sum_T M_{e_i}^{+}}{\sum_T M_{e_i}^{-}}$, the ratio of positive over negative classified mentions in a time frame $T$.

\item $berminghamsovn$: $\frac{\sum_T M_{e_i}^{-}}{\sum_{T} \sum_{E} M_{e_i}^{-}} $, the ratio of the negative classified mentions of entity $e_i$ over the total number of negative mentions of all entities in time frame $T$.

\item[-]$bermingham$ \cite{Bermingham}: $\log_{10} \frac{\sum_T M_{e_i}^{+} +1}{\sum_{T} \sum_{E} M_{e_i}^{-} +1}$

\item[-]$berminghamsovp$ \cite{Bermingham}: $\frac{\sum_T M_{e_i}^{+}}{\sum_{T} \sum_{E} M_{e_i}^{+}} $

\item[-]$connor$ \cite{pred10}: $\frac{\sum_T M_{e_i}^{+}}{\sum_{T}  M_{e_i}^{-}} $

\item[-]$gayo$ \cite{Gayo-Avello2012}: $\frac{\sum_T M_{e_i}^{+} +\sum_{E_{j \neq i}} M_{e_j}^{-}  }{\sum_{T} \sum_{E} M_{e_i}^{+} +M_{e_i}^{-} } $

\item[-]$polarity$:  $\sum_T M_{e_i}^{+} - \sum_T M_{e_i}^{-}$

\item[-]$polarityONeutral$: $\frac{\sum_T M_{e_i}^{+} - \sum_T M_{e_i}^{-}}{\sum_{T} M_{e_i}^{0}} $

\item[-]$polarityOTotal$: $\frac{\sum_T M_{e_i}^{+} - \sum_T M_{e_i}^{-}}{\sum_{T} M_{e_i} } $

\item[-]$subjOTotal$: $\frac{\sum_T M_{e_i}^{+} + \sum_T M_{e_i}^{-}}{\sum_{T} M_{e_i} } $

\item[-]$subjNeuv$: $\frac{\sum_T M_{e_i}^{+} + \sum_T M_{e_i}^{-}}{\sum_{T} M_{e_i}^{0}} $

\item[-]$subjSoV$: $\frac{\sum_T M_{e_i}^{+} + \sum_T M_{e_i}^{-}}{\sum_{T} \sum_{E} M_{e_i}^{+} + M_{e_i}^{-}}$

\item[-]$subjVol$: $\sum_T M_{e_i}^{+} + M_{e_i}^{-}$

\item[-]$share$ \cite{Bermingham}: $\frac{\sum_T M_{e_i}}{\sum_{T} \sum_{E} M_{e_i}} $

\item[-]$shareOfNegDistribution$: $\frac{ \frac{\sum_T M_{e_i}^{-}}{\sum_T M_{e_i}}}{\sum_{E} \frac{\sum_T M_{e_i}^{-}}{\sum_T M_{e_i}}}$, where $n$ is the number of political entities in the poll

\item[-]$normalized\_positive$: $\frac{\sum_T M_{e_i}^{+}}{\sum_T M_{e_i}}$

\item[-]$normalized\_negative$: $\frac{\sum_T M_{e_i}^{-}}{\sum_T M_{e_i}^{-}}$

\item[-]$normalized\_neutral$: $\frac{\sum_T M_{e_i}^{0}}{\sum_T M_{e_i}}$

\item[-]$normalized\_bermingham$: $\log_{10}{\frac{normalized\_positives +1}{normalized\_negatives +1}} $

\item[-]$normalized\_connor$: $\frac{normalized\_positives}{normalized\_negatives}$

\item[-]$normalized\_gayo$: \\ \\
$\frac{normalized\_positives + normalized\_others\_negatives}{normalized\_total\_positives + normalized\_total\_negatives}$

\item[-]$normalized\_polarity$: \\ \\ $normalized\_positives - normalized\_negatives$

\end{itemize}

The sentiment aggregate functions are used as features in the regression models.

\subsection{Data}\label{sec:data}

The data used in this work consists of tweets mentioning Portuguese political party leaders and polls from August 2011 to December 2013. This period corresponds to the Portuguese bailout when several austerity measures were adopted by the incumbent right wing governmental coalition of the PSD and CDS parties. 

\subsubsection{Twitter}
\begin{table}[h]
\centering
\caption{Distribution of positive, negative and neutral mentions per political party}\label{tab:twitter_data_dist}
    \begin{tabular}{ l | r | r | r | r |}
    \cline{2-5}
     &Negative & Positive & Neutral & Total Mentions \\ \cline{2-5} \hline
     \multicolumn{1}{ |c| }{PSD} & 69 723 & 121 & 37 133 &106 977\\ \hline
     \multicolumn{1}{ |c| }{PS} & 28 660 & 225 & 15 326 & 44 211 \\ \hline
     \multicolumn{1}{ |c| }{CDS} & 41 935 & 51 & 17 554 & 59 540 \\ \hline
     \multicolumn{1}{ |c| }{CDU} & 2 445 & 79 & 5 604 & 8 128 \\ \hline
     \multicolumn{1}{ |c| }{BE} & 9 603 & 306 & 4 214 & 14 123 \\ \hline
    \end{tabular}
\end{table} 

\begin{figure}
\centering
    \includegraphics[width=0.9\textwidth]{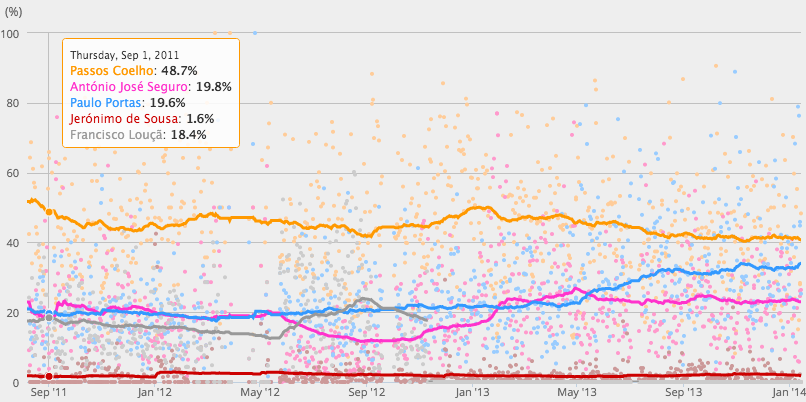}
\caption{Negatives share ($berminghamsovn$) of political leaders in Twitter.}
\label{fig:negshare}
\end{figure}

The Twitter data set contains 232,979 classified messages, collected from a network of 100 thousand different users classified as Portuguese. Table \ref{tab:twitter_data_dist} presents the distribution of positive, negative, and neutral mentions of the political leaders of the 5 most voted political parties in Portugal (PSD, PS, CDS, PCP and BE).
The negative mentions represent the majority of the total mentions, except for CDU where the number of negative mentions is smaller than the neutral ones. The positive mentions represent less than 1\% of the total mentions of each party, except for BE where they represent 2\% of the total mentions. The most mentioned parties are PS, PSD and CDS. The total mentions of these three parties represent 90\% of the data sample total mentions.  
Figure \ref{fig:negshare} depicts the time series of the $berminghamsovn$ (negatives share) sentiment aggregate function. The higher the value of the function the higher is the percentage of negative tweets mention a given political entity in comparison with the other entities. As expected, Pedro Passos Coelho (PSD) as prime-minister is the leader with the higher score throughout the whole time period under study. Paulo Portas (CDS) leader of the other party of the coalition, and also member of the government is the second most negatively mentioned in the period, while António José Seguro (PS) is in some periods the second higher.
PSD and CDS are the incumbent parties while PS is the main opposition party in the time frame under study.  PSD and CDS as government parties were raising taxes and cutting salaries. PS was the incumbent government during the years that led to the bailout and a fraction of the population considered responsible for the financial crisis. The bailout and the consequent austerity measures could explain the overwhelming percentage of negative mentions although we verified that in other time periods the high percentage of negatives mentions remains. We can say that Twitter users of this sample when mentioning political leaders on their tweets tend to criticize them.

\subsubsection{Political Opinion Polls}

\begin{figure}
\centering
    \includegraphics[width=0.8\textwidth]{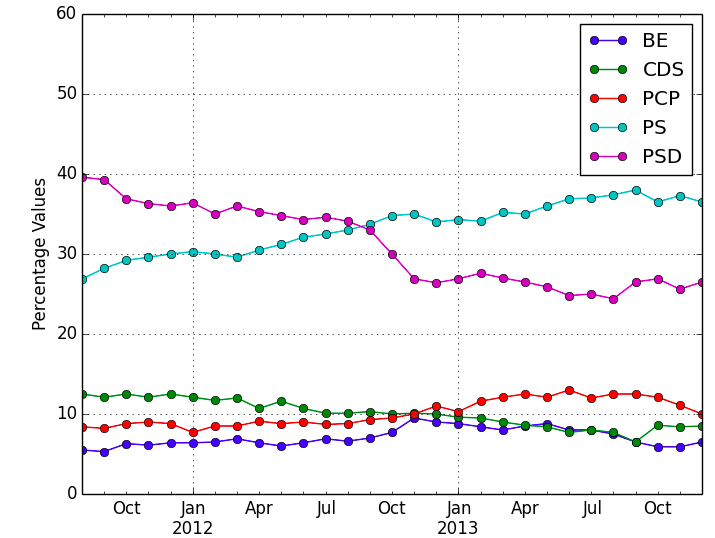}
\caption{Representation of the monthly poll results of each political candidate}
\label{fig:pollEuro}
\end{figure}

The polling was performed by Eurosondagem, a Portuguese private company which collects public opinion. This data set contains the monthly polls results of the five main Portuguese parties, from June 2011 to December 2013. Figure \ref{fig:pollEuro} represents the evolution of Portuguese polls results. We can see two main party groups: The first group, where both PSD and PS are included, has a higher value of vote intention (above 23\%). PSD despite starting as the preferred party in vote intention, has a downtrend along the time, losing the leadership for PS in September 2012. On the other hand, PS has in general an uptrend. The second group, composed by CDS, PCP and BE, has a vote intention range from 5\% to 15\%. While CDS has a downtrend in public opinion, PCP has an ascendant one. Although the constant tendencies (up and down trends), we noticed that the maximum variation observed between two consecutive months is 3\%. 
In June 2013 there was political crises in the government when CDS threaten to leave the government coalition due to the austerity measures being implemented and corresponds to the moment when PS takes the lead in the polls.

\subsection{Experimental Setup}\label{sec:experimental}
{\renewcommand{\arraystretch}{2}

We defined the period of 2011 to December 2012 as training set and the whole year of 2013 as test set. We applied a sliding window setting in which we predict the poll results of a given month using the previous 16 months as training set:

\begin{itemize}
\item Training set – containing the monthly values of the aggregators (both sentiment and buzz aggregator) for 16 months prior the month intended to be predicted. 
\item Test set - containing the values of the aggregators (both sentiment and buzz aggregator) of the month intended to be predicted.
\end{itemize}

We start by predicting the poll results of January 2013 using the previous 16 months as training set: 

\begin{enumerate}
\item We select the values of the aggregators of the 16 months prior January 2013 (September 2011 to December 2012). 
\item We use that data to train our regression model.
\item Then we input the aggregators' values of January 2013 - the first record of the test set - in the the trained model, to obtain the poll results prediction. 
\item We select the next month of the test set and repeat the process until all months are predicted.
\end{enumerate}

The models are created using two regression algorithms: a linear regression algorithm (Ordinary Least Squares - OLS) and a non-linear regression algorithm (Random Forests - RF). 
We also run an experiment using the derivative of the polls time series as gold standard, i.e., poll results variations from poll to poll. Thus, we also calculate the variations of the aggregate functions from month to month as features. 
Furthermore, we repeat each experiment including and excluding the lagged self of the polls, i.e., the last result of the poll for a given candidate ($y_{t-1}$)  or the last polls result variation ($\Delta y_{t-1}$) when predicting polls variations.
We use Mean Absolute Error (MAE) as evaluation measure, to determine the absolute error of each prediction. Then, we calculate the average of the twelve MAE's so we could know the global prediction error of our model. 

\begin{equation}
MAE = \frac{\sum_{i=1}^{n}|f_{i} - y_{i}|}{n}
\label{eq:mae}
\end{equation}
$n$ is the number of forecasts, $f_{i}$ is the model's forecast and $y_{i}$ the real outcome.

\subsection{Results and Discussion}\label{sec:results}
In this section we explain in detail the experiments and their results. We perform two different experiments: (1) using absolute values and (2) using monthly variations.

\subsubsection{Predicting Polls Results}
In this experiment, the sentiment aggregators take absolute values in order to predict the absolute values of polls results. Mathematically speaking, this experiment can be seen as:
$y \leftarrow$ \{$y_{t-1}$, $buzzAggregators$, $sentimentAggregators$\}. In Figure \ref{fig:expabs_mae} we see the global errors we obtained.

\begin{figure}
\centering
    \includegraphics[width=0.8\textwidth]{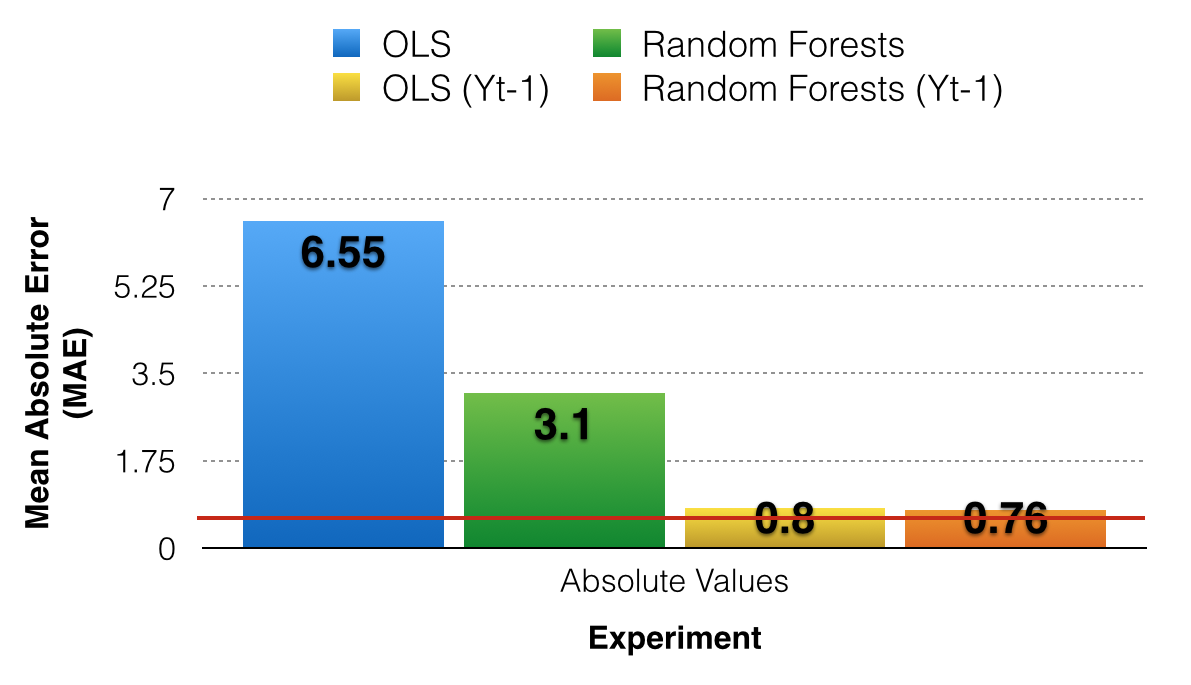}
\caption{Error predictions for polls results.}
\label{fig:expabs_mae}
\end{figure}

The results show that we obtain a MAE for the 5 parties poll results over 12 months of 6.55\% using Ordinary Least Squares and 3.1\% using Random Forests. The lagged self of the polls, i.e., assuming the last known poll result as prediction results in a MAE of 0.61 which was expectable since the polls exhibit slight changes from month to month. 
This experiment shows that the inclusion of the lagged self ($y_{t-1}$) produces average errors similar to the lagged self.

\subsubsection{Predicting Polls Results Variation}
According to our exploratory data analysis, the polls results have a small variation between two consecutive months. Thus, instead of predicting the absolute value of poll results, we tried to predict the variation, $\Delta y \leftarrow $ \{$\Delta (y_{t-1})$, $\Delta buzzAggregators$, $\Delta sentimentAggregators$\}

\begin{figure}
\centering
    \includegraphics[width=0.8\textwidth]{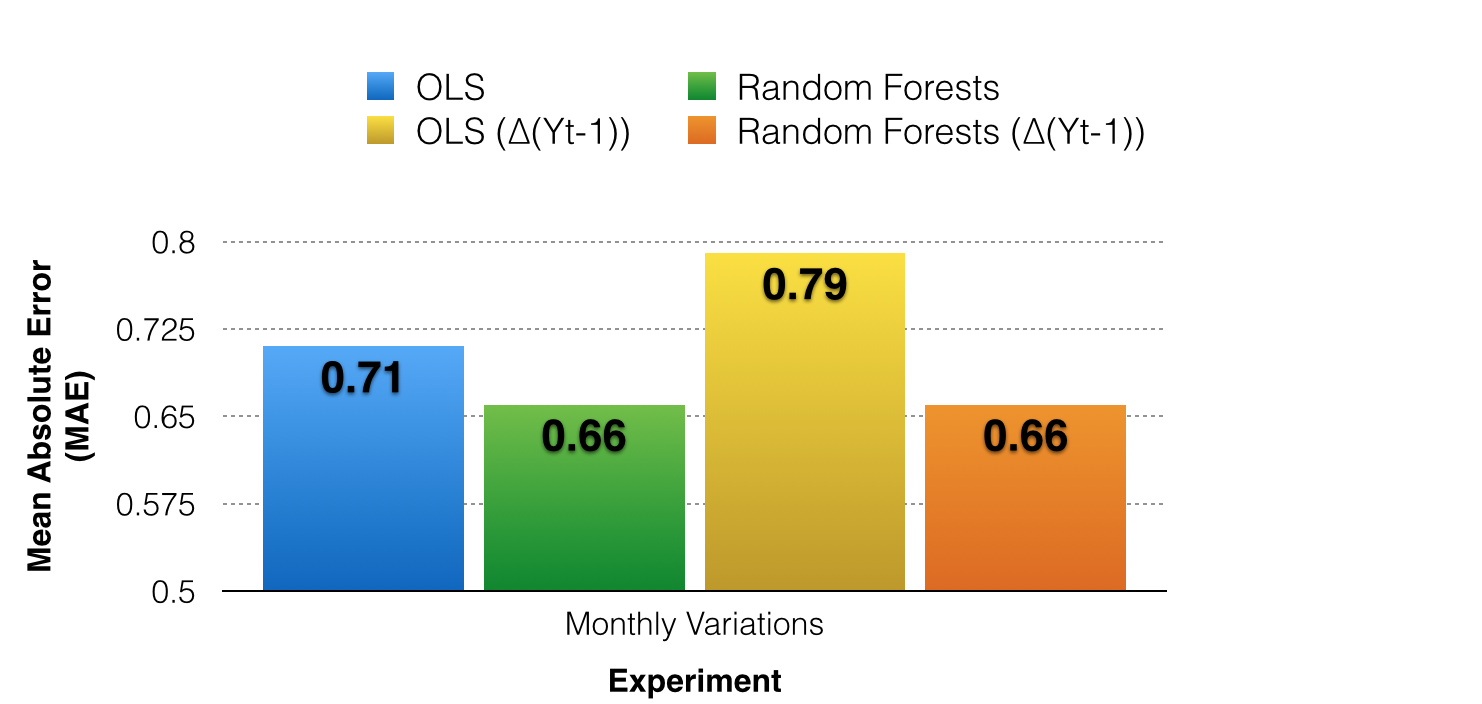}
\caption{Error predictions for polls results variation.}
\label{fig:exp_var_mae}
\end{figure}

In this particular experiment, the inclusion of the $\Delta y_{t-1}$ as feature in the regression model has not a determinant role (Figure \ref{fig:exp_var_mae}). Including that feature we could not obtain lower MAE than excluding it. It means that the real monthly poll variation is not constant over the year. In general, using a non-linear regression algorithm we obtain lower MAE. The results show that when leading with polls results with slight changes from poll to poll it makes sense to transform the dataset by taking differences between consecutive time-steps.

\paragraph{Buzz and Sentiment}

Several studies state that the buzz has predictive power and reflects correctly the public opinion on social media. Following that premise, we trained our models with buzz and sentiment aggregators separately to predict polls variations:
\begin{itemize}
\item $\Delta y \leftarrow$ \{$\Delta (y_{t-1})$, $\Delta buzzAggregators$\}
\item $\Delta y \leftarrow$ \{$\Delta (y_{t-1})$, $\Delta sentimentAggregators$\}
\end{itemize}
This experiment allowed us to compare the behavior of buzz and sentiment aggregators.

\begin{figure}
\centering
    \includegraphics[width=1.1\textwidth]{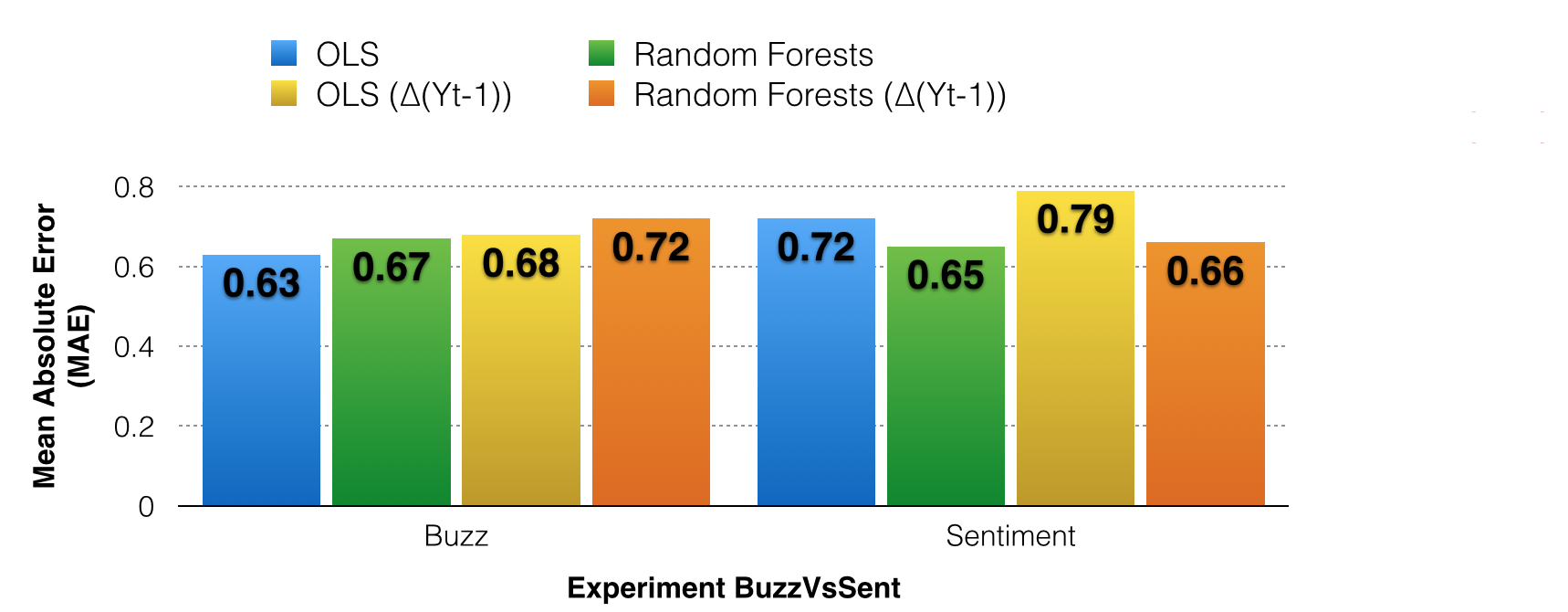}
\caption{Mean absolute error buzz vs sentiment.}
\label{fig:buzzvssent}
\end{figure}

According to Figure \ref{fig:buzzvssent}, buzz and sentiment aggregators have similar results. Although the OLS algorithm combined only with buzz aggregators has a slightly lower error than the other models, it is not a significant improvement. These results also show that Random Forests algorithm performs the best when combined only with sentiment aggregators.

\subsubsection{Feature Selection}
One of the main goals of our work is to understand which aggregator (or group of aggregators) better suits our case study. According to the previous experiments, we can achieve lower prediction errors when training our model with buzz and sentiment aggregators separately. However, when training our model with these two kinds of aggregators separately, we are implicitly performing feature selection. We only have two buzz features ($share$ and $total\_mentions$). Due to that small amount of features, it was not necessary to perform any feature selection technique within buzz features. Thus, we decided to apply a feature selection technique to the sentiment aggregators, in order to select the most informative ones to predict the monthly polls results variation. We use univariate feature selection, selecting 10\% of the sentiment features (total of 3 features). Using this technique, the Random Forests' global error rose from 0.65 to 0.73. However, OLS presents an MAE drop from 0.72 to 0.67. Another important fact to notice is that if we perform univariate feature selection to all aggregators (buzz and sentiment), we will achieve the same MAE value that when applied only to sentiment aggregators. It means that buzz aggregators are discarded by the feature selection technique.

We try a different approach and perform a recursive feature elimination technique. In this technique, features are eliminated recursively according to a initial score given by the external estimator. This method allows us to determine the number of features to select. Thus, also selecting 3 features, the OLS' MAE drops to 0.63. Once again, none of the buzz features were selected. 
Furthermore, both feature selection techniques select different features for each monthly prediction.

\subsection{Feature Importance}
\begin{figure*}[h]
\centering
    \includegraphics[width=\textwidth]{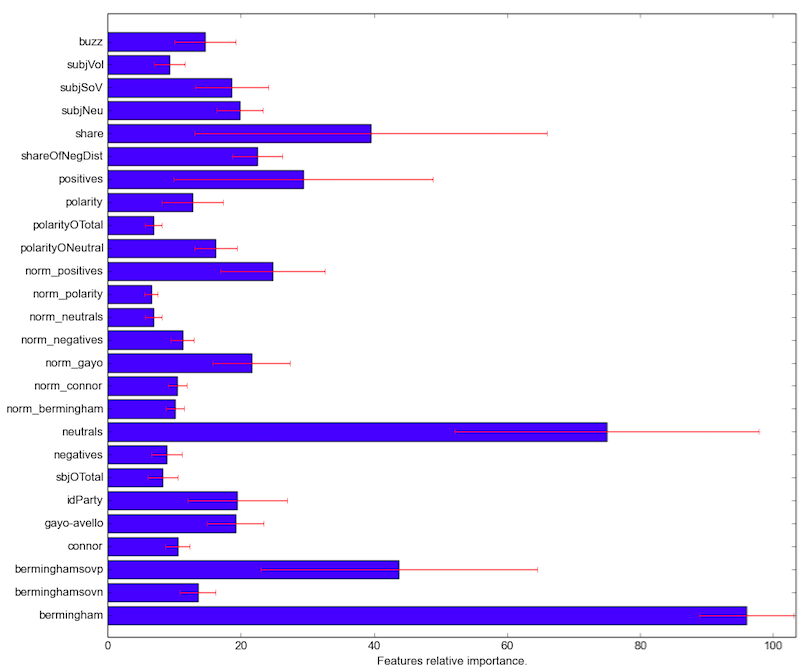}
\caption{Aggregate functions importance in the Random Forests models.}
\label{fig:importances}
\end{figure*}

We select the Random Forest model of monthly variations to study the features importance as depicted in Figure \ref{fig:importances}. 
The higher the score, the more important the feature is. The importance of a feature is computed as the (normalized) total reduction of the criterion brought by that feature. It is also known as the Gini importance. Values correspond to the average of the Gini importance over the different models trained in the experiments. The single most important feature is the $bermingham$ aggregate function, followed by $neutrals$. It is important to notice that when combining all the aggregate functions as features in a single regression model, the $buzz$ does not comprise a high Gini importance, even though when used as a single feature it produces similar results to the sentiment aggregate functions. In general, the standard deviation of the Gini importance is relatively high. This has to do with our experimental setup, as the values depicted in the bar chart correspond to the average of the Gini importance over 12 different models (12 months of testing set). Therefore, feature importances vary over time while the MAE tends to remain unchanged. We can say that different features have different informative value over time and consequently it is useful to combine all the sentiment aggregation functions as features of the regression models over time.

\subsection{Outlook}\label{sec:conclusions}

We studied a large set of sentiment aggregate functions to use as features in a regression model to predict political opinion poll results. The results show that we can estimate the polls results with low prediction error, using sentiment and buzz aggregators based on the opinions expressed on social media. We introduced a strong baseline for comparison, the lagged self of the polls.
In our study, we built a model where we achieve the lowest MAE using the linear algorithm (OLS), combined only with buzz aggregators, using monthly variations. The model has an MAE of 0.63\%.
We performed two feature selection techniques: (1) univariate feature selection and (2) recursive feature elimination. Applying the recursive technique to the sentiment features, we can achieve an MAE of 0.63, matching our best model. Furthermore, the chosen features are not the same in every prediction.
Regarding feature importance analysis our experiments showed that $bermingham$ aggregate function represents the highest Gini importance in the Random Forests model.


\section{Summary of the Contributions}

In this chapter we presented research work about entity-centric text-based prediction for ORM, making the following contributions:

\begin{itemize}
\item Analysis of the predictive power of online news regarding entity popularity on Twitter for entities that are frequently mentioned on the news.

\item Analysis of how to combine different sentiment aggregate functions to serve as features for predicting political polls.
\end{itemize}

\chapter{A Framework for Online Reputation Monitoring}\label{ch:framework}

In this chapter, we present a framework that puts together all the building blocks required to perform ORM. The framework is divided in two distinct components, one is dedicated to Entity Retrieval and the other to Text Mining. In practice these two components can act as two separate frameworks. Both are adaptable and can be reused in different application scenarios, from computational journalism to finance or politics.

We start with a framework overview description and then we focus specifically on each of the two components. The first component is RELink, a research framework for E-R retrieval. We carried the experiments on E-R retrieval, described in Chapter \ref{ch:er2} using RELink. Furthermore, since we did not have access to training data based on news articles, we describe a case study of using RELink for entity retrieval from a large news collection. We then describe the TexRep framework which is responsible for Text Mining related tasks for ORM, such as Entity Filtering, Sentiment Analysis or Predictive tasks. The experiments described in both Chapter \ref{ch:text} and Chapter \ref{ch:predict} were carried out using TexRep. We also provide further detail how TexRep was used as backend of the POPSTAR project. Finally, we perform an independent study of practical aspects of general purpose word embeddings from the Twitter stream to serve as resource for future users of TexRep.

\section{Framework Overview}

The framework provides Entity Retrieval and Text Mining functionalities that enable the collection, disambiguation, retrieval of entities and relationships, sentiment analysis, data aggregation, prediction and visualization of entity-centric information from heterogeneous Web data sources. Furthermore, given that both components are built using modular architectures providing abstraction layers and well defined interfaces, new functionalities or methods can be easily integrated.

The framework is divided in two components: RELink and TexRep. Both can work as independently dedicated frameworks using specific data sources or can be put together in a unifying setup for ORM. As depicted in Figure \ref{fram_high}, when working together, RELink and TexRep are connected through the {\it Entity Occurrences Warehouse}. This is the central module of our framework for ORM. The {\it Entity Occurrences Warehouse} contains extractions from occurrences of the entities of interest across the Web data sources.

\begin{figure}[h]
    \centering
        \centering
        \includegraphics[width= 1.0\textwidth]{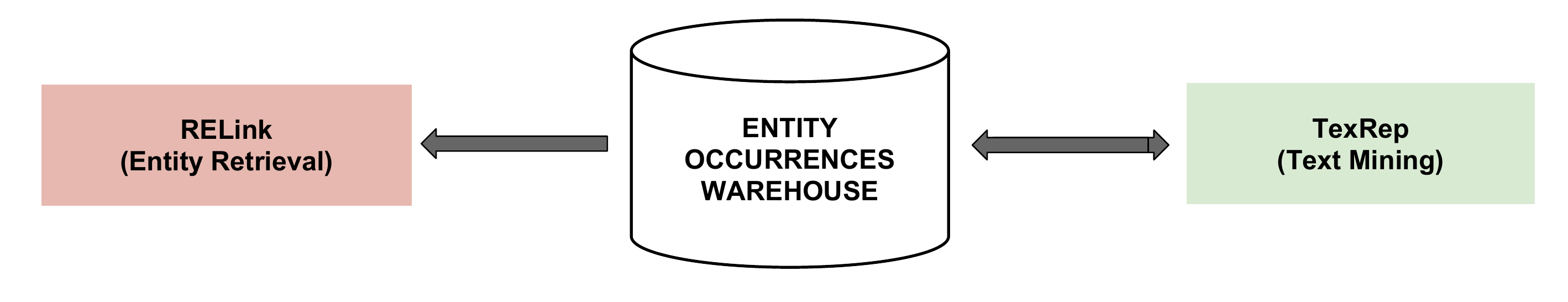}
        \caption{High-level overview on the ORM framework.} \label{fram_high}
\end{figure}

The data flow starts with TexRep collecting data from Web text data sources, extraction of text passages containing entity mentions and disambiguation. Entity-centric text passages are then stored in the {\it Entity Occurrences Warehouse}. This data can then be used for E-R retrieval indexing using RELink or for downstream Text Mining tasks (e.g. Sentiment Analysis) using other modules of TexRep. We now describe RELink and TexRep architectures and internal data flow.

\subsection{RELink}

The RELink framework is designed to facilitate experiments with E-R Retrieval query collections. The formulation of E-R queries in natural language and relational format $(Q^{E_{i-1}}$, $Q^{R_{i-1,i}}$, $Q^{E_i})$ provide opportunities to define and explore a range of query formulations and search algorithms. Although, RELink provides support for Late Fusion design patterns, it is mostly tailored for Early Fusion approaches where it is necessary to create entity and relationship representations at indexing time.

A typical Early Fusion E-R retrieval experimental setup would involve search over a free-text collection to extract relevant instances of entity tuples and then verify their correctness against the relevance judgments.  The key enabling components therefore are: (1) test collections of documents with annotated entity instances that could be extracted during E-R search, (2) an indexing facility, and (3) a retrieval module to process queries and rank results.

\begin{figure}[h]
    \centering
        \centering
        \includegraphics[width= 0.7\textwidth]{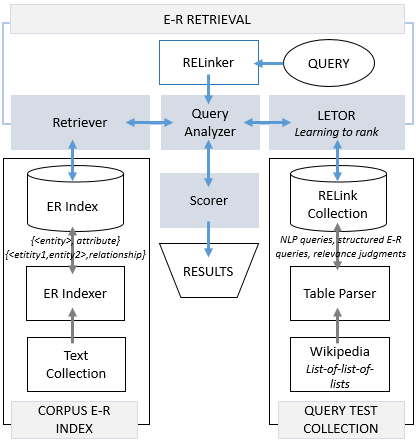}
        \caption{RELink Framework architecture overview.} \label{archrelink}
\end{figure}

Figure \ref{archrelink} depicts the architecture of RELink used in the experiments described in Chapter \ref{ch:er2}. We include the modules responsible for deriving relevance judgments from Wikipedia. The Table Parser module is described in Section \ref{sectables} in Chapter \ref{ch:er2}. Currently, the RELink Framework includes the ClueWeb-09-B\footnote{\url{http://www.lemurproject.org/clueweb09/}} collection combined with FACC1\cite{gabrilovich2013facc1} text span annotations with links to Wikipedia entities (via Freebase). The entity linking precision and recall in FACC1 are estimated at 85\% and 70-85\%, respectively \cite{gabrilovich2013facc1}. The RELink Extractor, part of E-R Indexer, applies an Open Information Extraction method \cite{schmitz2012open} over the annotated ClueWeb-09-B corpus. 
The two additional components are \textit{Corpus E-R Index} and \textit{E-R Retrieval}, both depicted in Figure \ref{archrelink}. The implementation of all modules in \textit{E-R Retrieval} and the Indexer module in \textit{Corpus E-R Index} are based on Apache Lucene and the Letor module serves as a wrapper for RankLib\footnote{\url{http://www.lemurproject.org/ranklib.php}}. 

\subsubsection{Indexing and Retrieval}
Based on the ClueWeb-09-B collection  we create two essential resources: \textit{entity index} and entity pair \textit{relationship index} for the entities that occur in the corpus. For a given entity instance, the ER Indexer identifies co-occuring terms within the same sentence and considers them as entity types for the observed entity instance. Similarly, for a given pair of entities, the ER Indexer verifies whether they occur in the same sentence and extracts the separating string. That string is considered a context term for the entity pair that describes their relationship type. We obtain 476M entity and 418M entity pair extractions with corresponding sentences that are processed by the Indexer. Once the inverted index (ER Index) is created, any instance of an entity or entity pair can be retrieved in response to the contextual terms, i.e., entity types and relationship types, specified by the users. 

\subsubsection{Search Process}
The  E-R retrieval process is managed by the RELinker module (Figure \ref{archrelink}). The Query Analyzer module processes information requests and passes queries in the structured format to the Retriever. Query search is performed in stages to allow for experimentation with different methods and parameter settings. First, the Retriever  provides an initial set of results using Lucene's default search settings and groups them by entity or entity pairs on query time using the Lucene's GroupingSearch. The Scorer then generates and applies feature functions of specific retrieval models with required statistics. Currently, the Scorer has implementations for Early Fusion variants EF-LM, EF-BM25 and ERDM. The RELinker is responsible for re-ranking and providing final results based on the scores provided by the Scorer and the parameter weights learned by Letor.

\subsection{TexRep}

TexRep is a research framework that implements Text Mining techniques to perform Online Reputation Monitoring (ORM) in various application domains, such as computational social sciences, political data science, computational journalism, computational finance or online marketing. 

TexRep was designed with two main challenges in mind: 1) it should be able to cope with the Text Mining problems underlying ORM and 2) it should be flexible, adaptable and reusable in order to support the specificities of different application scenarios. We define that a Text Mining based system for Online Reputation Monitoring must follow a set of technical and operational requirements:

\begin{itemize}

\item \textbf{Batch and real-time operation:} such a system must naturally be able to operate in real-time, i.e. collecting data as it is generated, processing it and updating indicators. However, it is also important to be able to operate in batch mode, in which it collects specific data from a period indicated by the user, if available, and then processes it. The system should use a distributed approach to deal with great volumes of data, (e.g. Hadoop). It should also be able to operate autonomously for long periods of time, measured in months.

\item \textbf{Adaptability:} the system should be able to adapt its models (e.g. polarity classification) through time as well as across different applications. Updating models often requires manually annotated data (e.g. NED). Therefore the system should provide a flexible annotation interface.

\item \textbf{Modularity:} researchers should be able to plug in specific modules, such as a new data source and respective crawler or a different visualization. The system interfaces should use REST APIs and JSON data format, which allow users to add new modules that interact with other data sources (e.g. Wikipedia or Facebook). 

\item \textbf{Reusability:} the system should enable repeatability of all experiments to allow the research community to obtain equal results. We will make the software package of a prototype publicly available as well as the data sources and configuration parameters used in experiments.

\item \textbf{Language independence:} each component of the system should apply a statistical language modeling completely agnostic to the language of the texts.

\end{itemize}

We decompose the use of Text Mining for ORM into four distinct but interconnected tasks: {\it Data Collection}, {\it Entity Filtering}, {\it Sentiment Analysis} and {\it Analytics}. Each task is accomplished by one or more software modules. For instance, {\it Analytics} tasks usually involves the use of the Aggregation, Prediction and Visualization modules. Figure~\ref{fram1} presents the TexRep architecture, including the data flow between modules.

\begin{figure}[t]
    \centering
    \includegraphics[width=1.0\linewidth]{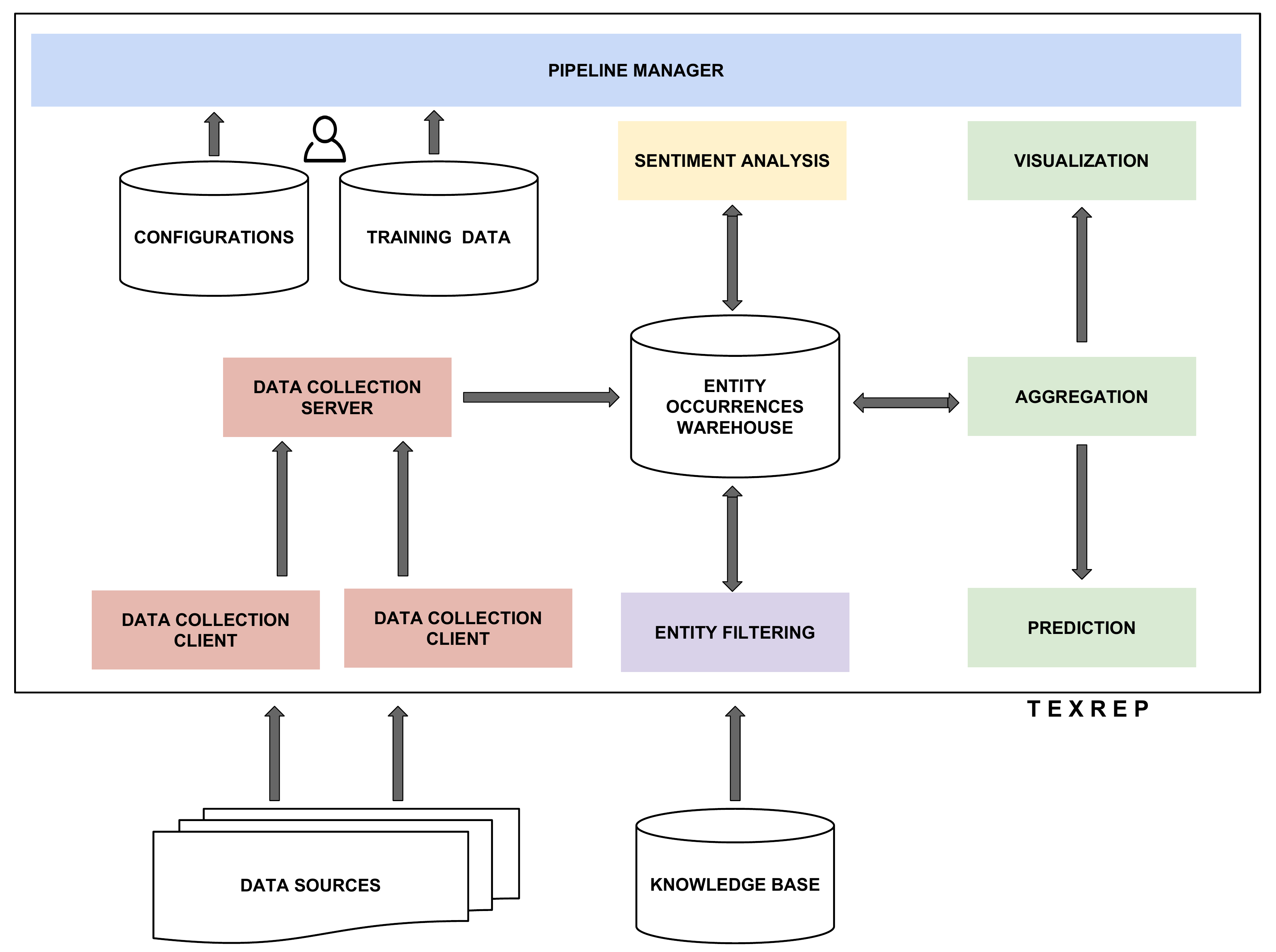}
    \caption{Architecture and data flows of the TexRep framework.}
    \label{fram1}
\end{figure}

Entity Filtering and Sentiment Analysis represent the most challenging Text Mining problems tackled in the TexRep framework. When tracking what is being said online about the target entities it is necessary to disambiguate mentions. When this is done incorrectly, the knowledge obtained by the other modules is negatively affected. Consequently, other Text Mining tasks, such as Sentiment Analysis, will benefit from filtering non relevant texts.

The current implementation of the {\it Entity Filtering} module uses the scikit-learn Python library as the Machine Learning library interface, providing access to TexRep users to the most suitable learning algorithm and parameter tuning for their specific needs. We studied a large set of features that describe the relationship between the target entity representation and a given text and we tried several different supervised learning algorithms that are available through the framework, such as Support Vector Machines (SVM) and Random Forests (RF). 

The {\it Sentiment Analysis} module also uses scikit-learn implementation of supervised learning algorithms in order to predict sentiment polarity and intensity in short texts using regression analysis. We use unsupervised learning of word embeddings~\cite{mikolov2013efficient} in short texts to construct syntactic and semantic representations of words. The {\it Sentiment Analysis} module combines word embeddings with traditional approaches, such as pre-processing techniques, bag-of-words and lexical-based features to train a classifier for sentiment polarity and a regressor for sentiment intensity. 

{\it Analytics} modules include Aggregation, Visualization and Prediction. These modules are application specific and depend on user configurations. For instance, in the political domain it is common to create aggregate functions that represent relative popularity indicators between political parties or candidates. These indicators are then used to predict elections. On the other hand, if we consider the financial domain, due to its high volatility, aggregation is usually performed with lower granularity (minutes instead of days) and target prediction variables are individual stock prices or variations. TexRep implements various aggregation functions and allows custom plug-in of tailored prediction models based on each application.

Therefore, TexRep is able to adapt itself to the specificities of different application scenarios by implementing a modular and flexible design through user configurations and abstraction layers. Data Collection depends on the specified data sources, thus TexRep decouples client-side implementations from the data collection process management using a REST API. If a user needs a different Data Collection Client from the ones provided by default, she is able to implement a specific client that is easily integrated into the framework. The same applies to the {\it Analytics} modules which are extensible by loading user-implemented methods through an abstraction layer. Furthermore, if users wish to extend TexRep with Topic Modeling, they only need to plug-in the new module and write topic assignments through the {\it Entity Occurrences Warehouse}. New aggregation functions could be implemented that use the topic of each mention as input in order to create entity-centric topic trends visualizations. 

The framework can be fully configured using configuration files that are processed in the {\it Pipeline Manager}, which is the module responsible for forwarding specific parameterization to the other modules. It is possible to specify the entities of interest, data sources, aggregate functions and prediction time windows. Module specific configurations are also specified in this module, such as which training data should be used by the modules that rely on machine learning. 

As explained, TexRep addresses the two aforementioned challenges of developing a Text Mining framework for ORM. 
The current version of the framework is implemented in Python, uses MongoDB as NoSQL database and implements the MapReduce paradigm for aggregations. The external and pluggable resources used are the scikit-learn library and the matplotlib for visualization, though users can replace these two resources by others of their preference. We provide the implementations of each module that we believe are the most generic as possible within the context of ORM. Nevertheless, users are also able to extend each module with the methods they see fit, such as, new features or data pre-processing steps. We now describe in detail how the different modules interact with each other, as well as, a detailed explanation of the current implementation of the {\it Entity Filtering}, {\it Sentiment Analysis} and {\it Analytics modules}.

\subsubsection{Data Flow}

TexRep collects data continuously and performs mini-batch processing and analytics tasks. The standard data flow is organized as follows. First the user defines the entities of interest in the configurations files, including canonical and alternative names. These configurations are processed by the Pipeline Manager and forwarded to the Data Collection clients to search for texts (e.g. news articles and tweets) using entity names as queries on each data source-specific API. The Data Collection Clients implement source-specific API clients, such as the case of Twitter and Yahoo Finance, for instance. If the user is interested in collecting RSS feeds of news outlets, then the Data Collection Client can be adapted to subscribe to those feeds and process them accordingly. 

Once collected, texts are stored in the Entity Occurrences Warehouse. Entity Filtering classifies each text as relevant or not for each target entity using a supervised learning approach. A knowledge base (e.g. Freebase) is used to extract target entity representations and to compute similarity features with extracted mentions contexts. Once the non-relevant texts are filtered, Sentiment Analysis takes place. The framework implements both polarity classification and sentiment regression for sentiment intensity detection. Then, Analytics modules are able to aggregate and create visualizations of trends in data or predictions of application specific dependent variables.

\subsubsection{Data Collection}

The Data Collection Server communicates with each Data collection Client using a REST API and therefore it allows modularity and a plugin approach for adapting to specific data sources. The task of data collection is based on user-defined entity configurations containing the list of entities under study.  Each data source has specific web interfaces (e.g. RSS feeds, Yahoo Finance API or Twitter API). The Data Collection Server manages the Data Collection Clients through specific interfaces (plugins) that are adequate for the corresponding source. For instance, collecting data from Twitter poses some challenges, namely due to the limits on the amount of data collected. We opted to create by default a Data Collection Client for SocialBus~\cite{bovsnjak2012twitterecho}, a distributed Twitter client that enables researchers to continuously collect data from particular user communities or topics, while respecting the established limits. 

Some data sources allow query by topics (e.g. entity names) while others do not (e.g. RSS feeds). Moreover, in the case of Twitter, we might be interested in continuously monitoring a fixed group of Twitter users (e.g., the accounts of the entities of interest). In such cases, when we cannot search directly by entity name in the specific data source, we use the list of entity names to process collected texts that might be relevant.
The Data Collection Server applies a sequential classification approach using a prefix tree to detect mentions. This method can be seen as first step of filtering but it is still prone to noisy mentions. For instance, a tweet with the word ``Cameron'' can be relative to several entities, such as a former UK prime minister, a filmmaker or a company. Consequently, this problem is later tackled by the Entity Filtering module.

Collected texts (e.g. news or tweets) are stored in a centralized document-oriented NoSQL database (e.g. MongoDB), the Entities Occurrence Warehouse. This setup provides modularity and flexibility, allowing the possibility of developing specific data collection components tailored to specific data sources and is completely agnostic to the data format retrieved from each data source. The Data Collection Server annotates each text with the target entity which will be used by the Entity Filtering module to validate that annotation.

\section{RELink Use Case}

In this section we present a use case of the RELink framework in the context of ORM applied to computational journalism. Never before has computation been so tightly connected with the practice of journalism. In recent years, the computer science community has researched \cite{demartini2010taer, matthews2010searching, balog2009sahara, alonso2010time, saleiro2015popmine, Teixeira2011, Sarmento2009, Abreu2015} and developed\footnote{NewsExplorer (IBM Watson): http://ibm.co/1OsBO1a} new ways of processing and exploring news archives to help journalists perceiving news content with an enhanced perspective.

We created a demo the TimeMachine, that brings together a set of Natural Language Processing, Text Mining and Information Retrieval technologies to automatically extract and index entity related knowledge from the news articles \cite{saleiro2013popstar, saleiro2013piaf, saleiro2015popmine, teixeira2011bootstrapping, Teixeira2011, Sarmento2009, Abreu2015}. It allows users to issue queries containing keywords and phrases about news stories or events, and retrieves the most relevant entities mentioned in the news articles through time. TimeMachine provides readable and user-friendly insights and a temporal perspective of news stories and mentioned entities. It visually represents relationships among public figures co-mentioned in news articles as a social network graph, using a force atlas algorithm layout \cite{jacomy2014forceatlas2} for the interactive and real-time clustering of entities.

\subsection{News Processing Pipeline}

The news processing pipeline, depicted in Figure \ref{fig:ecirpipeline}, starts with a news cleaning module which performs the boilerplate removal from the raw news files (HTML/XML). Once the news content is processed we apply the NERD module which recognizes entity mentions and disambiguates each mention to an entity using a set of heuristics tailored for news, such as job descriptors (e.g. ``Barack Obama, president of USA'') and linguistic patterns well defined for the journalistic text style. 
We use a bootstrap approach to train the NER system \cite{teixeira2011bootstrapping}. Our method starts by annotating entity names on a dataset of 50,000 news items. This is performed using a simple dictionary-based approach. Using such training set we build a classification model based on Conditional Random Fields (CRF). We then use the inferred classification model to perform additional annotations of the initial seed corpus, which is then used for training a new classification model. This cycle is repeated until the NER model stabilizes.
\begin{figure}[h!]
    \centering
    \includegraphics[width=0.8\columnwidth]{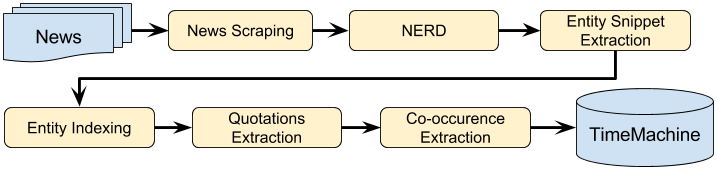}
    \caption{News processing pipeline.}
    \label{fig:ecirpipeline}
\end{figure}
The entity snippet extraction consists of collecting sentences containing mentions to a given entity. All snippets are concatenated generating an entity document, which is then indexed in the entity index. The entity index represents the frequency of co-occurrence of each entity with each term that it occurs with in the news. Therefore, by relying on the redundancy of news terms and phrases associated with an entity we are able to retrieve the most relevant entity to a given input keyword or phrase query. As we also index the snippet datetime it is possible to filter query results based on a time span. For instance, the keyword ``corruption'' might retrieve a different entity list results in different time periods. Quotations are typically short and very informative sentences, which may directly or indirectly quote a given entity. Quotations are automatically extracted (refer to "Quotations Extraction" module) using linguistic patterns, thus enriching the information extracted for each entity.
Finally, once we have all mentioned entities in a given news articles we extract entity tuples representing co-occurrences of entities in a given news article and update the entity graph by incrementing the number of occurrences of a node (entity) and creating/incrementing the number of occurrences of the edge (relation) between any two mentions.

\subsection{Demonstration}
The setup for demonstration uses a news archive of Portuguese news. It comprises two different datasets: a repository from the main Portuguese news agency (1990-2010), and a stream of online articles provided by the main web portal in Portugal (SAPO) which aggregates news articles from 50 online newspapers. The total number of news articles used in this demonstration comprises over 12 million news articles. The system is working on a daily basis, processing articles as they are collected from the news stream.
TimeMachine allows users to explore its news archive through an entity search box or by selecting a specific date. Both options are available on the website homepage and in the top bar on every page. There are a set of ``stories'' recommendations on the homepage suited for first time visitors. The entity search box is designed to be the main entry point to the website as it is connected to the entity retrieval module of TimeMachine. 
\begin{figure}[H]
    \centering
    \includegraphics[width=0.5\textwidth]{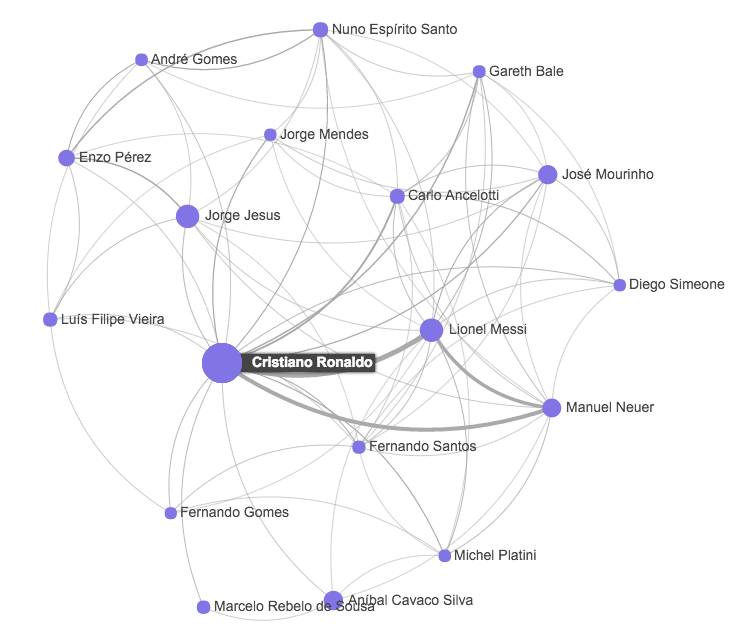}
    \caption{Cristiano Ronaldo egocentric network.}
    \label{fig:architecture}
\end{figure}
Users may search for surface names of entities (e.g. ``Cristiano Ronaldo'') if they know which entities they are interested to explore in the news, although the most powerful queries are the ones containing keywords or phrases describing topics or news stories, such as ``Eurozone crisis'' or  ``Ballon d'Or nominees''. When selecting an entity from the ranked list of results, users access the entity profile page which contains a set of automatically extracted entity specific data: name, profession, a set of news articles, quotations from the entity and related entities. 
An entity timeline is also provided to allow users to navigate entity specific data through time. By selecting a specific period, different news articles, quotations and related entities are retrieved.  Furthermore, users have the option of ``view network'' which consists in a interactive network depicting connections among entities co-mentioned in news articles for the selected time span. An example of such visualization is depicted in Figure \ref{fig:architecture}, 
and it is implemented using the graph drawing library Sigma JS, together with "Force Atlas" algorithm for the clustered layout of entities. Nodes consist of entities and edges represent a co-occurrence of mentioned entities in the same news articles. The size of the nodes and the width of edges is proportional to the number of mentions and co-occurrences, respectively. Different node colors represent specific news topics where entities were mentioned. By selecting a date interval on the homepage, instead of issuing a query, users get a global interactive network of mentions and co-occurrences of the most frequent entities mentioned in the news articles for the selected period of time.


\section{TexRep Use Case}

This section describes the design and implementation of the POPmine system, an use case of the proposed framework, developed in the scope of the POPSTAR project. It is an open source platform which can be used and extended by researchers interested in tracking reputation of political entities on the Web. POPmine operates either in batch or online mode and is able: to collect texts from web-based conventional media (news items in mainstream media sites) and social media (blogs and Twitter); to process those texts, recognizing topics and political entities; to analyze relevant linguistic units; to generate indicators of both frequency of mention and polarity (positivity/negativity) of mentions to political entities across sources, types of sources, and across time. As a proof of concept we present these indicators in a web application tailored for tracking political opinion in Portugal, the POPSTAR website. The system is available as an open source software package that can be used by other researchers from social sciences but also from any other area that is interested in tracking public opinion on the web.

We opted to use data from news articles, tweets and blog posts and each of these data sources requires its specific crawler. News articles and blog posts are collected using RSS feeds which eases the implementations of a specific crawler. Collecting data from Twitter poses some challenges. The need for large amounts of data, coupled with Twitter’s imposed limits demand for a distributed system. We opted to use SocialBus\footnote{\url{http://reaction.fe.up.pt/socialbus/}} which enables researchers to continuously collect data from particular user communities, while respecting Twitter’s imposed limits.

The data collection components crawl data from specific data sources which implement specific web interfaces (e.g. RSS feeds, Twitter API). Each data source must have its own data collection module which in turn connects to the POPmine system using REST services. POPmine stores data collected in a document oriented NoSQL database (MongoDB). This configuration allows modularity and flexibility, allowing the possibility of developing specific data collection components tailored to specific data sources. 

The default setting of data collection modules comprise the following components:

\begin{itemize}

\item \textbf{News:} Data from online news are provided by the service Verbetes e Notícias from Labs Sapo. This service handles online news from over 60 Portuguese news sources and is able to recognize entities mentioned in the news.

\item \textbf{Blogs:} Blog posts are provided by the blogs’ monitoring system from Labs Sapo, which includes all blogs with domain sapo.pt, blogspot.pt (Blogger) and Wordpress (blogs written in Portuguese).

\item \textbf{Twitter:} Tweets are collected using the platform SocialBus, responsible for the compilation of messages from 100.000 Portuguese users of Twitter. Tweets are collected in real time and submitted to a language classification. In our experiments we opted to collect the tweets written in Portuguese.

\end{itemize}

The information extraction component comprises a knowledge base containing metadata about entities, e.g., names or jobs. Using a knowledge base is crucial to filter relevant data mentioning politicians, such as news, tweets and blog posts. In our application scenario, we opted to use Verbetes, a knowledge base which comprises names, alternative names, and professions of Portuguese people mentioned often in news articles. 

The Information Extraction components address two tasks: Named Entity Recognition and Named Entity Disambiguation. We envision an application scenario where we need to track political entities. Usually this type of entities are well known therefore we opted to use a knowledge base to provide metadata about the target entities, namely the most common surface forms of their names. Once we had the list of surface forms to search for we applied a sequential classification approach using a prefix tree to detect mentions. This method is very effective on news articles and blog posts but can result in noisy mentions when applied to Twitter. For instance, a tweet containing the word ``Cameron'' can be related with several entities, such as the former UK prime minister, a filmmaker or a company. Furthermore, tweets are short which results in a reduced context for entity disambiguation. We then apply the Entity Filtering approach of TexRep.

The opinions warehouse contains the messages filtered by the information extraction component and applies polarity classification to those messages using an external resource - the Opinionizer classifier \citep{amirSemEval2014}. One of the requirements of the Opinionizer is to use manually labeled data to train the classifier. We developed an online annotation tool for that effect.

We create opinion and polls indicators using the aggregator which is responsible to apply aggregation functions and smoothing techniques. Once we obtain the aggregated data we make available a set of web services that can be consumed by different applications such as the POPSTAR website or other research experiences, such as polls predictions using social media opinions.

\subsection{Data Aggregation}
\label{indicators}

Buzz is the daily frequency with which political leaders are mentioned by Twitter users, bloggers and online media news. We use two types of indicators. The first type is the relative frequency with which party leaders are mentioned by each medium (Twitter, Blogs and News), on each day. This indicator is expressed, for each leader of each party, as a percentage relative to the total number of mentions to all party leaders. The second indicator is the absolute frequency of mentions, a simple count of citations for each political leader. 

To estimate trends in Buzz, we use the Kalman Filter. We allow users to choose the smoothing degree for each estimated trend. Users can choose between three alternatives: a fairly reactive one, where trend is highly volatile, allowing close monitoring of day-by-day variations; a very smooth one, ideal to capture long term trends; and an intermediate option, displayed by default.

 After identifying the polarity in each of the tweets,  there are several ways to quantify the overall sentiment regarding political leaders. We can, for instance, look at each target independently or in relative terms, compare positive with negative references or simply look at one side of the polarity, or look at daily, weekly or monthly data records.

In this first prototype we opted to present two separate indicators and their evolution across time, using in both cases the day as reference period. The fist indicator is the logarithm of the ratio of positive and negative tweets by political leader (party leaders and the president). In other words, a positive sign means that the political leader under consideration received more positive than negative tweets that day, while a negative result means that he received more negative than positive tweets. In mathematical notation:

\[ log sentiment_i =  log (\frac{positives_i + 1}{negatives_i +1})\]

The second approach is to simply look at the negative tweets (the vast majority of tweets in our base classifier) and calculate their relative frequency for each leader. In this way it is possible to follow each day which party leaders were, in relative terms, more or less subject to tweets with negative polarity. In mathematical notation:

\[ negatives share = \frac{negatives_{i,d}}{\sum_{negatives_{d}}} \]

\begin{figure}[h]
    \centering
    \includegraphics[width=0.7\textwidth]{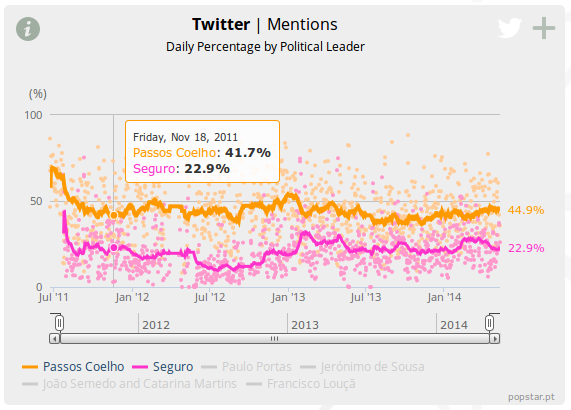}
    \caption{Twitter buzz share of political leaders.}
    \label{fig:site_buzz_share}
\end{figure}

\subsection{Visualization}

We created a website\footnote{\url{http://www.popstar.pt}} to allow interactive visualization of the data collected and processed in real time by the POPmine platform. The site was developed within the scope of the POPSTAR project (Public Opinion and Sentiment Tracking, Analysis, and Research) and presents the following data: a) mentions to Portuguese party leaders in Twitter, in the blogosphere and in online news; b) sentiment conveyed through tweets regarding party leaders, c) voting intentions for the main political parties, measured by traditional polls; and d) evaluation of the performance of said party leaders, measured by polls. An example chart is depicted in Figure \ref{fig:site_buzz_share}. 

Besides providing our indicators in the form of charts, the website also has a dashboard offering a more compact view of trends across indicators for all politicians.


\section{Learning Word Embeddings for ORM\footnote{The material contained in this section was published in P. Saleiro, L. Sarmento, E. M. Rodrigues, C. Soares, E. Oliveira, ``Learning Word Embeddings from the Portuguese Twitter Stream: A Study of some Practical Aspects'' \citep{saleiro-epia}.}}

Word embeddings have great practical importance since they can be used as pre-computed high-density \emph{features} to ML models, significantly reducing the amount of training data required in a variety of Text Mining tasks. We aim to provide general purpose pre-trained word embeddings for the Text Mining tasks in ORM. We are particularly interested in learning word embeddings from the Twitter stream due to the specificities of user generated content. It is relatively easy to get access to word embeddings trained from well formed texts such as Wikipedia or online news. However, to the best of our knowledge there are no publicly available word embeddings learned from the Portuguese Twitter stream.

There are several inter-related challenges with computing and consistently distributing word embeddings concerning the:
\begin{itemize}
\item \textbf{intrinsic properties of the embeddings}. How many dimensions do we actually need to store all the ``useful" semantic information? How big should the embedded vocabulary be to have practical value? How do these two factors interplay?
\item \textbf{type of model} used for generating the embeddings. There are multiple possible models and it is not obvious which one is the ``best", both in general or in the context of a specific type of applications.
\item the size and properties of \textbf{training data}: What is the minimum amount of training data needed? Should we include out of vocabulary words in the training?
\item optimization techniques to be used, \textbf{model hyperparameter} and \emph{training parameters}.
\end{itemize}

Not only the space of possibilities for each of these aspects is large, there are also challenges in performing a consistent large-scale evaluation of the resulting embeddings \cite{levy2015improving}. This makes systematic experimentation of alternative word-embedding configurations extremely difficult. 

In this work, we make progress in trying to find good combinations of some of the previous parameters. We focus specifically in the task of computing word embeddings for processing the Portuguese Twitter stream. User-generated content (such as twitter messages) tends to be populated by words that are specific to the medium, and that are constantly being added by users. These dynamics pose challenges to NLP systems, which have difficulties in dealing with out of vocabulary words. Therefore, learning a semantic representation for those words directly from the user-generated stream - and as the words arise - would allow us to keep up with the dynamics of the medium and reduce the cases for which we have no information about the words.

Starting from our own implementation of a neural word embedding model, which should be seen as a flexible baseline model for further experimentation, our research tries to answer the following practical questions:

\begin{itemize}
\item how large is the vocabulary the one can realistically embed given the level of resources that most organizations can afford to buy and to manage (as opposed to large clusters of GPU's only available to a few organizations)?
\item how much data, as a function of the size of the vocabulary we wish to embed, is enough for training meaningful embeddings?
\item how can we evaluate embeddings in automatic and consistent way so that a reasonably detailed systematic exploration of the previously describe space of possibilities can be performed?
\end{itemize}

By answering these questions based on a reasonably small sample of Twitter data (5M), we hope to find the best way to proceed and train embeddings for Twitter vocabulary using the much larger amount of Twitter data available (300M), but for which parameter experimentation would be unfeasible. This work can thus be seen as a \emph{preparatory study} for a subsequent attempt to produce and distribute a large-scale database of embeddings for processing Portuguese Twitter data.

\subsection{Neural Word Embedding Model}

The neural word embedding model we use is the Continuous Bag-of-words (CBOW) \citep{mikolov2013efficient}. Given a sequence of 5 words - $w_{i-2}$ $w_{i-1}$ $w_i$ $w_{i+1}$ $w_{i+2}$, the task the model tries to perform is that of predicting the middle word, $w_i$, based on the two words on the left - $w_{i-2}$  $w_{i-1}$ - and the two words on the right - $w_{i+1}$ $w_{i+2}$: $P(w_i |w_{i-2}, w_{i-1}, w_{i+1}, w_{i+2})$. This should produce embeddings that closely capture distributional similarity, so that words that belong to the same semantic class, or which are synonyms and antonyms of each other, will be embedded in ``close'' regions of the embedding hyper-space.

The neural model is composed of the following layers:
\begin{itemize}
\item a \textbf{Input Word Embedding Layer}, that maps each of the 4 input words represented by a 1-hot vectors with $|V|$ dimensions (e.g. 32k) into a low dimension space (64 bits). The projections matrix - $W_{input}$ - is shared across the 4 inputs. This is \emph{not} be the embedding matrix that we wish to produce.
\item a \textbf{Merge Layer} that \emph{concatenates} the 4 previous embeddings into a single vector holding all the context information. The concatenation operation ensures that the rest of the model has explicit information about the \textbf{relative position} of the input words. Using an \emph{additive} merge operation instead would preserve information only about the presence of the words, not their sequence.
\item a \textbf{Intermediate Context Embedding Dense Layer} that maps the preceding representation of 4 words into a lower dimension space, still representing the entire context. We have fixed this context representation to 64 dimensions. This ultimately determines the dimension of the resulting embeddings. This intermediate layer is important from the point of view of performance because it isolates the still relatively high-dimensional input space (4 x 64 bits input word embeddings) from the very high-dimensional output space.    
\item a final \textbf{Output Dense Layer} that maps the takes the previous 64-bit representation of the entire input context and produces a vector with the dimensionality of the word output space ($|V|$ dimensions). This matrix - $W_{output}$ - is the one that stores the word embeddings we are interested in.
\item A \textbf{Softmax Activation Layer} to produces the final prediction over the word space, that is the $P(w_i |w_{i-2}, w_{i-1}, w_{i+1}, w_{i+2})$ distribution
\end{itemize}
All neural activations in the model are sigmoid functions. The model was implemented using the Syntagma\footnote{https://github.com/sarmento/syntagma} library which relies on Keras~\cite{chollet2015} for model development, and we train the model using the built-in ADAM~\cite{kingma2014adam} optimizer with the default parameters. 


\subsection{Experimental Setup} \label{expsetup}
We are interested in assessing two aspects of the word embedding process. On one hand, we wish to evaluate the semantic quality of the produced embeddings. On the other, we want to quantify how much computational power and training data are required to train the embedding model as a function of the size of the vocabulary $|V|$ we try to embed. These aspects have fundamental practical importance for deciding how we should attempt to produce the large-scale database of embeddings we will provide in the future. All resources developed in this work are publicly available\footnote{https://github.com/saleiro/embedpt}.

Apart from the size of the vocabulary to be processed ($|V|$), the hyperparamaters of the model that we could potentially explore are i) the dimensionality of the input word embeddings and  ii) the dimensionality of the output word embeddings. As mentioned before, we set both to 64 bits after performing some quick manual experimentation. Full hyperparameter exploration is left for future work. 

Our experimental testbed comprises a desktop with a nvidia TITAN X (Pascal), Intel Core Quad i7 3770K 3.5Ghz, 32 GB DDR3 RAM and a 180GB SSD drive.

\subsubsection{Training Data}
We randomly sampled 5M tweets from a corpus of 300M tweets collected from the Portuguese Twitter community \cite{bovsnjak2012twitterecho}. The 5M comprise a total of 61.4M words (approx. 12 words per tweets in average). From those 5M tweets we generated a database containing 18.9M distinct 5-grams, along with their frequency counts. In this process, all text was down-cased. To help anonymizing the n-gram information, we substituted all the twitter handles by an artificial token ``T\_HANDLE". We also substituted all HTTP links by the token ``LINK". We prepended two special tokens to complete the 5-grams generated from the first two words of the tweet, and we correspondingly appended two other special tokens to complete 5-grams centered around the two last tokens of the tweet.

Tokenization was perform by \emph{trivially} separating tokens by blank space. No linguistic pre-processing, such as for example separating punctuation from words, was made. We opted for not doing any pre-processing for not introducing any linguistic bias from another tool (tokenization of user generated content is not a trivial problem). The most direct consequence of not performing any linguistic pre-processing is that of increasing the vocabulary size and diluting token counts. However, in principle, and given enough data, the embedding model should be able to learn the correct embeddings for both actual words (e.g. ``ronaldo") and the words that have punctuation attached (e.g. ``ronaldo!"). In practice, we believe that this can actually be an advantage for the downstream consumers of the embeddings, since they can also relax the requirements of their own tokenization stage. Overall, the dictionary thus produced contains approximately 1.3M distinct entries. Our dictionary was sorted by frequency, so the words with lowest index correspond to the most common words in the corpus.

We used the information from the 5-gram database to generate all training data used in the experiments. For a fixed size $|V|$ of the target vocabulary to be embedded (e.g. $|V|$ = 2048), we scanned the database to obtain \emph{all} possible 5-grams for which all tokens were among the top $|V|$ words of the dictionary (i.e. the top $|V|$ most frequent words in the corpus).  Depending on $|V|$, different numbers of valid training 5-grams were found in the database: the larger $|V|$ the more valid 5-grams would pass the filter. The number of examples collected for each of the values of $|V|$ is shown in Table \ref{size_training_data}. 
 
\begin{table}[t]
\centering
\caption{Number of 5-grams available for training for different sizes of target vocabulary $|V|$}
\label{size_training_data}
\begin{tabular}{|l|l|}
\hline
$|V|$    & \# 5-grams   \\ \hline
2048   & 2,496,830 \\ \hline
8192   & 6,114,640 \\ \hline
32768  & 10,899,570 \\ \hline
\end{tabular}
\end{table}

Since one of the goals of our experiments is to understand the impact of using different amounts of training data, for each size of vocabulary to be embedded $|V|$ we will run experiments training the models using 25\%, 50\%, 75\% and 100\% of the data available. 

\subsubsection{Metrics Related with the Learning Process}

We tracked metrics related to the learning process itself, as a function of the vocabulary size to be embedded $|V|$ and of the fraction of training data used (25\%, 50\%, 75\% and 100\%). For all possible configurations, we recorded the values of the training and validation loss (cross entropy) after each epoch. Tracking these metrics serves as a minimalistic sanity check: if the model is not able to solve the word prediction task with some degree of success (e.g. if we observe no substantial decay in the losses) then one should not expect the embeddings to capture any of the distributional information they are supposed to capture.

\subsubsection{Tests and Gold-Standard Data for Intrinsic Evaluation}
Using the gold standard data (described below), we performed three types of tests:

\begin{itemize}
\item \textbf{Class Membership Tests}: embeddings corresponding to members of the same semantic class (e.g. ``Months of the Year", ``Portuguese Cities", ``Smileys") should be close, since they are supposed to be found in mostly the same contexts.
\item \textbf{Class Distinction Test}: this is the reciprocal of the previous Class Membership test. Embeddings of elements of different classes should be different, since words of different classes ere expected to be found in significantly different contexts.
\item \textbf{Word Equivalence Test}: embeddings corresponding to \emph{synonyms}, \emph{antonyms}, \emph{abbreviations} (e.g. ``porque" abbreviated by ``pq") and \emph{partial references} (e.g. ``slb and benfica") should be almost equal, since both alternatives are supposed to be used be interchangeable in all contexts (either maintaining or inverting the meaning).
\end{itemize}

Therefore, in our tests, two words are considered: 
\begin{itemize}
\item \emph{distinct} if the cosine of the corresponding embeddings is lower than \textbf{0.70} (or \textbf{0.80}). 
\item to \emph{belong to the same class} if the cosine of their embeddings is higher than \textbf{0.70} (or \textbf{0.80}).
\item equivalent if the cosine of the embeddings is higher that \textbf{0.85} (or \textbf{0.95}).
\end{itemize}
We report results using different thresholds of cosine similarity as we noticed that cosine similarity is skewed to higher values in the embedding space, as observed in related work \cite{dinu2014improving,faruqui2016problems}.
We used the following sources of data for testing Class Membership:
\begin{itemize}
\item AP+Battig data. This data was collected from the evaluation data provided by \cite{rodrigues2016lx}. These correspond to 29 semantic classes.
\item Twitter-Class - collected manually by the authors by checking top most frequent words in the dictionary and then expanding the classes. These include the following 6 sets (number of elements in brackets): smileys (13), months (12), countries (6), names (19), surnames (14) Portuguese cities (9).
\end{itemize}
For the Class Distinction test, we pair each element of each of the gold standard classes, with all the other elements from other classes (removing duplicate pairs since ordering does not matter), and we generate pairs of words which are supposed belong to different classes. For Word Equivalence test, we manually collected equivalente pairs, focusing on abbreviations that are popular in Twitters (e.g. ``qt" $\simeq$ ``quanto" or ``lx" $\simeq$ ``lisboa" and on frequent acronyms  (e.g. ``slb" $\simeq$ ``benfica"). In total, we compiled 48 equivalence pairs.

For all these tests we computed a \emph{coverage} metric. Our embeddings do not necessarily contain information for all the words contained in each of these tests. So, for all tests, we compute a \emph{coverage} metric that measures the fraction of the gold-standard pairs that could actually be tested using the different embeddings produced. Then, for all the test pairs actually covered, we obtain the success metrics for each of the 3 tests by computing the ratio of pairs we were able to correctly classified as i) being distinct (cosine $<$ 0.7 or 0.8), ii) belonging to the same class (cosine $>$ 0.7 or 0.8), and iii) being equivalent (cosine $>$ 0.85 or 0.95).

It is worth making a final comment about the gold standard data. Although we do not expect this gold standard data to be sufficient for a wide-spectrum evaluation of the resulting embeddings, it should be enough for providing us clues regarding areas where the embedding process is capturing enough semantics, and where it is not. These should still provide valuable indications for planning how to produce the much larger database of word embeddings.

\subsection{Results and Analysis}

We run the training process and performed the corresponding evaluation for 12 combinations of size of vocabulary to be embedded, and the volume of training data available that has been used. Table \ref{tab1} presents some overall statistics after training for 40 epochs.

\begin{table}[h]
\centering
\caption{Overall statistics for 12 combinations of models learned varying $|V|$ and volume of training data. Results observed after 40 training epochs.}
\label{tab1}
\fontsize{9}{8}\selectfont
\begin{tabular}{|p{0.2\textwidth}|l|l|l|l|}
\hline
Embeddings                & \# Training Data Tuples & Avg secs/epoch & Training loss & Validation loss \\ \hline
$|V|$ = 2048  & 561,786  (25\% data)     & 4                  & 3.2564        & 3.5932          \\ \hline
$|V|$ = 2048  & 1,123,573 (50\% data)    & 9                  & 3.2234        & 3.4474          \\ \hline
$|V|$ = 2048    & 1,685,359 (75\% data)    & 13                 & 3.2138        & 3.3657          \\ \hline
$|V|$ = 2048   & 2,496,830  (100\% data)   & 18                 & 3.2075        & 3.3074          \\ \hline
$|V|$ = 8192    & 1,375,794 (25\% data)    & 63                 & 3.6329        & 4.286           \\ \hline
$|V|$ = 8192    & 2,751,588 (50\% data)     & 151                & 3.6917        & 4.0664          \\ \hline
$|V|$ = 8192    & 4,127,382 (75\% data)     & 187                & 3.7019        & 3.9323          \\ \hline
$|V|$ = 8192  & 6,114,640 (100\% data)    & 276                & 3.7072        & 3.8565          \\ \hline
$|V|$ = 32768   & 2,452,402 (25\% data)    & 388                & 3.7417        & 5.2768          \\ \hline
$|V|$ = 32768   & 4,904,806 (50\% data)    & 956                & 3.9885        & 4.8409          \\ \hline
$|V|$ = 32768   & 7,357,209 (75\% data)    & 1418               & 4.0649        & 4.6             \\ \hline
$|V|$ = 32768  & 10,899,570 (100\% data)    & 2028               & 4.107         & 4.4491          \\ \hline
\end{tabular}
\end{table}
\begin{figure}[h]
  \begin{minipage}[b]{0.50\textwidth}
    \includegraphics[width=\textwidth]{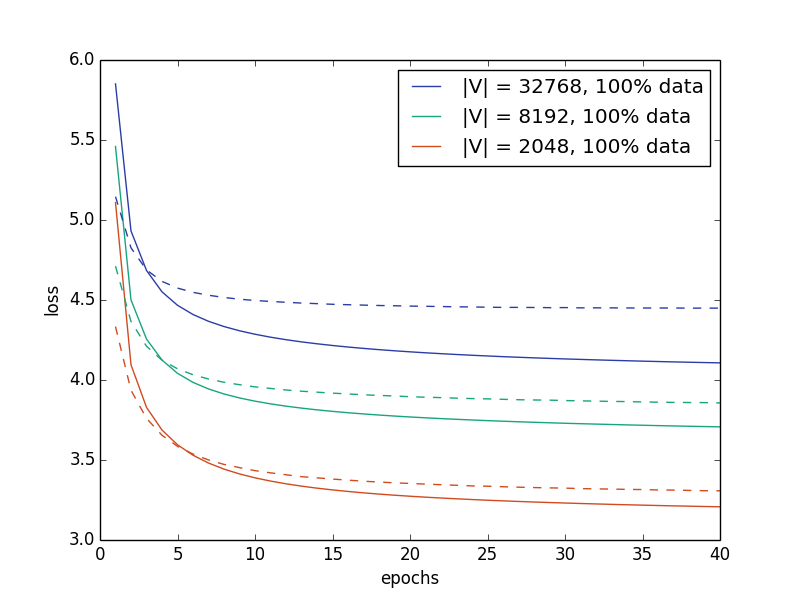}
  \end{minipage}
  \hfill
  \begin{minipage}[b]{0.50\textwidth}
    \includegraphics[width=\textwidth]{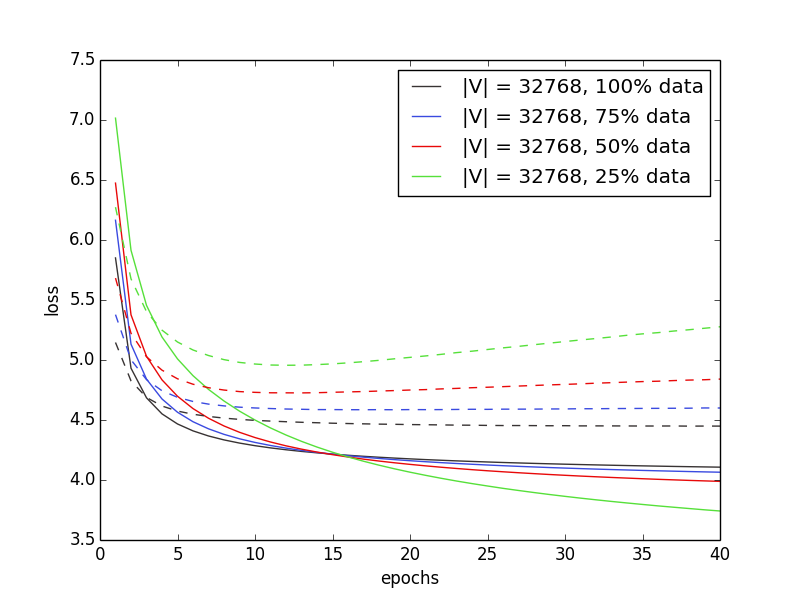}
  \end{minipage}
  \caption{Continuous line represents loss in the training data while dashed line represents loss in the validation data. Left side: effect of increasing $|V|$ using 100\% of training data. Right side: effect of varying the amount of training data used with $|V|$ = 32768.}
\label{fig1}
\end{figure} 
 The average time per epoch increases first with the size of the vocabulary to embed $|V|$ (because the model will have more parameters), and then, for each $|V|$, with the volume of training data. Using our testbed (Section \ref{expsetup}), the total time of learning in our experiments varied from a minimum of 160 seconds, with $|V|$ = 2048 and 25\% of data, to a maximum of 22.5 hours, with $|V|$ = 32768 and using 100\% of the training data available (extracted from 5M tweets). These numbers give us an approximate figure of how time consuming it would be to train embeddings from the complete Twitter corpus we have, consisting of 300M tweets.

We now analyze the learning process itself.  We plot the training set loss and validation set loss for the different values of $|V|$ (Figure \ref{fig1} left) with 40 epochs and using all the available data. As expected, the loss is reducing after each epoch, with validation loss, although being slightly higher, following the same trend. When using 100\% we see no model overfitting. We can also observe that the higher is $|V|$ the higher are the absolute values of the loss sets. This is not surprising because as the number of words to predict becomes higher the problem will tend to become harder. Also, because we keep the dimensionality of the embedding space constant (64 dimensions), it becomes increasingly hard to represent and differentiate larger vocabularies in the same hyper-volume. We believe this is a specially valuable indication for future experiments and for deciding the dimensionality of the final embeddings to distribute.

On the right side of Figure \ref{fig1} we show how the number of training (and validation) examples affects the loss. For a fixed $|V|$ = 32768 we varied the amount of data used for training from 25\% to 100\%. Three trends are apparent. As we train with more data, we obtain better validation losses. This was expected. The second trend is that by using less than 50\% of the data available the model tends to overfit the data, as indicated by the consistent increase in the validation loss after about 15 epochs (check dashed lines in right side of Figure \ref{fig1}). This suggests that for the future we should not try any drastic reduction of the training data to save training time. Finally, when not overfitting, the validation loss seems to stabilize after around 20 epochs. We observed no phase-transition effects (the model seems simple enough for not showing that type of behavior). This indicates we have a practical way of safely deciding when to stop training the model.

\subsubsection{Intrinsic Evaluation}
Table \ref{tabres} presents results for the three different tests described in Section \ref{expsetup}. The first (expected) result is that the coverage metrics increase with the size of the vocabulary being embedded, i.e., $|V|$. Because the Word Equivalence test set was specifically created for evaluating Twitter-based embedding, when embedding $|V|$ = 32768 words we achieve almost 90\% test coverage. On the other hand, for the Class Distinction test set - which was created by taking the cross product of the test cases of each class in Class Membership test set - we obtain very low coverage figures. This indicates that it is not always possible to re-use previously compiled gold-standard data, and that it will be important to compile gold-standard data directly from Twitter content if we want to perform a more precise evaluation. 

The effect of varying the cosine similarity decision threshold from 0.70 to 0.80 for Class Membership test shows that the percentage of test cases that are classified as correct drops significantly. However, the drop is more accentuated when training with only a portion of the available data. The differences of using two alternative thresholds values is even higher in the Word Equivalence test. 

The Word Equivalence test, in which we consider two words equivalent word if the cosine of the embedding vectors is higher than 0.95, revealed to be an extremely demanding test. Nevertheless, for $|V|$ = 32768 the results are far superior, and for a much larger coverage, than for lower $|V|$. The same happens with the Class Membership test. 

On the other hand, the Class Distinction test shows a different trend for larger values of $|V|$ = 32768 but the coverage for other values of $|V|$ is so low that it would not make sense to hypothesize about the reduced values of True Negatives (TN) percentage obtained for the largest $|V|$. It would be necessary to confirm this behavior with even larger values of $|V|$. One might hypothesize that the ability to distinguish between classes requires larger thresholds when $|V|$ is large. Also, we can speculate about the need of increasing the number of dimensions to be able to encapsulate different semantic information for so many words.

\begin{table}[h]
\caption{Evaluation of resulting embeddings using Class Membership, Class Distinction and Word Equivalence tests for different thresholds of cosine similarity.}
\label{tabres}
\centering
\fontsize{9}{8}\selectfont
\begin{tabular}{|p{2cm}|p{0.5cm}|p{0.5cm}|p{0.5cm}|p{0.5cm}|p{0.3cm}|p{0.3cm}|p{0.5cm}|p{0.5cm}|p{0.5cm}|}
\hline
\multicolumn{1}{|c|}{Embeddings}      & \multicolumn{3}{c|}{Class Membership}                                                                                                                                                    & \multicolumn{3}{c|}{Class Distinction}                                                                                                                                            & \multicolumn{3}{c|}{Word Equivalence}                                                                                                                                                  \\ \hline
\multicolumn{1}{|c|}{$|V|$, \%data} & \multicolumn{1}{c|}{\small coverage} & \multicolumn{1}{c|}{\begin{tabular}[c]{@{}c@{}}Acc.\\ @0.70\end{tabular}} & \multicolumn{1}{c|}{\begin{tabular}[c]{@{}c@{}}Acc.\\  @0.80\end{tabular}} & \multicolumn{1}{c|}{\small coverage} & \multicolumn{1}{c|}{\begin{tabular}[c]{@{}c@{}}TN\\ @0.70\end{tabular}} & \multicolumn{1}{c|}{\begin{tabular}[c]{@{}c@{}}TN\\ @0.80\end{tabular}} & \multicolumn{1}{c|}{\small coverage} & \multicolumn{1}{c|}{\begin{tabular}[c]{@{}c@{}}Acc.\\ @0.85\end{tabular}} & \multicolumn{1}{c|}{\begin{tabular}[c]{@{}c@{}}Acc.\\ @0.95\end{tabular}} \\ \hline
2048, 25\%                            & \multirow{4}{*}{12.32\%}      & 30.71\%                                                                    & 4.94\%                                                                      & \multirow{4}{*}{1.20\%}       & 100\%                                                                   & 100\%                                                                   & \multirow{4}{*}{31.25\%}      & 26.67\%                                                                    & 2.94\%                                                                     \\ \cline{1-1} \cline{3-4} \cline{6-7} \cline{9-10} 
2048, 50\%                            &                               & 29.13\%                                                                    & 12.69\%                                                                     &                               & 100\%                                                                   & 100\%                                                                   &                               & 26.67\%                                                                    & 2.94\%                                                                     \\ \cline{1-1} \cline{3-4} \cline{6-7} \cline{9-10} 
2048, 75\%                            &                               & 29.13\%                                                                    & 18.12\%                                                                     &                               & 100\%                                                                   & 100\%                                                                   &                               & 33.33\%                                                                    & 2.94\%                                                                     \\ \cline{1-1} \cline{3-4} \cline{6-7} \cline{9-10} 
2048, 100\%                           &                               & 32.28\%                                                                    & 26.77\%                                                                     &                               & 100\%                                                                   & 100\%                                                                   &                               & 33.33\%                                                                    & 6.67\%                                                                     \\ \hline
8192, 25\%                            & \multirow{4}{*}{29.60\%}      & 14.17\%                                                                    & 4.94\%                                                                      & \multirow{4}{*}{6.54\%}       & 100\%                                                                   & 100\%                                                                   & \multirow{4}{*}{70.83\%}      & 14.71\%                                                                    & 2.94\%                                                                     \\ \cline{1-1} \cline{3-4} \cline{6-7} \cline{9-10} 
8192, 50\%                            &                               & 22.41\%                                                                    & 12.69\%                                                                     &                               & 99\%                                                                    & 100\%                                                                   &                               & 20.59\%                                                                    & 2.94\%                                                                     \\ \cline{1-1} \cline{3-4} \cline{6-7} \cline{9-10} 
8192, 75\%                            &                               & 27.51\%                                                                    & 18.12\%                                                                     &                               & 99\%                                                                    & 100\%                                                                   &                               & 20.59\%                                                                    & 2.94\%                                                                     \\ \cline{1-1} \cline{3-4} \cline{6-7} \cline{9-10} 
8192, 100\%                           &                               & 33.77\%                                                                    & 21.91\%                                                                     &                               & 97\%                                                                    & 100\%                                                                   &                               & 29.41\%                                                                    & 5.88\%                                                                     \\ \hline
32768, 25\%                           & \multirow{4}{*}{47.79\%}      & 17.73\%                                                                    & 5.13\%                                                                      & \multirow{4}{*}{18.31\%}      & 98\%                                                                    & 100\%                                                                   & \multirow{4}{*}{89.58\%}      & 16.28\%                                                                    & 2.33\%                                                                     \\ \cline{1-1} \cline{3-4} \cline{6-7} \cline{9-10} 
32768, 50\%                           &                               & 52.30\%                                                                    & 21.06\%                                                                     &                               & 83\%                                                                    & 98\%                                                                    &                               & 34.88\%                                                                    & 9.30\%                                                                     \\ \cline{1-1} \cline{3-4} \cline{6-7} \cline{9-10} 
32768, 75\%                           &                               & 85.15\%                                                                    & 49.41\%                                                                     &                               & 44\%                                                                    & 88\%                                                                    &                               & 58.14\%                                                                    & 23.26\%                                                                    \\ \cline{1-1} \cline{3-4} \cline{6-7} \cline{9-10} 
32768, 100\%                          &                               & 95.59\%                                                                    & 74.80\%                                                                     &                               & 13\%                                                                    & 57\%                                                                    &                               & 72.09\%                                                                    & 34.88\%                                                                  \\ \hline
\end{tabular}

\end{table}

\subsubsection{Further Analysis regarding Evaluation Metrics}

Despite already providing interesting practical clues for our goal of trying to embed a larger vocabulary using more of the training data we have available, these results also revealed that the intrinsic evaluation metrics we are using are overly sensitive to their corresponding cosine similarity thresholds. This sensitivity poses serious challenges for further systematic exploration of word embedding architectures and their corresponding hyper-parameters, which was also observed in other recent works \cite{faruqui2016problems}. 

By using these absolute thresholds as criteria for deciding the similarity of words, we create a dependency between the evaluation metrics and the \emph{geometry} of the embedded data. If we see the embedding data as a graph, this means that metrics will change if we apply scaling operations to certain parts of the graph, even if its structure (i.e. relative position of the embedded words) does not change. 

For most practical purposes (including training downstream ML models) absolute distances have little meaning. What is fundamental is that the resulting embeddings are able to capture topological information: similar words should be \emph{closer to each other} than they are to words that are dissimilar to them (under the various criteria of similarity we care about), independently of the absolute distances involved.

It is now clear that a key aspect for future work will be developing additional performance metrics based on topological properties. We are in line with recent work \cite{gladkova2016intrinsic}, proposing to shift evaluation from absolute values to more exploratory evaluations focusing on weaknesses and strengths of the embeddings and not so much in generic scores. For example, one metric could consist in checking whether for any given word, all words that are known to belong to the same class are closer than any words belonging to different classes, independently of the actual cosine. Future work will necessarily include developing this type of metrics.

\subsection{Concluding Remarks}

Producing word embeddings from tweets is challenging due to the specificities of the vocabulary in the medium. We implemented a neural word embedding model that embeds words based on n-gram information extracted from a sample of the Portuguese Twitter stream, and which can be seen as a flexible baseline for further experiments in the field. Work reported in this paper is a preliminary study of trying to find parameters for training word embeddings from Twitter and adequate evaluation tests and gold-standard data.

Results show that using less than 50\% of the available training examples for each vocabulary size might result in overfitting. The resulting embeddings obtain reasonable performance on intrinsic evaluation tests when trained a vocabulary containing the 32768 most frequent words in a Twitter sample of relatively small size. Nevertheless, results exhibit a skewness in the cosine similarity scores that should be further explored in future work. More specifically, the Class Distinction test set revealed to be challenging and opens the door to evaluation of not only similarity between words but also dissimilarities between words of different semantic classes without using absolute score values. 

Therefore, a key area of future exploration has to do with better evaluation resources and metrics. We made some initial effort in this front. However, we believe that developing new intrinsic tests, agnostic to absolute values of metrics and concerned with topological aspects of the embedding space, and expanding gold-standard data with cases tailored for user-generated content, is of fundamental importance for the progress of this line of work.

Furthermore, we plan to make public available word embeddings trained from a large sample of 300M tweets collected from the Portuguese Twitter stream. This will require experimenting with and producing embeddings with higher dimensionality (to avoid the cosine skewness effect) and training with even larger vocabularies. Also, there is room for experimenting with some of the hyper-parameters of the model itself (e.g. activation functions, dimensions of the layers), which we know have impact on final results.

\section{Summary of the Contributions}

The work reported in this chapter makes the following contributions:

\begin{itemize}
\item A framework that supports research in Entity Retrieval and Text Mining tasks in the context of Online Reputation Monitoring. This framework is composed by two major components that can act as independent frameworks: RELink and TexRep.

\item The RELink framework that supports comprehensive research work in E-R retrieval, supporting the semi-automatic creating of test queries, as well as, Early Fusion based approaches for E-R retrieval.

\item The TexRep framework that is able to collect texts from online media, such as Twitter or online news, and identify entities of interest, classify sentiment polarity and intensity. The framework supports multiple data aggregation methods, as well as visualization and modeling techniques that can be used for both descriptive analytics, such as analyze how political polls evolve over time, and predictive analytics, such as predict elections.

\item A study of some practical aspects, namely vocabulary size, training data size and intrinsic evaluation for the training and publishing word embeddings from the Portuguese Twitter stream that can be later used for ORM related tasks.
\end{itemize}

\chapter{Conclusions}\label{ch:conclusions}

In this thesis we have addressed two computational problems in Online Reputation Monitoring: Entity Retrieval and Text Mining. Entities are the gravitational force that drives the ORM process and consequently the work reported in this thesis gravitates around entities and their occurrences across the Web. We researched and developed methods for text-based extraction, entity-relationship retrieval, analysis and prediction of entity-centric information spread across the Web. 

The main objectives of this thesis were achieved resulting in several contributions to the problem of Online Reputation Monitoring. Several competitive baselines were developed which we believe represent significant progress in a research area where open source work is scarce. However, there are still many issues to be addressed in the future. Recent developments in Deep Neural Networks create opportunities to improve performance in several tasks we addressed in this thesis. Once we have access to larger quantities of training data it will be possible to easily adapt our research framework to include these techniques. 

\section{Summary and Main Contributions}

\subsubsection{Entity-Relationship Retrieval}
We have established that ORM benefits from entity retrieval capabilities and should not be constrained to classic data analytics reports. Users ought to be able to search for entity-centric information from Social Media and online news. Furthermore, reputation is not an isolated asset and depends also of the reputation of ``neighboring'' entities. We studied the problem of Entity-Relationship Retrieval using a IR-centric perspective and we made several contributions to this line of research:

\begin{itemize}
\item Generalization of the problem of entity-relationship search to cover entity types and relationships represented by any attribute and predicate, respectively, rather than a predefined set.
\item A general probabilistic model for E-R retrieval using Bayesian Networks.
\item Proposal of two design patterns that support retrieval approaches using the E-R model.
\item Proposal of a Entity-Relationship Dependence model that builds on the basic Sequential Dependence Model (SDM) to provide extensible entity-relationship representations and dependencies, suitable for complex, multi-relations queries. 
\item Proposal of an indexing method that supports a retrieval approach to the above problem.
\item A semi-automatic method for generating E-R test collections, which resulted in the RELink Query Collection comprising 600 E-R queries.
\item Results of experiments at scale, with a comprehensive set of queries and corpora.
\end{itemize}

Entity-Relationship (E-R) Retrieval is a complex case of Entity Retrieval where the goal is to search for multiple unknown entities and relationships connecting them. Contrary to entity retrieval from structured knowledge graphs, IR-centric approaches to E-R retrieval are more adequate in the context of ORM. This happens due to the dynamic nature of the data sources which are much more transient than other more stable sources of information (e.g Wikipedia) used in general Entity Retrieval. Consequently, we developed E-R retrieval methods that do not rely on fixed and predefined entity types and relationships, enabling a wider range of queries compared to Semantic Web-based approaches.

We started by presenting a formal definition of E-R queries where we assume that a E-R query can be decomposed as a sequence of sub-queries each containing keywords related to a specific entity or relationship. Then we adopted a probabilistic formulation of the E-R retrieval problem. When creating specific representations for entities (e.g. context terms) and for pairs of entities (i.e. relationships) it is possible to create a graph of probabilistic dependencies between sub-queries and entity plus relationship representations. We use a Bayesian network to depict these dependencies in a probabilistic graphical model. To the best of our knowledge this represents the first probabilistic model of E-R retrieval.  

However, these conditional probabilities cannot be computed directly from raw documents in a collection. In fact, this is a condition inherent to the problem of Entity Retrieval. Documents serve as proxies to entities and relationship representations and consequently, we need to fuse information spread across multiple documents to be able to create those representations. We proposed two design patterns, Early Fusion and Late Fusion, inspired from Model 1 and Model 2 of \citet{balog2006formal}. However, in the context of ORM, we are only interested in Early Fusion.

Early Fusion aggregates context terms of entity and relationship occurrences to create two dedicated indexes, the \textit{entity} index and the \textit{relationship} index. Once we have the two indexes it is possible to apply any retrieval method to compute the relevance scores of entity and relationship documents (i.e. representations) given the E-R sub-queries. The joint probability to retrieve the final entity tuples is computed using a factorization of the conditional probabilities, i.e., the individual relevance scores.

On the other hand, Late Fusion consists in matching the E-R sub-queries directly on a standard document index alongside a set of entity occurrence in each document. Once we compute the individual relevance scores of each document given a E-R sub-query, we then aggregate the entity occurrences of the top k results to compute the final joint probability. When using traditional retrieval models, such as Language Models or BM25, these design patterns can be used to create unsupervised baselines for E-R retrieval.

Since our objective was to explore an Early Fusion approach to E-R retrieval we developed a novel supervised Early Fusion-based model for E-R retrieval, the Entity-Relationship Dependence Model (ERDM). It uses Markov Random Field to model term dependencies of E-R sub-queries and entity/relationship documents. ERDM can be seen as an extension of the Sequential Dependence Model (SDM) \citep{metzler2005markov} for ad-hoc document retrieval in a way that it relies on query term dependencies but creates a more complex graph structure that connects terms of multiple (sub-)queries and multiple documents to compute the probability mass function under the MRF. 

One of the difficulties we faced while researching E-R retrieval was the lack of test collections. We therefore decided to contribute to this research problem by creating a semi-automatic method for creating test collections. We realized that web tabular data often include implicit relationships between entities that belong to the same row in a table. We developed a table parser that extracts tuples of related entities from Wikipedia Lists-of-lists-of-lists tables. We then extract metadata, such as table title or column name, and provide it to editors, together with the list of entity tuples. We asked editors to create E-R queries in which the list of entity tuples could serve as relevance judgments. This process resulted in the creation and publication of the RELink Query Collection comprising 600 E-R queries. We believe RELink QC will foster research work in E-R retrieval.

We performed experiments at scale using the ClueWeb-09B Web corpus from which we extracted and indexed more than 850 million entity and relationship occurrences. We evaluated our methods using four different query sets comprising a total of 548 E-R queries. As far as we know, this is the largest experiment in E-R retrieval, considering the size of the query set and the data collection. Results show consistently better performance of the ERDM model over three proposed baselines. When comparing Language Models and BM25 as feature functions we observed variance on the performance depending on the query set. Furthermore, using unsupervised Early Fusion proved to be very competitive when compared to ERDM, suggesting that it can be used in some application scenarios where the overhead of computing sequential dependencies might be unfeasible.

\subsubsection{Entity Filtering and Sentiment Analysis}

Entity Filtering and Sentiment Analysis are two fundamental Text Mining problems in ORM. We participated in two well known external benchmark competitions in both tasks resulting in state-of-the-art performance. We made the following contributions to these two problems:

\begin{itemize}
\item  A supervised learning approach for Entity Filtering on tweets, achieving state-of-the-art performance using a relatively small training set. 

\item Created and made available word embeddings trained from financial texts.

\item A supervised learning approach for fine-grained sentiment analysis of financial texts.
\end{itemize}

Entity Filtering can be seen as targeted named entity disambiguation. We developed a supervised method that classifies tweets as relevant or non-relevant to a given target entity. This task is fundamental in ORM as downstream tasks, such as prediction, can be highly affected by noisy input data. We implemented a large set of features that can be generated to describe the relationship between a tweet mentioning a entity and a reference entity representation. 

We relied on metadata, such as entity categories, text represented with TF-IDF, similarity between tweets and Wikipedia entity articles, Freebase entities disambiguation, feature selection of terms based on frequency and feature matrix transformation using SVD. Although our approach can be perceived as relatively simple and low cost, we achieved first place with an Accuracy over 0.90 at the Filtering Task of RepLab 2013, in a test set containing more than 90 thousand tweets and 61 different target entities. 

Regarding Sentiment Analysis, we decided to focus our efforts in a not so well explored sub-area, namely financial texts. We participated in SemEval 2017 Task 5 which focused on fine-grained sentiment analysis of financial news and microblogs. The task consisted in predicting a real continuous variable from -1.0 to +1.0 representing the polarity and intensity of sentiment concerning companies/stocks mentioned in short texts. We modeled it as a regression analysis problem. 

Previous work in this domain showed that financial sentiment is often depicted in an implicit way. We created financial-specific word embeddings in order to obtain domain specific syntactic and semantic relations between words in this context. We combined traditional bag-of-words, lexical-based features and bag-of-embeddings to train a regressor of both sentiment and intensity. Results showed that different combination of features attained different performances on each sub-tasks. Nevertheless, we were able to obtain cosine similarities above 0.65 in both sub-tasks and mean average errors below 0.2 in a scale range of 2.0, representing less than 10\% of the maximum possible error.

\subsubsection{Text-based Entity-centric Prediction}

We explored two text-based prediction problems in the context of ORM, performing analysis of the predictive power of entity-centric information on the news to predict entity popularity on Twitter, as well, as a study of sentiment aggregate functions to predict political opinion. We made the following contribution in this research area:

\begin{itemize}
\item Analysis of the predictive power of online news regarding entity popularity on Twitter for entities that are frequently mentioned on the news.

\item Analysis of how to combine different sentiment aggregate functions to serve as features for predicting political polls.
\end{itemize}

We are aware that entity popularity on social media can be influenced by endogenous and exogenous factors but we are only interested in exploring the interplay between online news and social media reactions. This could be useful for anticipating public relations damage control or even for editorial purposes to maximize attention and consequently revenue. We explored different sets of signal extracted from online news mentioning entities that are frequently mentioned on the news such as politicians of footballers. These signals could influence or at least are correlated with future popularity of those entities on Twitter.

Results show that performance varies depending on the target entity. In general, results are better in the case of predicting popularity of politicians, due to the high unpredictability of live events associated with sports. This is a general conclusion of this study as online news do not have predictive power for live events as Twitter reactions happen quickly than the publication of the news for such cases. Results also show that the time of prediction affects the performance of the models. For instance, in the case of politicians F1 score is higher when time of prediction occurs after lunch time, which is an evidence that in politics most of the news events that trigger social media reactions are reported in the morning news. 

The second predictive studied we carried out consisted in using entity-centric sentiment polarity extracted from tweets to predict political polls. There is no consensus on previous research work on what sentiment aggregate functions is more adequate to predict political results. We explored several sentiment aggregate functions described in the literature to assess which one or combination would be more effective on predicting polls during the Portuguese bailout (2011-2013). In our study, we achieved the lowest mean average error using a combination of buzz aggregation functions to predict monthly poll variations instead of absolute values. On the other hand, the most important individual feature was an aggregate function consisting on the logarithm of the ration positive and negative classified tweets.

\subsubsection{A Framework for ORM}

We also created a framework specifically tailored for ORM that puts together the sub-tasks we tackled throughout this thesis. We believe this framework represents a significant contribution and paves the way to future research in the computational problems inherent to the process of monitoring reputation online. More precisely we make the following contributions:

\begin{itemize}
\item A framework that supports research in Entity Retrieval and Text Mining tasks in the context of Online Reputation Monitoring. This framework is composed by two major components that can act as independent frameworks: RELink and TexRep.

\item The RELink framework that supports comprehensive research work in E-R retrieval, supporting the semi-automatic creating of test queries, as well as, Early Fusion based approaches for E-R retrieval.

\item The TexRep framework that is able to collect texts from online media, such as Twitter or online news, and identify entities of interest, classify sentiment polarity and intensity. The framework supports multiple data aggregation methods, as well as visualization and modeling techniques that can be used for both descriptive analytics, such as analyze how political polls evolve over time, and predictive analytics, such as predict elections.

\item A study of some practical aspects, namely vocabulary size, training data size and intrinsic evaluation for the training and publishing word embeddings from the Portuguese Twitter stream that can be later used for ORM related tasks.
\end{itemize}

The framework is divided in two distinct components, one is dedicated to Entity Retrieval and the other to Text Mining. In practice these two components can act as two separate frameworks. Both are adaptable and can be reused in different application scenarios, from computational journalism to finance or politics. RELink framework is designed to facilitate experiments with E-R Retrieval query collections. 
TexRep was designed with two main challenges in mind: 1) it should be able to cope with the Text Mining problems underlying ORM and 2) it should be flexible, adaptable and reusable in order to support the specificities of different application scenarios. We also presented two use cases of our framework for ORM. In the first we use RELink in the context of computational journalism while in the second we described the design and the implementation of the POPmine system, an use case of the proposed framework in the scope of the POPSTAR project. 

Furthermore, we presented a study of the practical aspects of learning word embeddings from the Twitter stream. Our goal was to try to assess the feasibility of producing and publishing general purpose word embeddings for ORM. Results showed that using less than 50\% of the available training examples for each vocabulary size might result in over-fitting. We obtained interesting performance on intrinsic evaluation when trained a vocabulary containing 32768 most frequent words in a Twitter sample of relatively small size. We proposed a set of gold standard data for intrinsic evaluation of word embeddings from user generated content. Nevertheless, we realized that evaluation metrics using absolute values as thresholds might not be suitable due to the cosine skewness effect on large dimensional embedding spaces. We propose to develop topological intrinsic evaluation metrics in future work.

\section{Limitations and Future Work}

One of the major obstacles we faced during the course of this thesis was the limited availability of labeled data for training and evaluation of the different tasks we tackled. This is a common limitation in the scope of Online Reputation Monitoring. Due to this obstacle we did not have the chance to perform extensive experimentation using more than one data source and language for each task. This aspect reduces the generalization of the results obtained since they might be biased towards the available datasets we had access to. Therefore, we leave for future work experimentation on each task with multiple datasets using different data sources and languages to perform comparable evaluations. 

We also recognize that we tried to address many different tasks which reduced our capability of addressing every task with the same level of depth. Nevertheless, we believe that exploring several new tasks in the scope of ORM constitutes a strong contribution to foster future research work in this area. During the course of this thesis, we did not have the possibility of performing user studies to assess the global usefulness of our framework for ORM. We would like to leave that as future work. 

While we had the objective of applying E-R retrieval in online news and social media which represent the natural data sources for ORM, it was not possible to evaluate our approaches using these type of data sources. Research work in E-R retrieval is still in its early stages and we believed it was necessary to first contribute to general E-R retrieval and leave for future work specific evaluation in the context of ORM. We implemented and created a demo of the Early Fusion approach since it is unsupervised. However, it was not possible to apply ERDM to online news due to the lack of training queries and relevance judgments for parameter tuning. In either cases, we aim to conduct an user experience in a near future to collect queries and relevance judgments in the context of ORM. 

Recent work in Deep Neural Networks makes the opportunity to beat the baselines we created in this thesis however, most of the tasks we addressed do not have enough labeled data to use these techniques. One of the most interesting avenues we would like to explore would be the use of neural networks as feature functions of the ERDM model. Since we have a dataset of more than 850 million entity and relationship extractions this represents an ideal scenario for Deep Learning. We propose to use a window based prediction task similar to the CBOW model for training word embeddings. Given a fixed window size, one would learn a neural network that would provide a ranked list of entities/relationships given an input query. We believe this approach would reduce the computational costs of the current ERDM feature functions since we would not need to keep two huge indexes at query time.

We would like also to explore different priors in entity and relationship documents within ERDM. For instance, creating source and time sensitive rankings would be useful when using transient information sources. Another promising avenue is transfer learning, specially due to the lack of training resources in the context of ORM. The possibility of bilingual training or cross-domain (e.g. politics to finance) transfer knowledge would constitute a major progress in this area.

\backmatter

\begin{spacing}{0.9}


\bibliographystyle{unsrtnat} 
\cleardoublepage
\bibliography{References/references} 



\end{spacing}

\printthesisindex 

\end{document}